École Polythecnique Federale de Lausanne
School of Life Sciences

Master Project in Life Sciences Engineering

# Evaluating Speaker Identity Coding in Self-supervised Models and Humans

---

Carried out in the Senseable Intelligence Group
at Harvard Medical School and MIT
Under the supervision of Satrajit S. Ghosh, PhD

---

Done by
**Gasser Elbanna**

Under the direction of
Antoine Bosselut, PhD
In the NLP Group at EPFL

This page intentionally left blank.

# Acknowledgements

First and foremost, I would like to thank my advisor Dr. Satrajit Ghosh for his relentless support and mentorship throughout my thesis that started with philosophical questions about voice identity and ended with empirical results and more philosophical questions :). I had the pleasure to discuss science with him and partake in thought-provoking conversations and to him I am immensely grateful.

I would also like to thank the Bertarelli foundation for funding my thesis and granting me this lifetime opportunity to carry out my MSc. thesis in Boston. I wish to thank Dr. Antoine Bosselut for supporting me on the other side of the Atlantic.

Special thanks for all the lab members of the Senseable Intelligence Group for the genuine companionship and all the fruitful feedback.

Last but not least, I would like to thank my parents and siblings for encouraging me to pursue my dream even if it was thousands of miles away from them. I am very grateful for their unconditional love and support and to them I dedicate this and future work.

# Contents

# List of Figures

# Abbreviations

**ANN** Artificial Neural Networks.

**ASpD** Automatic Speaker Discrimination.

**ASpR** Automatic Speaker Recognition.

**BIDS** Brain Imaging Data Structure.

**BOLD** Blood Oxygenation Level Dependent.

**BYOL-I** Bootstrap Your Own Latent for Identity.

**BYOL-A** Bootstrap Your Own Latent for Audio.

**BYOL-S** Bootstrap Your Own Latent for Speech.

**CIFTI** Connectivity Informatics Technology Initiative.

**CKA** Centered Kernel Alignment.

**CNN** Convolution Neural Network.

**CPD** Spearman Correlation between Pairwise Distances.

**CSF** Cerebral Spinal Fluid.

**CV** Cross-Validation.

**CvT** Convolution Vision Transformer.

**DCT** Discrete Cosine Transform.

**DL** Deep Learning.

**DTI** Diffusion Tensor Imaging.

**F0** Fundamental Frequency.

**FD** Formant Dispersion.

**fMRI** Functional Magnetic Resonance Imaging.

**gw-MRF** gradient-weighted Markov Random Field.

**HNR** Harmonics-to-Noise Ratio.

**HRF** Hemodynamic Response Function.

**HSIC** Hilbert-Schmidt Independence Criterion.

**IFC** Inferior Frontal Cortex.

**KNN** k-Nearest Neighbors.

**MLP** Multi-layer Perceptron.

**MS-HBM** Multi-Session Hierarchical Bayesian Model.

**PaCMAP** Pairwise Controlled Manifold Approximation and Projection.

**PCA** Priciple Component Analysis.

**ROI** Region of Interest.

**RSFC** Resting-State Functional Connectivity.

**SSM** Self-Supervised Model.

**STG/S** Superior Temporal Gyrus/Sulcus.

**SVC** Support Vector Classifier.

**T1w** T1-weighted Imaging.

**T2w** T2-weighted Imaging.

**TR** Time to Repeat.

**t-SNE** t-Distributed Stochastic Neighbor Embedding.

**TERA** Transformer Encoder Representations from Alteration.

**TVA** Temporal Voice Area.

**UMAP** Uniform Manifold Approximation and Projection.

**VC1** First version of VoxCeleb Dataset.

**VFs** Vocal Folds.

# Abstract

Speaker identity plays a significant role in human communication and is being increasingly used in societal applications, many through advances in machine learning. Speaker identity perception is an essential cognitive phenomenon that can be broadly reduced to two main tasks: recognizing a voice or discriminating between voices. Several studies have attempted to identify acoustic correlates of identity perception to pinpoint salient parameters for such a task. Unlike other communicative social signals, most efforts have yielded inefficacious conclusions. Furthermore, current neurocognitive models of voice identity processing consider the bases of perception as acoustic dimensions such as fundamental frequency, harmonics-to-noise ratio, and formant dispersion. However, these findings do not account for naturalistic speech and within-speaker variability. Representational spaces of current self-supervised models have shown significant performance in various speech-related tasks. In this work, we demonstrate that self-supervised representations from different families (e.g., generative, contrastive, and predictive models) are significantly better for speaker identification over acoustic representations. We also show that such a speaker identification task can be used to better understand the nature of acoustic information representation in different layers of these powerful networks. By evaluating speaker identification accuracy across acoustic, phonemic, prosodic, and linguistic variants, we report similarity between model performance and human identity perception. We further examine these similarities by juxtaposing the encoding spaces of models and humans and challenging the use of distance metrics as a proxy for speaker proximity. Lastly, we show that some models can predict brain responses in Auditory and Language regions during naturalistic stimuli. These empirical findings provide both enhanced interpretability to these representational spaces and also support using this family of models as candidates to study speaker identity perception in humans.

# Chapter 1

# Introduction

## 1.1 Speech Processing

The human voice is a ubiquitous yet mystifying signal that plays a crucial role in our daily interactions. This signal is an outcome of an intricate motor action which most of us execute dexterously. Furthermore, the scientific investigation of voices and, in particular speech, is guided by two fundamental processes [1], as illustrated in Figure 1.1. First, **Speech Production** requires orchestrating a number of physiological structures with precise spatial and temporal co-ordination. The sound production from this physical apparatus can be modeled as a three-fold dynamical system, often referred to as the **Power-Source-Filter** model. The power component, in this context, refers to the lungs that creates a pressure difference generating an airflow through the vocal folds (VFs) or through other upstream constructions. Air flows through the VFs causes them to oscillate with a certain frequency and creates a waveform called the *glottal pulse* making the VFs ipso facto the source component in this system. Upstream constrictions can give rise to other sound sources as well. The vocal tract (filter) acts as a resonance chamber that determines the spectral shape of the generated sound and, in most cases, its phonemic content. Together these components operate in synergy yielding a rich and dynamic time-varying pressure wave. Second, **Speech Perception** is the process by which the signal is sensed and subsequently decoded and analyzed. On the perception side, a hypothesis is that the acoustic signal is transformed into a perceptual space, which allows us to discern information about the speaker such as demographics (e.g. age, sex, dialect,..etc.), linguistic information (e.g. language spoken and words uttered), paralinguistic information (e.g. emotional, mental and physical state), social attributes (e.g. attractiveness, trustworthiness and competence) as well as the identity of the speaker [2–11]. Throughout the speech chain, from production to perception, the signal transforms across multiple related representations (e.g. motor, acoustic, neural, and perceptual) with the goal of carrying out daily communications.

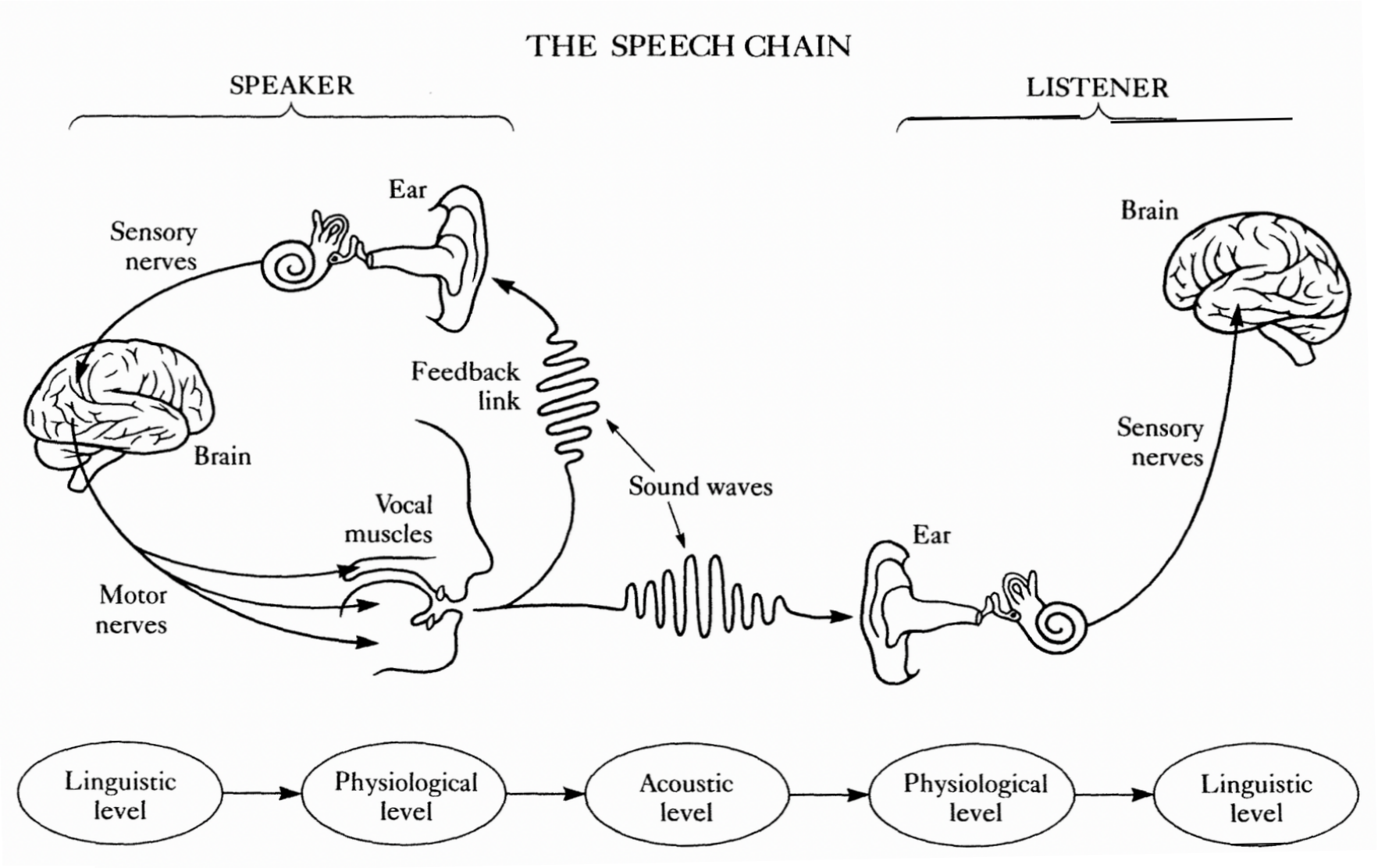

Figure 1.1: The Speech Chain (Figure taken from [12]).

## 1.2 Speaker Identity Perception

Identifying an individual from their voice or vocal patterns is an inherent biological ability that has developed across several animal species [13]. Research in evolutionary biology reports that personal identity signaling via vocal patterns is evident in non-human primates, amphibians, birds, and mammals [14–16]. For instance, songbirds and territorial frogs can discriminate companions from strangers by their voices [17, 18]. Also, the male American bullfrog *"Rana catesbeiana"* responds less aggressively to familiar voices compared to unfamiliar ones [19]. Furthermore, additional evidence suggests that the faculty of voice recognition is evident early within a species. Human newborns have shown their ability to identify their mothers' voices prenatally as well as after birth [20–23]. Additionally, maternal bats leverage genetically coded vocal signatures to recognize their offsprings [24]. These findings imply that auditory recognition evolved early and can be traced back 250 million years ago, predating language development. This ability has been ramified phylogenetically and ontogenetically to subserve the essence of social communication.

In the realm of voice perception, the term *voice recognition* has been associated with mul-tiple connotations. Several studies have referred to the task of training subjects to identify unfamiliar voices as a recognition task [25–27]. These studies conflated discrimination and recognition tasks as being modulated by a single cognitive ability. Nevertheless, the seminal work by Van Lancker and Kreiman on *Phonagnosia* [28], a neurological deficit characterized by impaired ability to recognize voices [29], has mapped out and fine-tuned our fundamental understanding of speaker identity perception and the associated cognitive processes. They observed that brain lesions might have differential effect on speaker recognition and speaker discrimination meaning that one ability can be impaired without necessarily affecting the other. Moreover, they found that the ability to recognize a voice remains intact even if the ability to discriminate between unfamiliar voices is compromised. These findings suggest that speaker identity processing can be considered as two separate and unordered cognitive tasks: recognizing a familiar voice and discriminating between unfamiliar voices [28–30]. Both tasks are vital parts in our daily interactions such as: recognizing a speaker on the telephone, or discriminating between voices in an online video meeting.

Although every voice has its own unique anatomic individuality [31], humans can still misidentify speakers due to intrinsic or extrinsic variations in the perceived speech signal. There are several culprits that might contribute to such variability in performance. One major factor is familiarity with the speaker's voice. From a neurological standpoint, as discussed above, the impact of familiarity has been demonstrated to be a functional demarcation between recognition and discrimination abilities. Acknowledging this familiarity-based partitioning dimension leads us to consider the differences in perceptual processing modes for familiar and unfamiliar voices. For example, in face perception research, it has been shown that unfamiliar faces are processed in a more attributal and featural fashion while familiar faces are processed as holistic patterns instead of as bundles of features [32–35]. In a similar vein, as shown in Figure 1.2, one can imagine similar modes for voice processing for two main reasons. First, several studies showed that recognizing a familiar voice relies on different set of attributes depending on the individual [36], and see [31] p. 179. Additionally, altering some acoustic patterns (e.g. speech rate or backward speech) tends to show speaker idiosyncratic effects in the context of recognition [27, 37, 38]. Second, it might be reasonable to not consult an exhaustive list of acoustic attributes to recognize a familiar vocal pattern whereas more acoustic detail is needed to discriminate between unfamiliar voices. Further evidence supporting featural and configural processing can be deduced from the language-familiarity effect [39]. This phenomenon implies that listeners are better at recognizing voices in their native language. Interestingly, the effect size was shown to be significantly larger in recognition tasks compared to discrimination. This means that changing the spoken language impaired the ability of listeners to recognize a familiar voice whereas this was not the case for discriminating between unfamiliar voices. This might highlight the vulnerability of the holistic-configurational mode for familiar voices to alterations in the linguistic patterns while the low-level featural mode for unfamiliar voices seems to be invariant to such high-level changes. All of these findings again converge to the neurological studies [28] stating that recognizing familiar voices and discriminating unfamiliar voices are different yet related cognitive tasks. We have also seen behavioral and perceptual evidence from voice sorting tasks that unfamiliar listeners tend to perceive significantly more identities than familiar ones [40–42]. As these behavioral findings might indicate the contribution of familiarity to identity coding, they also highlight the difficulty of telling people *together* compared to telling people *apart*. The faculty to perform the latter task is modulated by between-speaker variability, whereas the former is modulated by within-speaker variability which is disrupted in case of unfamiliar listeners [43].

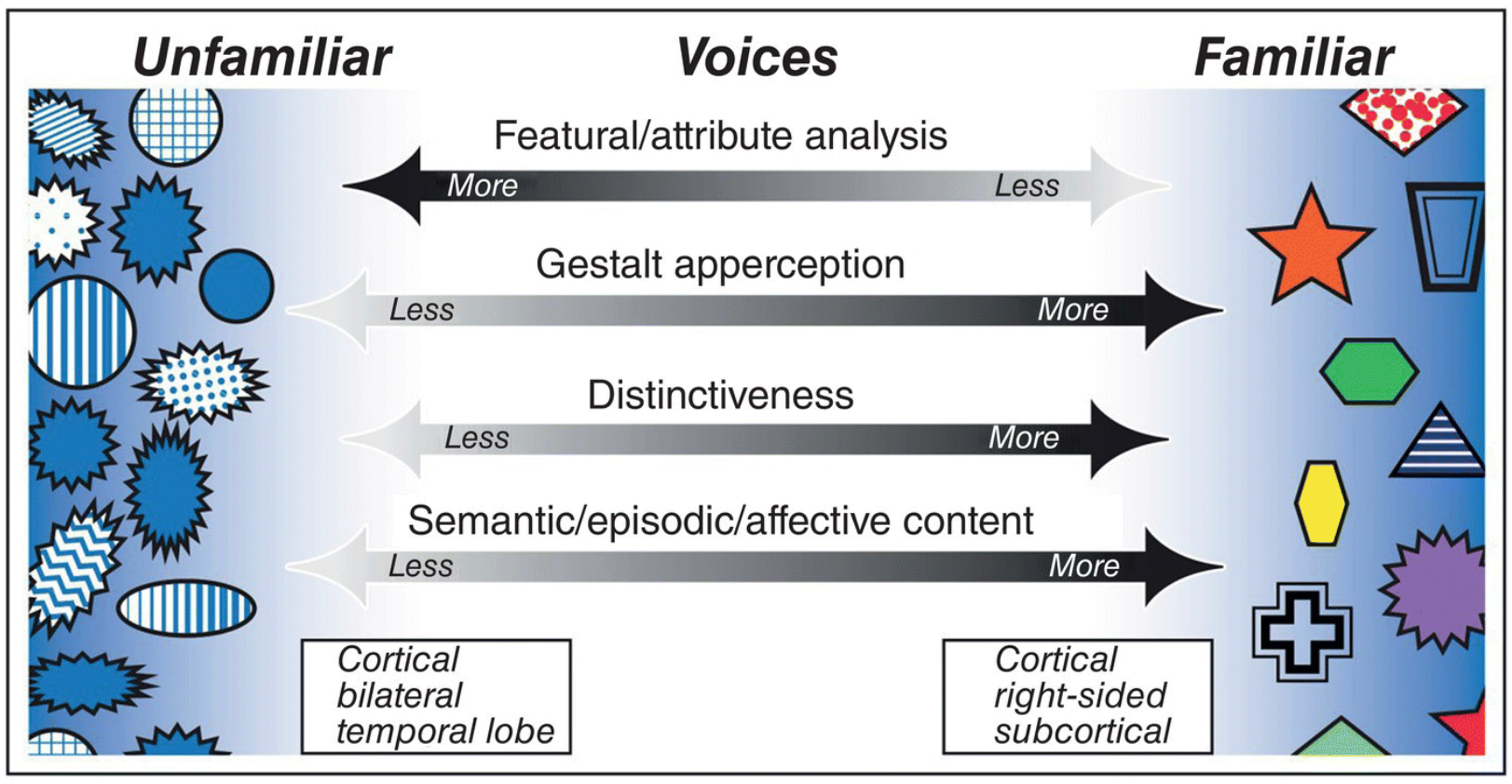

Figure 1.2: Featural versus Configural Voice Processing (Figure taken from [44]).

Within-speaker variability can be considered as a second factor, besides familiarity, that impacts speaker identity processing. The importance of within-subject variability has been accentuated in face perception literature [45, 46], followed by recent impetus in voice perception studies to acknowledge the criticality of intra-speaker variability on our understanding to speaker identity perception [43]. Within-speaker variability is evident in our daily settings as we accommodate and adapt our voices as a response to the surrounding environments and situations [47]. This allows us to modulate our voices and convey emotions, thoughts, and intentions. These modulations can be volitional and indeed affect speaker identity coding such as child-directed or pet-directed speech [48], elderly-directed speech [49], whispered speech [50], singing [51–53], vocal disguises [54, 55] and even variations in non-speech vocalizations such as laughter [56]. Moreover, it has been shown that incongruency in speaking styles might impair identity processing such as reading aloud versus spontaneous speaking [57–59]. We also can visually notice the differences in speaking patterns as illustrated in Figure 1.3. On the other hand, there are involuntary modulations in voices that occur due to aging. Children tend to have high F0 compared to adults which starts changing for some during puberty [60, 61]. Voice changes might also occur depending on the physical health [6] and the mental health [62] status.

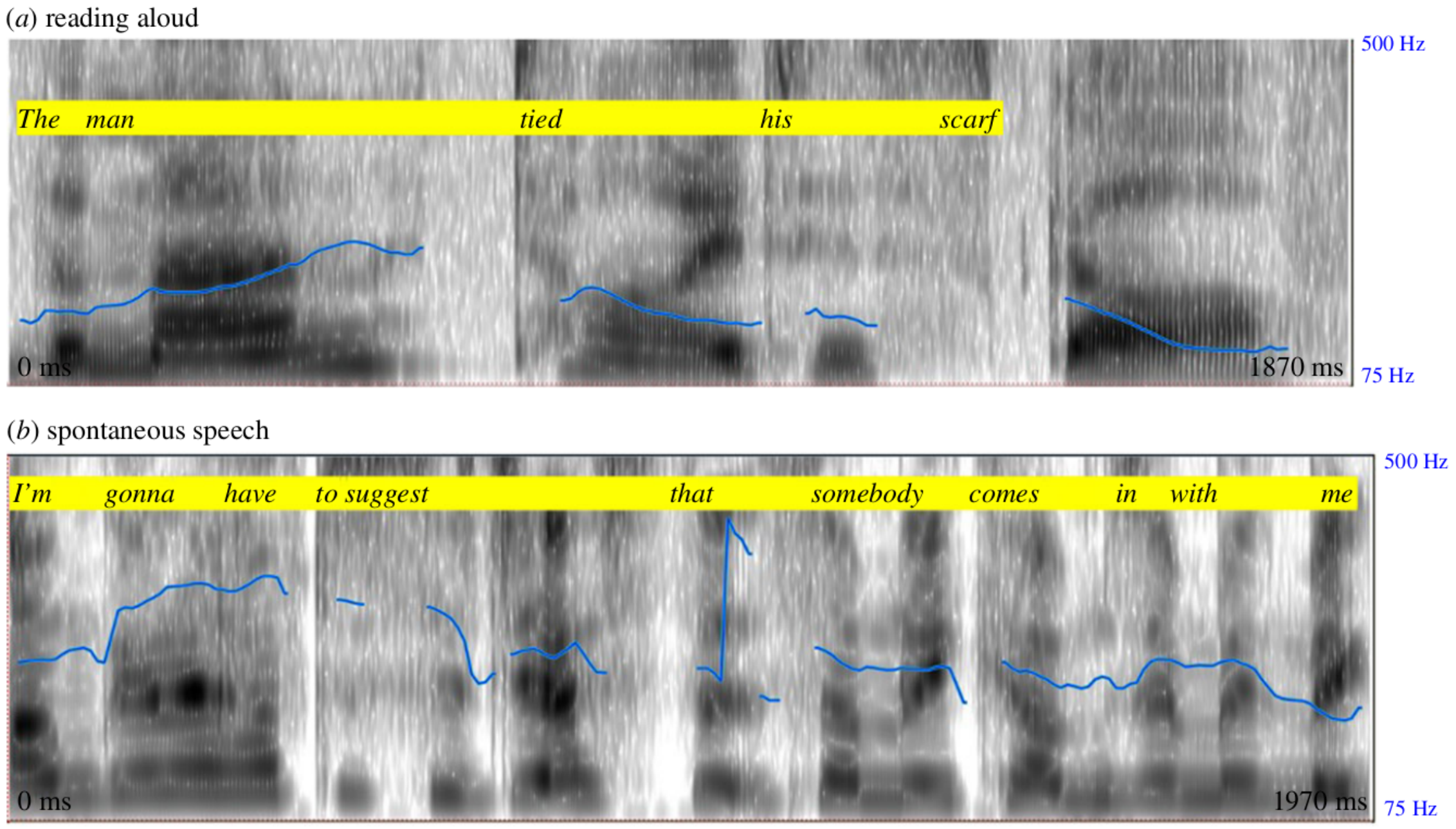

Figure 1.3: (a) Spectrogram of an adult reading a scripted sentence. (b) Spectrogram of the same talker in a spontaneous conversation (Figure taken from [63]).

## 1.3 Why study speaker identity perception?

Studying identity processing and the perceptual strategies incorporated in this phenomenon is crucial to advancing our understanding of voice-identity coding in the brain. Additionally, there has been an impetus towards studying voice and speech as biomarkers for physiological disorders such as respiratory-related failures [64], cardiac-related failures [65], and mental disorders (see review in [62]), heralding a plethora of voice datasets of healthy subjects and patients. While such work leverages the ubiquitous signal of voice as a window into mental and physical states, public availability of such corpora may jeopardize the speakers' identity.

In the digital era of recorded virtual meetings and social media platforms, the task of identifying a voice and accessing their medical records using these corpora metadata has become a much less effortful task. To that end, studying speaker identity perception to delineate the identity signal in the voice allows us to build better anonymization algorithms. The main challenge is to suppress the identity component of speech yet preserve the linguistic and paralinguistic information such the anonymized speech corpus can be publicly shared with much reduced potential for re-identification of the speakers. Recent challenges have led to novel algorithms for speaker anonymization [66].

Another line of research that might benefit from this work is audio forensics. There has been growing evidence suggesting that listeners exposed to incongruent speaking styles such as read and spontaneous speech might impair speaker identity processing [57–59]. These studies discuss legal and forensic implications for the reliability of ear witness testimony in court. Consequently, studying the degree of variability in one's voice and its effect on identity processing are important to the current audio forensic regulations and settings.

Moreover, studying identity perception might be beneficial for improving the current state-of-the-art models for automatic speaker recognition (ASpR) [67]. One sensitive and important use-case for ASpR models is voice biometrics and we have recently witnessed unexplained gender and racial disparities [68] similar to the disparities observed in facial recognition softwares [69]. Consequently, a thorough experimentation on the encoding spaces of the current models is needed (see blog[1].

Lately, a text-to-speech synthesis model has been proposed called Vall-E [70]. This model purports to transfer vocal identity characteristics into synthesized speech from roughly 3-second enrolled utterance. These data-driven approaches are featuring impressive results without a clear understanding of their encoding spaces. Such ambiguity in this technology might have ethical implications. Thus, understanding speaker identity coding might help us propose more hypothesis-driven approaches to better disentangle identity signal from other aspects of speech.

## 1.4 Neurocognitive Models of Speaker Identity Coding

We have juxtaposed multiple times, in this discussion, voice perception research with face perception due to the plethora of work investigating the commonalities between both domains

[1] https://sensein.group/2022/09/26/discrimination_in_artificial_intelligence_for_voice_applications.html

[71–73]. As a result neurocognitive frameworks for voice identity processing have been in contrast to face identity models. A recent neurocognitive model for voice identity has been outlined through the lens of lesion studies (phonagnosia) and compared to face representations processing [74]. Figure 1.4 shows the stages of voice processing suggesting a parallel yet interacting pipelines to process three main components of a vocal sound; affect, speech, and identity. Regarding identity processing, vocal sounds undergo a general processing step that might separate and analyze different types of sounds (e.g. speech, music, environmental sounds). Then, the first stage includes structural encoding of voices [75] in which invariant acoustic features are extracted from the signal creating a perceptual construct of the voice. They hypothesize that the yielded construct is represented in a space relative to an average/prototype voice (see Section 1.5 for details). The potential brain candidates associated with this stage are assumed to be predominantly in the right hemisphere of the posterior and mid regions of superior temporal gyrus/sulcus (STG/S) [76–80] as well as auditory regions such as Heschl's gyrus [81, 82] and planum temporale [78, 83]. The neurological deficit that might emerge in this stage is hypothesized to be *Apperceptive Phonagnosia* which is characterized by the inability to construct and extract voice representations. Subsequently, the second stage involves matching the input voice, if familiar to the listener, to a stored vocal patterns for recognition with potential brain candidates in the anterior and mid regions of STG/S and aspects of the anterior temporal lobe [84, 85]. It is proposed that a damage in this stage would be called *Familiarity-Associative Phonagnosia* which impairs the ability to recognize voices. The final stage of this model describes the multi-modal processing of several representations to extract semantic information regarding the person's identity and a deficit in this stage might imply *Semantic-Associative Phonagnosia*. It is important to note that the aforementioned deficits were suggested in their model as a correspondence to visual processing stages and deficits *"Prosopagnosia"*.

This framework is an extension to the seminal functional face and voice perception model proposed by [7], shown in Figure 1.5. These models proposed a sequential multi-stage framework for identity processing starting with a structural encoding phase for representing voices in a perceptual space followed by a pattern recognition task. Based on this model, one might argue that the discrimination task might be a sub-element within the recognition pipeline. Additionally, any damage in an early stage might affect the subsequent stages. However, that contradicts the evidence discussed in Section 1.2 regarding the unordered and separated cognitive abilities [28].

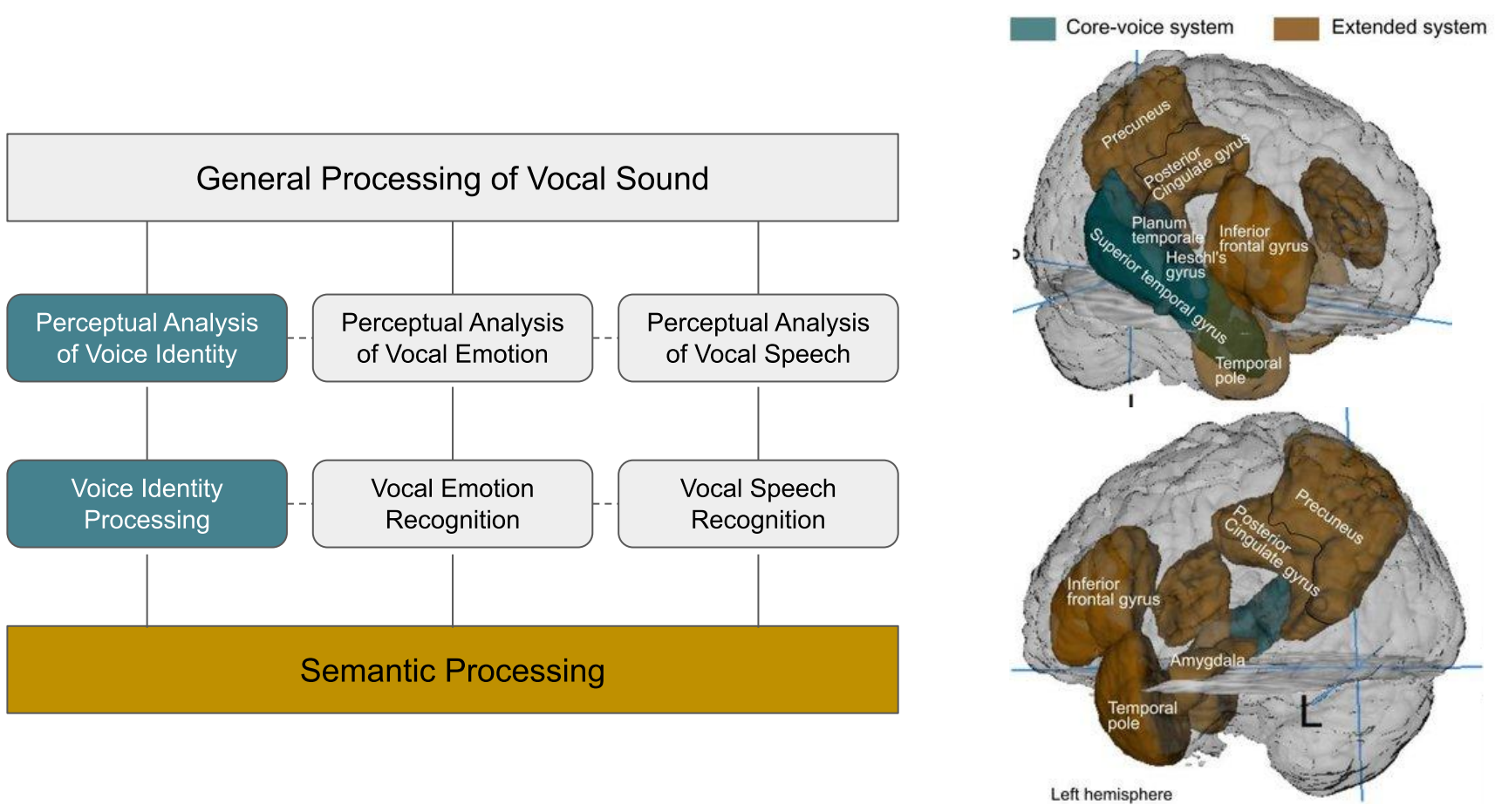

Figure 1.4: Neurocognitive framework for speaker identity coding. This figure is adapted from [74].

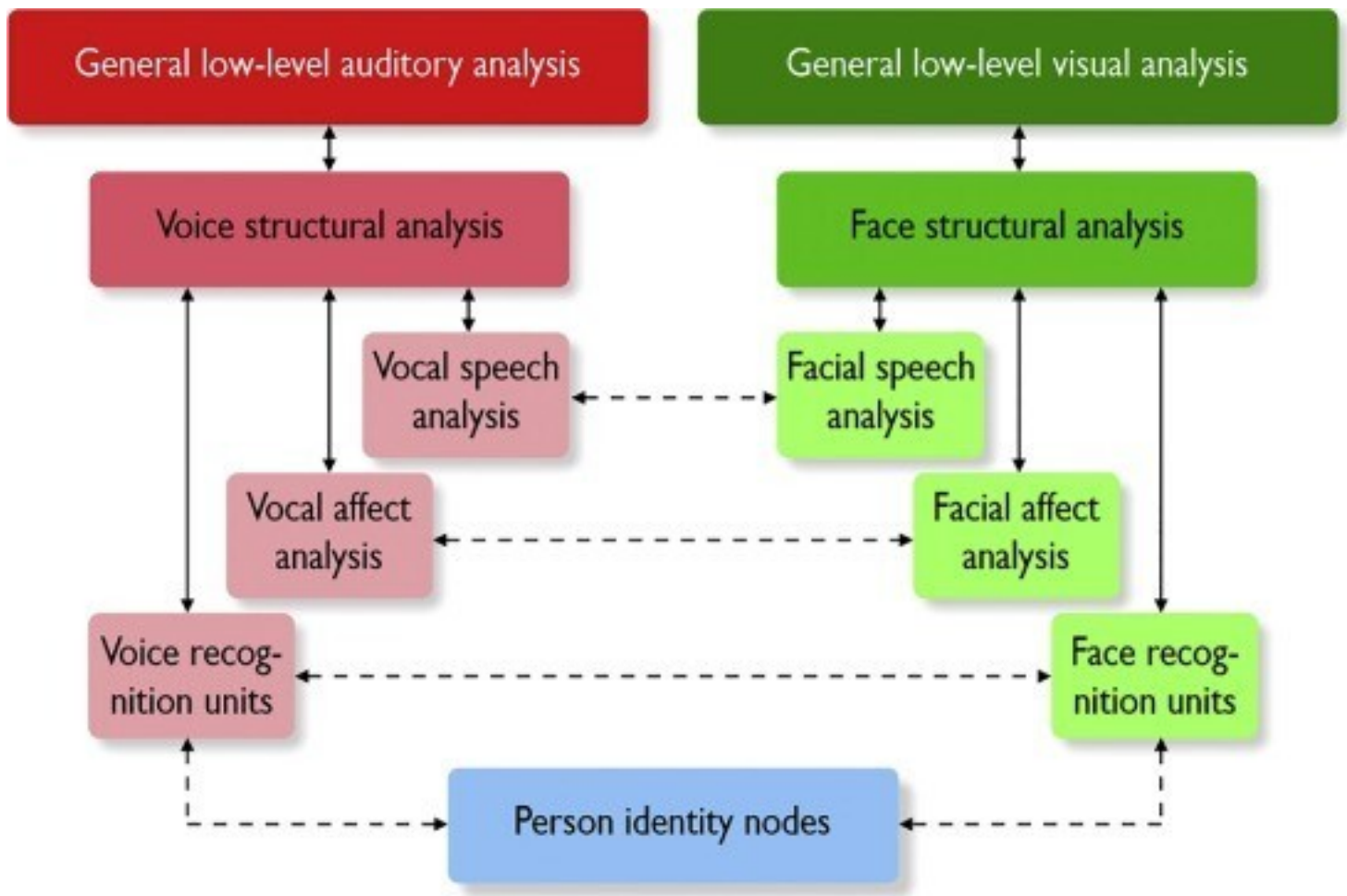

Figure 1.5: Functional Model of Face and Voice Perception. This figure is taken from [86] which is adapted from [7].

Furthermore, a more recent model was proposed by [86], as shown in Figure 1.6, which revised the previous models and highlighted the differences between voice and face perception. The model still accounts for some hierarchical dependency between voice discrimination and recognition similar to the previous one [7]. However, they also included modality-specific processing in addition to the conventional parallel functional organization. They highlighted the uniqueness of both modalities (faces and voices) and the need for considering separate processing pipelines. This modification was based on neurological evidence suggesting that *Phonagnosia* and *Prosopagnosia* are modality-specific deficits impairing only voice and face recognition respectively [3, 87, 88].

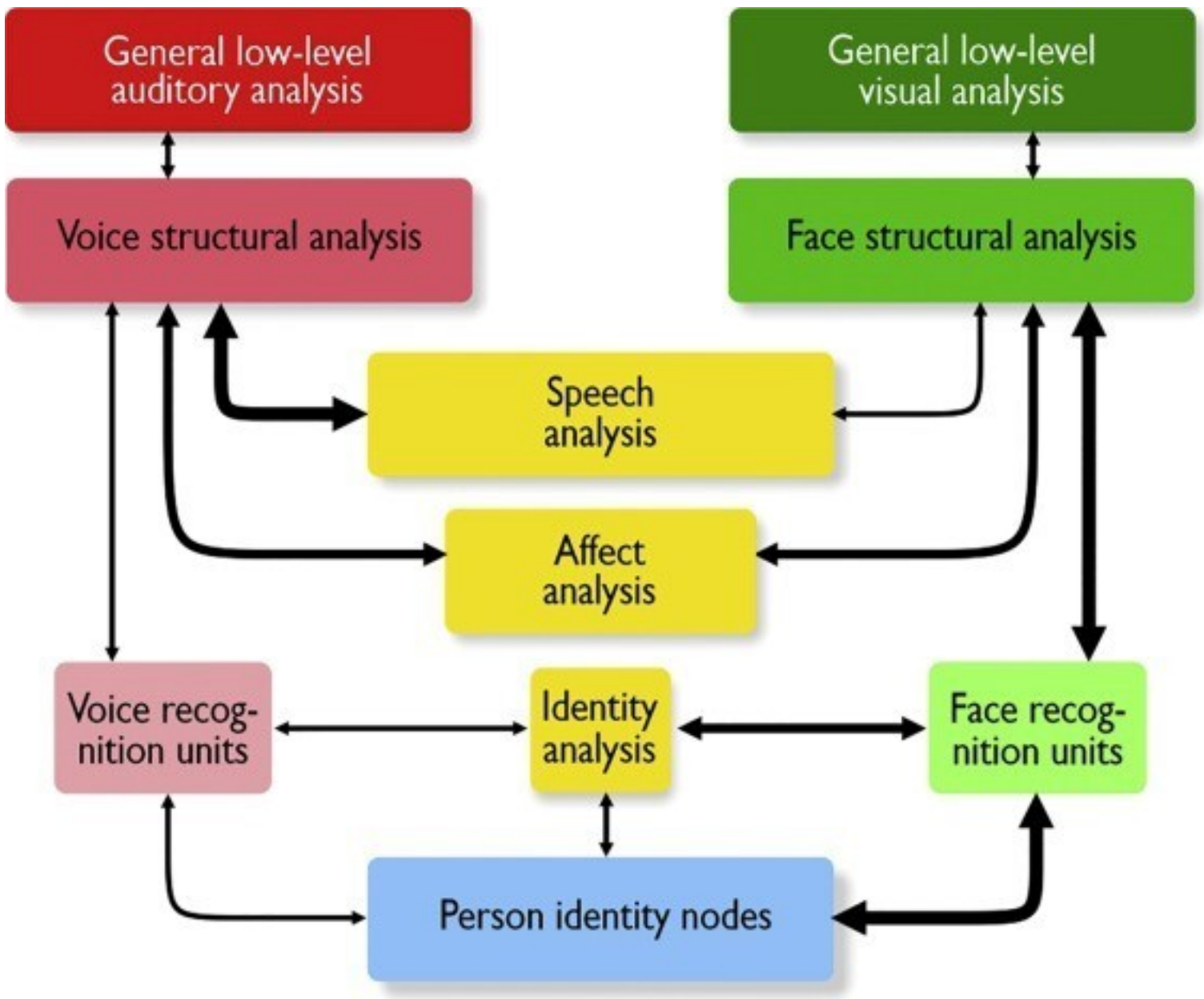

Figure 1.6: Modified Functional Model of Face and Voice Perception. This figure is taken from [86].

## 1.5 Modeling Speaker Identity Coding

Several studies have attempted to identify acoustic correlates of identity perception to pinpoint salient parameters for such a task. Unlike other communicative social signals, most efforts have yielded inefficacious conclusions [89]. Thus, the underlying coding mechanisms for extracting identity from voice is still unclear. Nevertheless, there has been a proposed hypothesis claiming that voice identity representations are encoded in temporal voice areas (TVA) as a function of acoustical distance relative to a prototype/average voice indicating that the identity space is following a norm-based coding mechanism [8, 90–94]. Furthermore, Latinus et al. suggested that the bases of identity perception are acoustic dimensions such as fundamental frequency (F0), harmonics-to-noise ratio (HNR), and formant dispersion (FD) [95]. Together these three features make up a perceptual space in which each speakers' identity is encoded separately and the average of all voices in the space define the prototype/average voice of this population. In their experiment [95], they recorded 64 speakers (32F/32M) uttering the syllable *"had"*. Then, they extracted several acoustic features from the utterances and selected the three aforementioned features (F0, FD, and HNR) as candidate dimensions for the perceptual space after running principle component analysis (PCA) that resulted in HNR, FD, and F0 explaining around $20.7\%$, $16.1\%$, and $11.3\%$ of the variance, respectively. They averaged the recorded voices across the three features using voice morphing for each gender creating a gender-based voice prototype, as shown in Figure 1.7. Subsequently, they computed the euclidean distance of voice samples to the average voice *"distance-to-mean"* and they showed cerebral correlates with the computed distance-to-mean implying that brain activity in the mid-STG/S in both hemispheres correlates with the acoustical distances. These findings support the candidacy of acoustic space as a model for the perceptual space. Further studies have shown behavioral correlates supporting the prototype-based acoustic model [8, 93, 96].

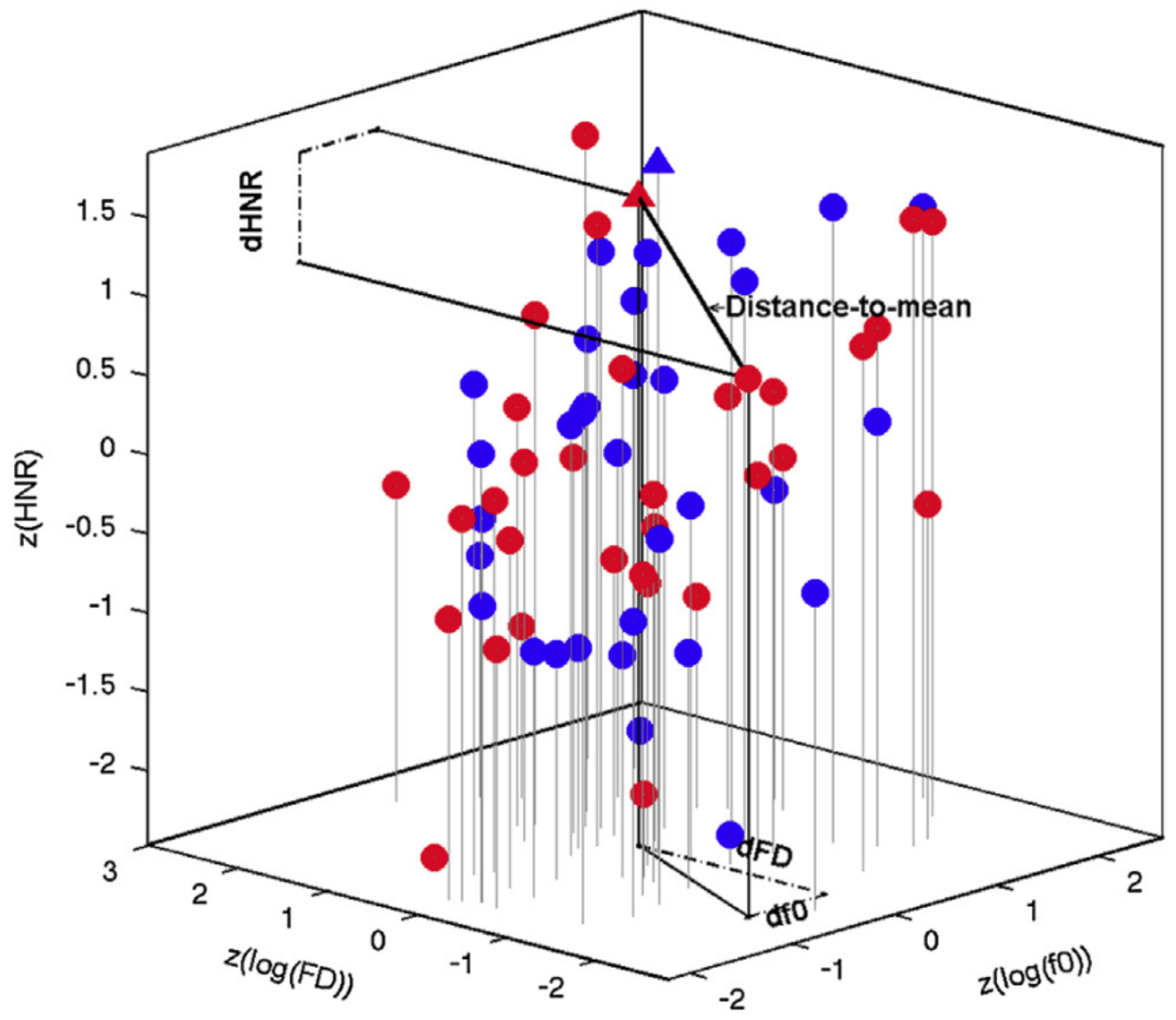

Figure 1.7: Speakers' utterances (*red*:females, *blue*:males) encoded in a 3D acoustic space along with a gender-based voice prototype (*triangles*). This figure is taken from [95].

It is worth-noting that these studies share multiple limitations. First, most of these studies did not account for naturalistic speech (i.e., they only used vowels or syllable-like words). Thus, the proposed acoustic-based perceptual models can't explain the language-familiarity effect in speaker recognition [39] which focuses on high-level changes in linguistic and prosodic components of speech while maintaining same vocal signature (i.e. same speaker). Another limitation is that most studies considered single sample per speaker discarding the effect of within-speaker variability. Although recent work has studied within-speaker variability in the context of norm-based coding, yet the sample size used was very small (N=3) to generalize the findings to large populations [97]. In conclusion, we argue that current hypotheses and acoustic-based models can't outline or fully capture the underlying non-linearities of the perceptual system in addition to speaker identity representational manifold. These non-linearities might be a result of the adaptive and dynamical nature of the perceptual space making it infeasible to create a direct mapping between fixed acoustic spaces and perceptual space.

Accordingly, we might need to consider more complex models and data-driven spaces for speaker identity processing that may explain identity coding in humans. We have witnessed a surge in deep learning (DL) models to solve sophisticated tasks in several domains such as audio, computer vision and language. Recently, a specific family of DL models, self-supervised models (SSMs), has generated robust speech representations without access to training labels. SSMs demonstrated high performance on several audio-related tasks (e.g. speech recognition, music genre classification, emotion recognition) [98].

## 1.6 Current Project

In this project, we propose to study data-driven representations, in particular SSMs, as systems that encode voice-related information and relate their performance and representational space to human perception. In addition, the interpretability of what kind of physiological or linguistic information is contained in these encoding spaces remains a challenge. We hypothesize that such data-driven encoding spaces might provide better precision for speaker identity perception compared to currently used acoustic feature spaces. We further posit that a systematic analysis across such encoding models will also provide insight into the information embedded in these representational spaces.

To that end, in Chapter 2, we demonstrate the predictive power of SSMs on ASpR tasks in contrast to acoustic representations. We also highlight the equivariances and limitations of identity representation using SSMs through a set of experiments that manipulate specific aspects of speech to demonstrate their impact on preserving identity and capture speaker-specific representations. In Chapter 3, we examine the notion of studying distances across encoded voices and its correlation to human perceptual space. We investigate the validity of the distance metrics used as a proxy for voice proximity through conducting a voice discrimination experiment. Finally, in Chapter 4, we take a step further and map the embeddings from different models to brain responses by fitting an encoding model to predict brain signal captured during a naturalistic fMRI study.

# Chapter 2

# Speaker Identity Coding in SSMs

## 2.1 Introduction

The first chapter of the project is aimed to assess the suitability of SSMs for the purpose of speaker identification. In order to achieve this aim, we address the following research questions to carry out this assessment:

- Are self-supervised models good candidates to study speaker identity coding?
- What aspects of speech do self-supervised models encode?
- What are the models' invariances and equivariances when recognizing a speaker?

This assessment is conducted via a three-fold approach. First, we evaluate the performance of these models on a large-scale ASpR task. This allows us to determine the extent to which SSMs encode speaker-related information in contrast to low-level handcrafted/acoustic features, refer to Section 2.2. To thoroughly examine a model's performance, in Section 2.3, we evaluate intermediate representations across layers to determine the model's optimal layer for speaker identification task. Subsequently, we used Centered Kernel Alignment (CKA) in Section 2.4 to quantify and examine the similarities across model representations. Lastly, we examined the models' equivariances and invariances through a set of experiments compiled in Section 2.5. Each experiment studies the impact of altering the following elements on the ASpR performance of each model on an ASpR task: phonemic structure, dynamics of fundamental frequency, linguistic and para-linguistic information, speaking style, and background noise. These alterations provide acoustic insight into the representational characteristics in these complex non-linear models. Moreover, we visualize the models' representations in a compressed low-dimensional space using dimensionality reduction methods. That way, we highlight different acoustic and demographic labels for all input utterances to compare the encoded information in all representational spaces.

Figure 2.1 lists the candidate representations studied in this work. Both, handcrafted (acoustic) and data-driven (SSMs) are used (for details see Section 5.2). We selected a diverse set of SSMs as candidate models for speaker identity perception. We explored representations from different families of self-supervised models. Also, we included acoustic and low-level features as examples for handcrafted representations and a baseline for our analysis. All models incorporated in this work are mentioned in Table 5.1, along with their embeddings dimensions, pretraining datasets, and Python package used for implementation.

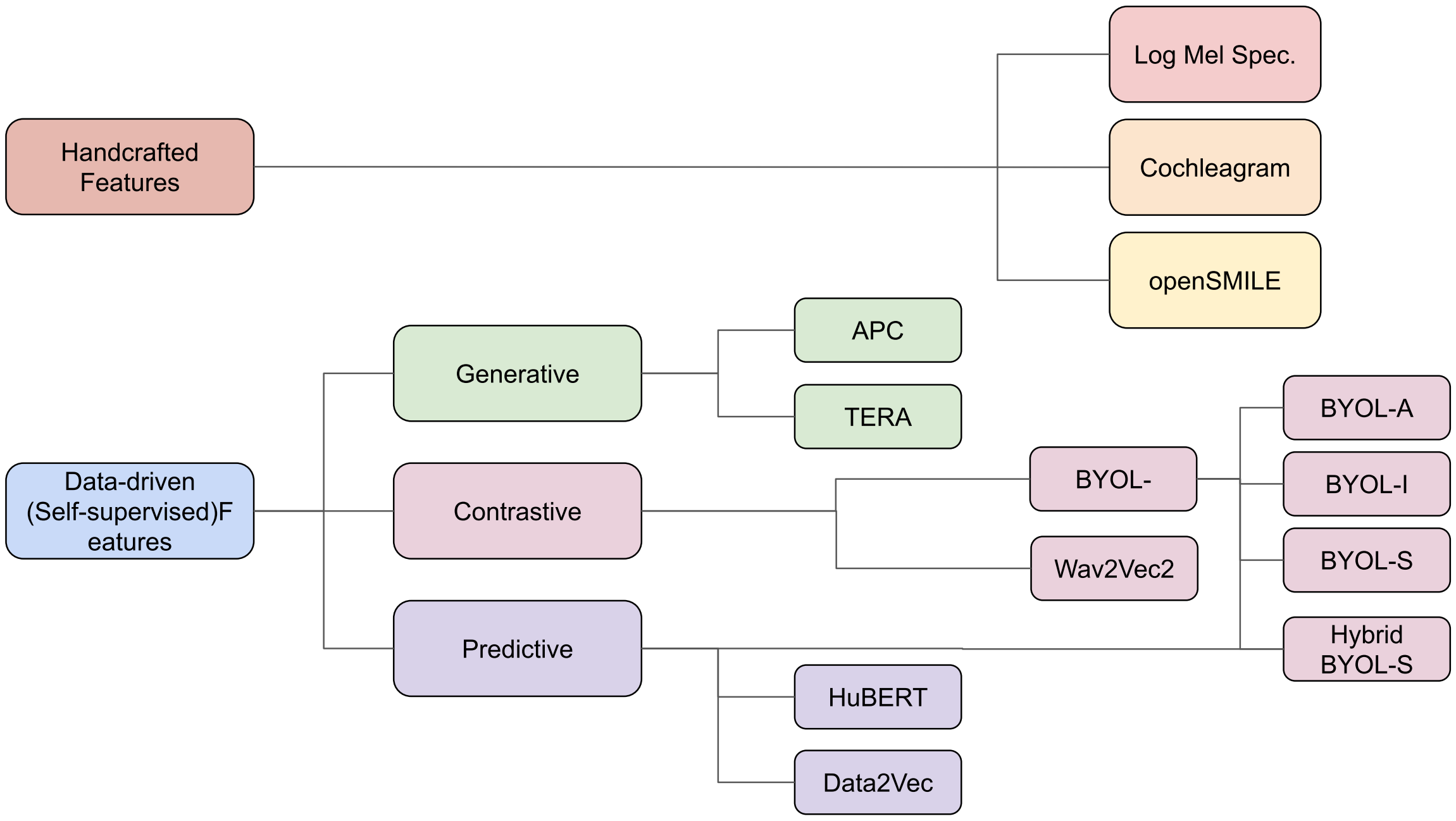

Figure 2.1: List of candidate models and representations.

## 2.2 ASpR Benchmark

To evaluate the SSMs' predictive power compared to handcrafted features, we benchmarked their performance on a large-scale ASpR task using VoxCeleb 1 (VC1) corpus [99].

### 2.2.1 Methods

VC1 comprises $153,514$ utterances from $1,251$ speakers/celebrities with average duration of $8.2$s. These utterances were converted to single-channel audio stream with sampling rate of $16$ kHz. We opted for this corpus as our benchmark because of the considerable intrinsic and extrinsic variability in the audio signal. For instance, for each speaker/celebrity, the utterances are sourced from interviews conducted in various environments (e.g. street, studio,..etc.). Consequently, this benchmark will evaluate the goodness of all candidate models to capture speaker vocal characteristics that are transferable across diverse background environments.

For the benchmark, the dataset is divided into three splits; train, validation and test sets. We followed the splits provided on their website [1] in which we have $138,361$, $6,904$ and $8,251$ utterances, respectively.

The benchmark evaluation is adapted from [100] with their code implementation[2] for reproducibility. We modified the code to follow our feature extraction implementation. The evaluation pipeline, as shown in Figure 2.2 includes extracting the features of VC1 utterances from each model (handcrafted or self-supervised). We generated a 1D vector for each utterance by temporally pooling the extracted features using *mean+max* and standardized these vectors before training. Then, we trained a single shallow linear layer classifier/decoder with 100 nodes to fit the features from the training set of VC1 using Adam optimizer and testing on the validation set for early stopping with patience of 20 epochs. The classifier trains for 200 epochs with learning rate being tuned to optimize the results on the validation set between 0.0001 to 0.01. We manually checked lower values that yielded worst results. We ran this process three times for each model and reported the unweighted average recall percentage (UAR %) to account for the unbalanced labels in the dataset. We used the `scikit-learn` package to for the benchmark.

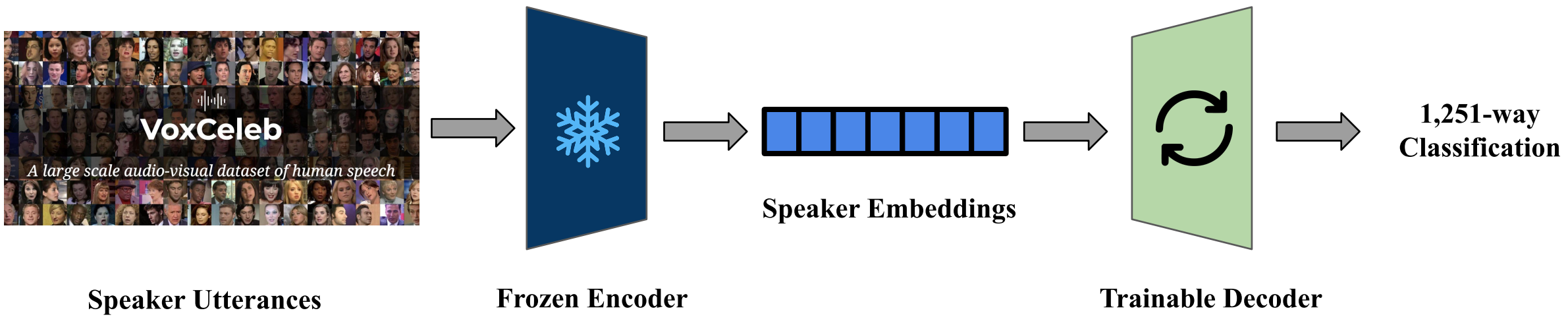

Figure 2.2: ASpR Benchmark Evaluation Pipeline.

[1]https://www.robots.ox.ac.uk/~vgg/data/voxceleb/vox1.html
[2]https://github.com/nttcslab/\gls{byola}

We also explored representations from three different layers within the Wav2Vec2, HuBERT, and Data2Vec models, which use a similar architectural structure as shown in Figure 2.3. Thus, we extracted latent representations, i.e., representations from the final convolution layer, and contextual representations which are generated from the last layer that is usually a transformer layer. Best-performing layer for each of the three models is also reported and the process for selecting the best layer is discussed in Section 2.3.

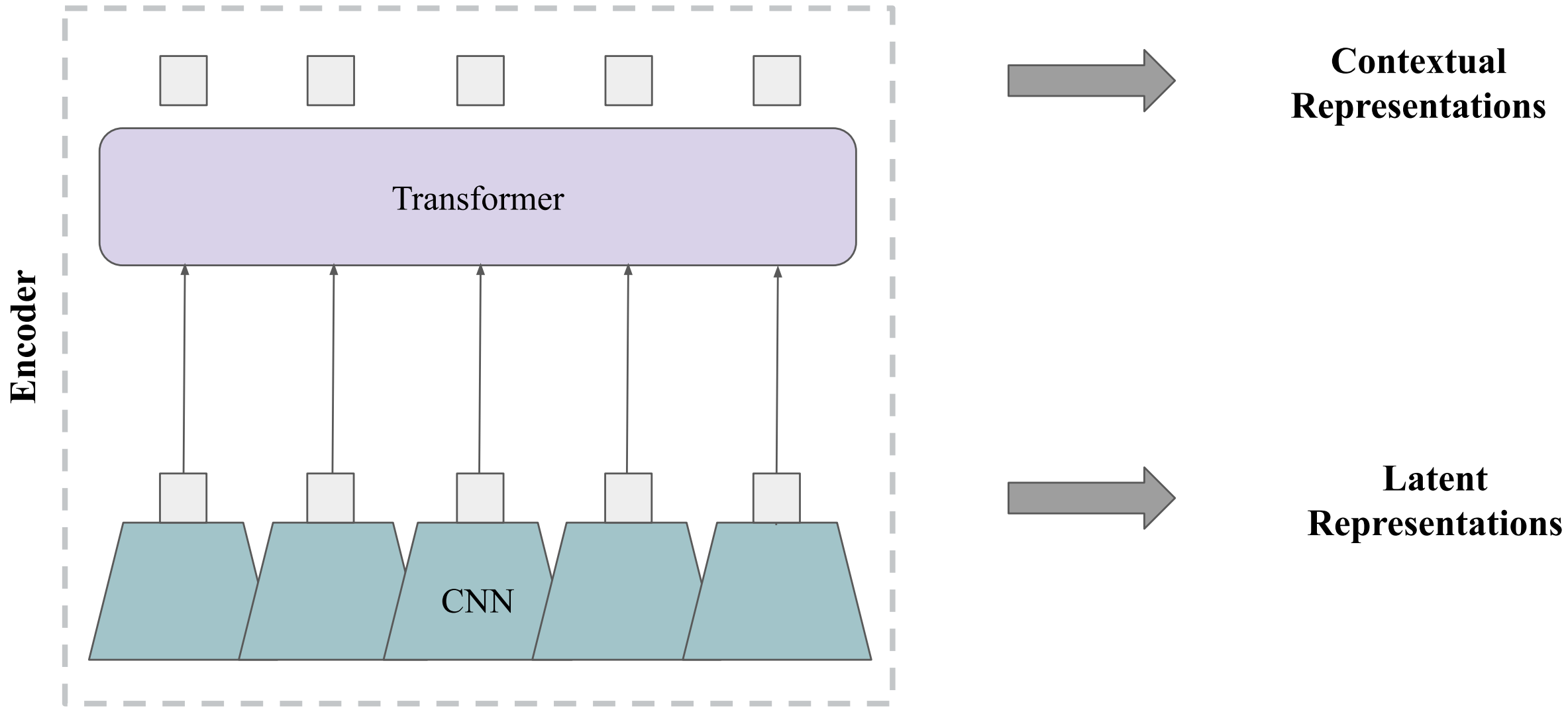

Figure 2.3: Architecture structure for Wav2Vec2, HuBERT, and Data2Vec.

### 2.2.2 Results

We reported the UAR % on the test set for each model as shown in Figure 2.4. This figure displays the variation in speaker identification accuracy (unweighted average recall - UAR) using embeddings from different models. The shades of colors in the figure refer to different categories of representations. For handcrafted representations, we reported their average performance as horizontal lines as illustrated in Figure 2.4.

### 2.2.3 Discussion

Almost all models outperform low-level acoustic features (openSMILE features, spectrograms, and cochleagrams) in speaker identification accuracy. These data-driven features show significantly higher performance than chance-level. These results advocate for SSMs as potential models to study speaker identity coding in humans.

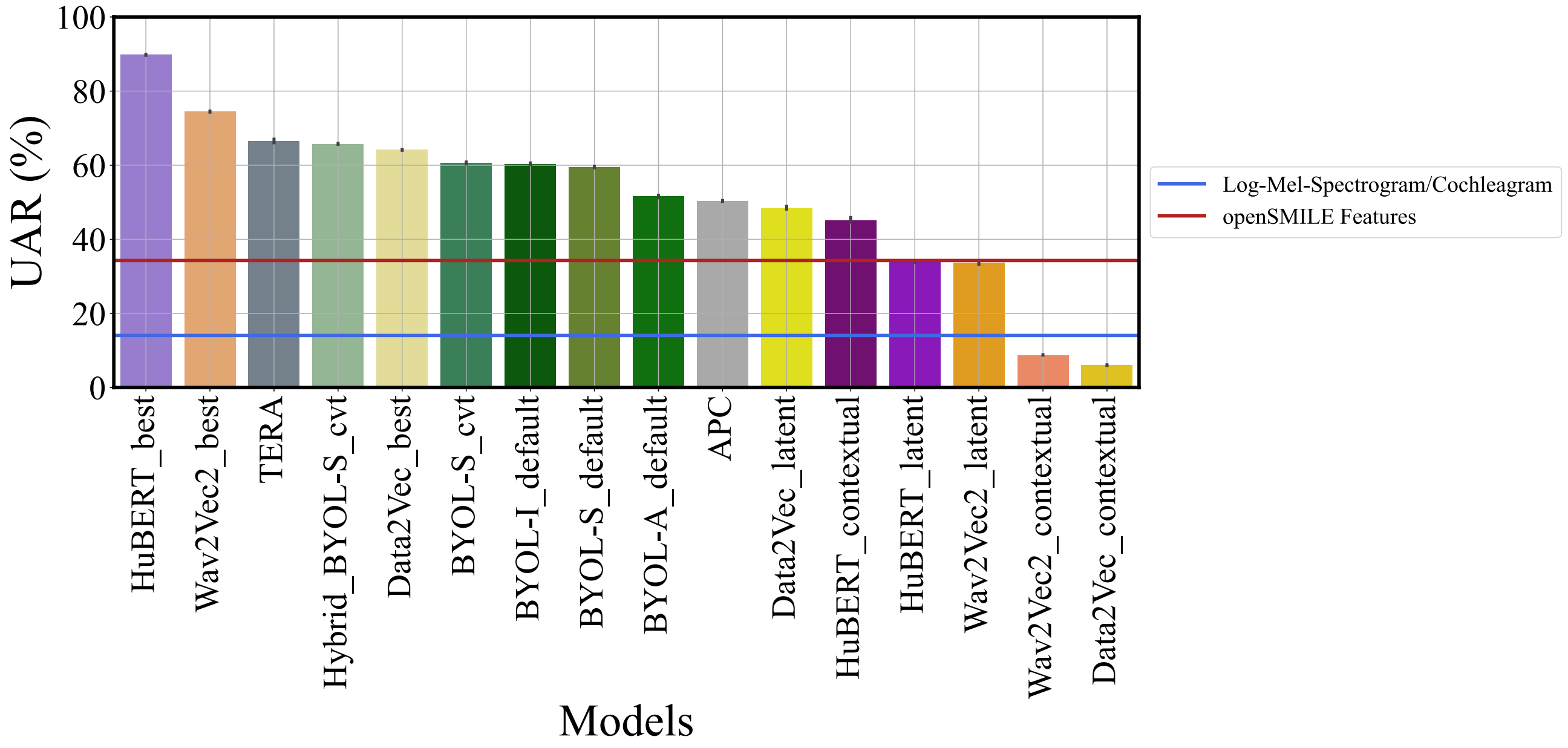

Figure 2.4: ASpR performance benchmark on VC1 dataset using single linear layer. Models are sorted descendingly (from left to right) based on their mean UAR %, the error bars shows the standard deviation in performace across three repeated runs. The red horizontal line indicates average performance of openSMILE features (both sets). The blue horizontal line indicates the average performance ofspectro/cochlea-grams. The bar plot colors indicate the category of the models. BYOL- models are in shades of green, generative models are in shades of grey, Wav2Vec layers are in shades of orange, HuBERT layers are in shades of purple, and Data2Vec layers are in shades of yellow.

The optimal layer in HuBERT [101] model (89.7%) is the best-performing representations on this task with around 15% increase in UAR % compared to the subsequent model in ranking and outperforming openSMILE acoustic features with more than 40%. Before considering the best layers of the three models, we noticed that TERA [102] and Hybrid BYOL-S [103] were the best-performing models on this task. Incorporating the best layers to the benchmark altered the perception of the extent to which data-driven representations can capture speaker identity statistics. On the other hand, significant discrepancies in performance between latent and contextual/final representations are observed. In general, the latent features outperform the contextual except for HuBERT [101] model. The final/contextual layer of Wav2Vec2 [104] and Data2Vec [105] models are worse than spectro/cochlea-grams and openSMILE features.

Furthermore, it is noteworthy to observe that models such as BYOL-A/default (BYOL for Audio) [106], BYOL-S/default (BYOL for Speech) [107], and BYOL-I/default (BYOL for Identity), which share the same encoder architecture, learning paradigm, and training hyper-parameters show around 8% difference in performance, due to differences in their training data.

BYOL-A was pretrained on all Audioset corpus [108] while BYOL-S was pretrained on the speech subset in Audioset. The discrepancy in performance between these models reflects the contribution of the input diet to the model's predictive power. This suggests that BYOL-S and BYOL-I captured more speech-related features that improved the encoder's performance. We also see that using same model but different encoder architecture can further improve performance as demonstrated in BYOL-S/default (CNN) and BYOL-S/CvT (Convolution vision transformer). These findings align with the experimentation carried out by [109] on BYOL models. Additionally, further improvement is observed when augmenting the task objective as in the case of BYOL-S/CvT and Hybrid BYOL-S/CvT. Hybrid BYOL-S/CvT, in addition to sharing the same learning paradigm as BYOL-S/CvT, is trained to predict acoustic features of the original input as well. Plugging in this auxiliary task during training contributed to around 5% increase more than BYOL-S/CvT.

Lastly, it is worth noting that BYOL-I was partially pre-trained on VC1 meaning that this model had seen the utterances used in the benchmark yet in a different task (self-supervised). However, the performance didn't change compared to the other BYOL models suggesting that the task objective design is critical for representation learning.

Table 2.1: Datasets used for training BYOL- models

| Model Name | Dataset | Duration (h) |
|---|---|---|
| BYOL-A | AudioSet | 5800 |
| BYOL-S | AudioSet (Speech subset) | 2190 |
| BYOL-I | VoxCeleb 1&2 | 2000 |

In order to validate the reproducibility of our findings, we conducted the same benchmark evaluation on a different dataset, called TIMIT [110] which is discussed in detail in Section 5.2 along with the benchmark results in Figure 5.1.

## 2.3 Layer-wise Analysis

Given the discrepancies in performance between latent and contextual representations for Wav2Vec2, HuBERT, and Data2Vec models, we sought to further report the performance of each layer in the models' architecture.

### 2.3.1 Methods

We randomly sampled a gender-balanced group of 100 speakers from VC1 dataset and followed the same linear evaluation pipeline explained in Section 2.2. We ran the linear evaluation once on each layer.

### 2.3.2 Results

Figure 2.5 shows the UAR % of each embedding layer for each of the three models. Both Wav2Vec2 and Data2Vec share the same architecture, comprising 7 CNN layers followed by 24 transformers layers, whereas, HuBERT has more transformer layers (48), as shown in the figure. The best-performing layer for each model (highlighted with vertical lines) are transformer layer 4, 7, and 3 for Wav2Vec2, HuBERT, and Data2Vec respectively.

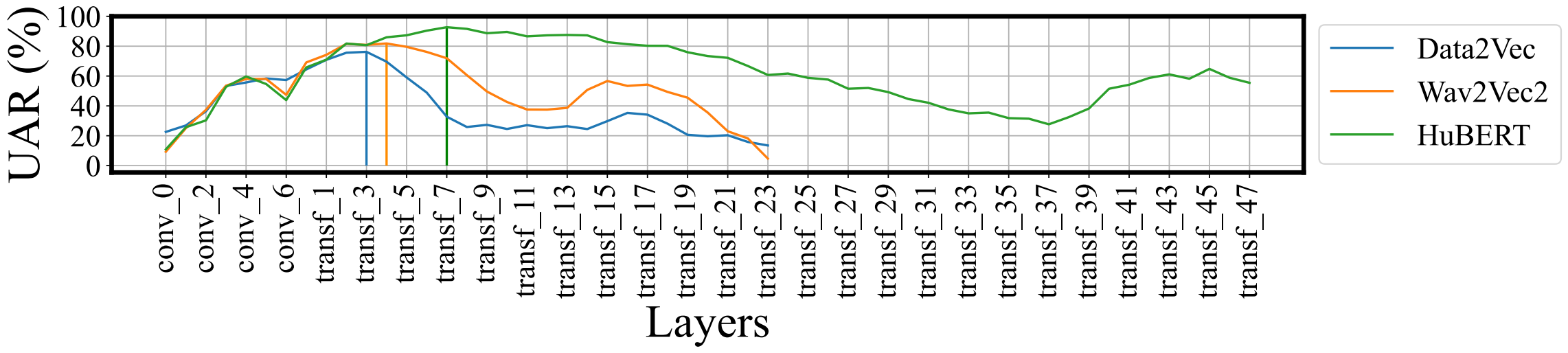

Figure 2.5: Layer-wise analysis for Wav2Vec2, HuBERT and Data2vec on ASpR task with 100 speakers. The vertical lines highlight the best-performing layer for each model. conv: convolution layer, transf: transformer layer.

### 2.3.3 Discussion

As shown in Figure 2.5, we noticed similar trends in performance across layers. The performance increases gradually until it reaches the peak at an intermediate layer then starts dropping. Thus, these models learn to maximize speaker identity information in earlier layers. We hypothesize that such behavior might be due to the objective tasks used for these models. For instance, these models were trained to predict hidden speech units [101] which might suggest making the final layers learn statistics related to the phonetic and linguistic information of the input utterance and discard speaker-related information. We provide further evidence regarding our claim in Section 2.5. To better visualize the discrepancy in performance, Figure 2.6 shows the performance of 3 layers (latent, best, and contextual/final) in each model. This result reflects the heterogeneity of captured content across layers within one model. Additionally, we ran similar layerwise analysis on TIMIT benchmark and found reproducible patterns, see Figure 5.2

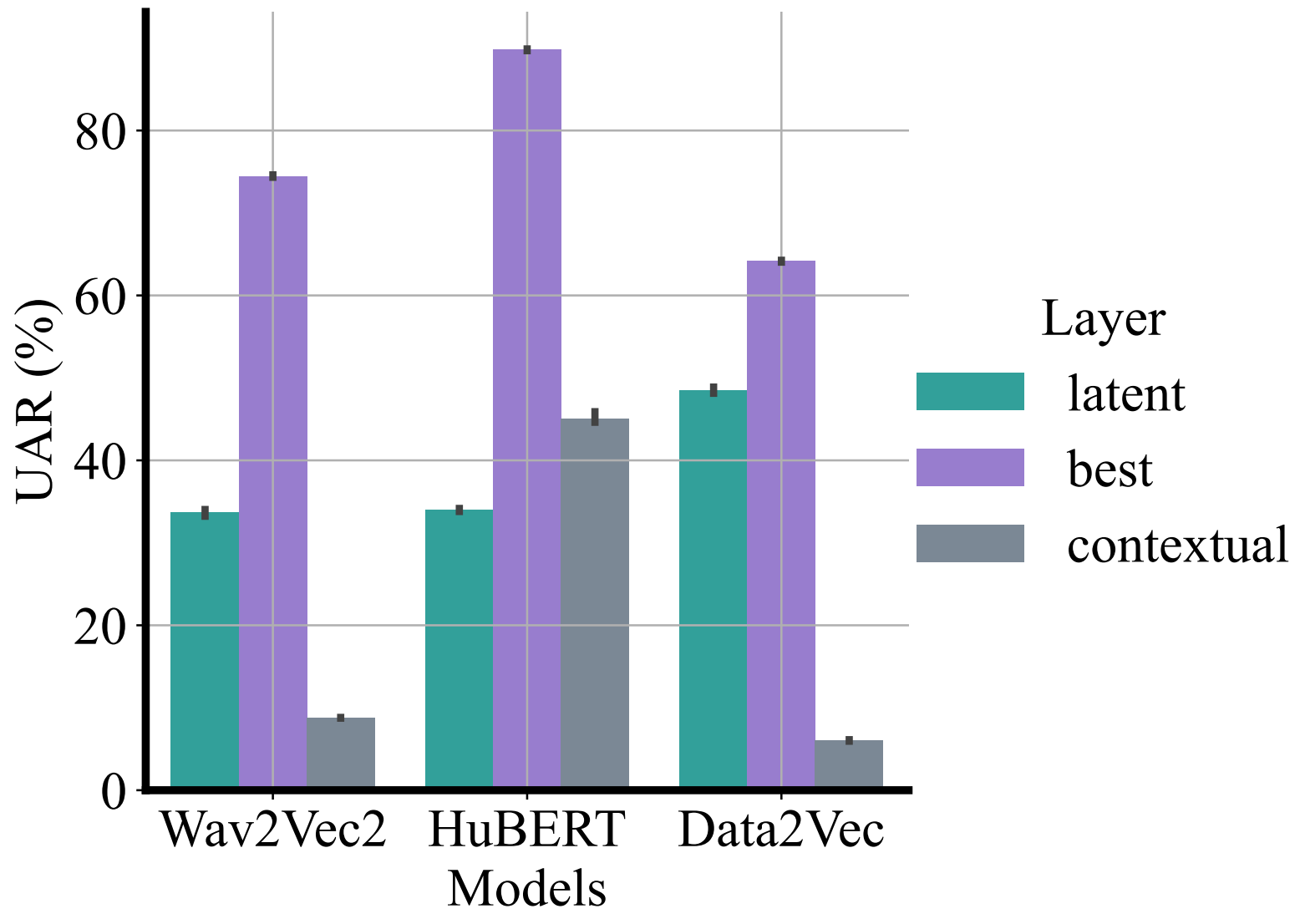

Figure 2.6: Comparing latent, best and contextual/final representations of Wav2Vec2, HuBERT, and Data2Vec.

## 2.4 Similarity Analysis

After reporting the predictive power of candidate models on a ASpR task, we measured the similarities across models' representations. We used *centered kernel alignment* (CKA; [111]), a dot-product-based similarity metric that has been shown to be robust for computing similarities across neural network representations [111].

### 2.4.1 Methods

To explain this method, imagine we extracted the embeddings of audio samples from a particular model yielding a matrix $X$ with a size of $NxM$ where $N$ is the number of samples/examples and $M$ is the number of features/dimensions. Then, we compute a kernel between all possible pairs of samples/examples resulting a kernel matrix where $K_{ij} = k(x_i, x_j)$. Thus, the kernel matrix is now of size $NxN$. In a similar vein, a kernel matrix is computed on another model of interest $Y$ to generate $L_{ij} = l(y_i, y_j)$. Subsequently, we center the rows and the columns of both matrices and compute the Hilbert-Schmidt Independence Criterion (HSIC), as shown in Equation 2.1.

$$HSIC(K, L) = \frac{tr(KHLH)}{(n-1)^2} \tag{2.1}$$

where $H$ is the centering matrix and $n$ is the number of samples. To make the similarity notion invariant to scaling, we normalize the HSIC estimator to compute a CKA score as shown in Equation 2.2.

$$CKA(K, L) = \frac{HSIC_0(K, L)}{\sqrt{HSIC_0(K, K) HSIC_0(L, L)}} \tag{2.2}$$

For each model, we extracted the embeddings/features from the test set of VC1 (8,251 utterances). Then, we standardized the generated features to compute a CKA score across all models. We used the code implementation[3] proposed by [111]. The pipeline for computing CKA between a pair of models is visualized in Figure 2.7.

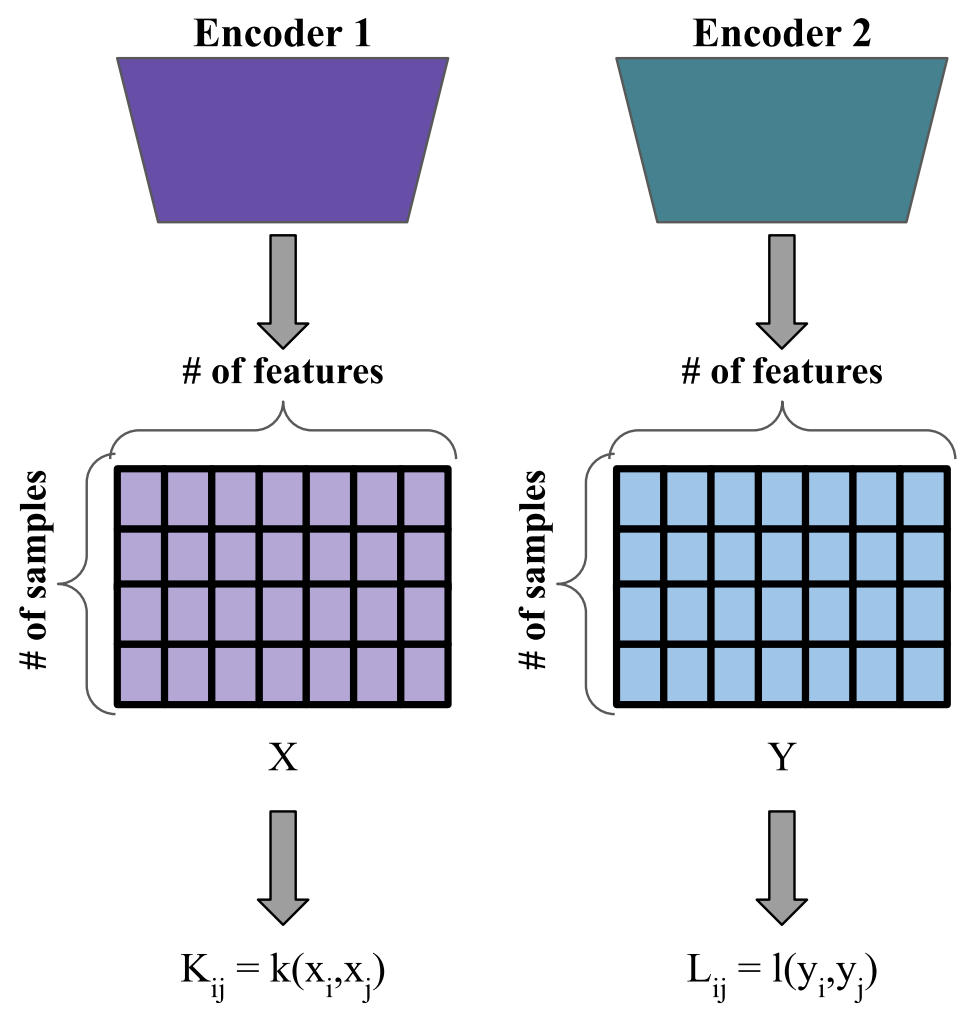

Figure 2.7: Pipeline for computing kernel matrices from two encoders that will be fed to Equation 2.2.

## 2.4.2 Results

The heatmap in Figure 2.8 shows the CKA scores across all models. The scores vary from 0 to 1 (dark to light colors).

[3] https://colab.research.google.com/github/google-research/google-research/blob/master/representation_similarity/Demo.ipynb

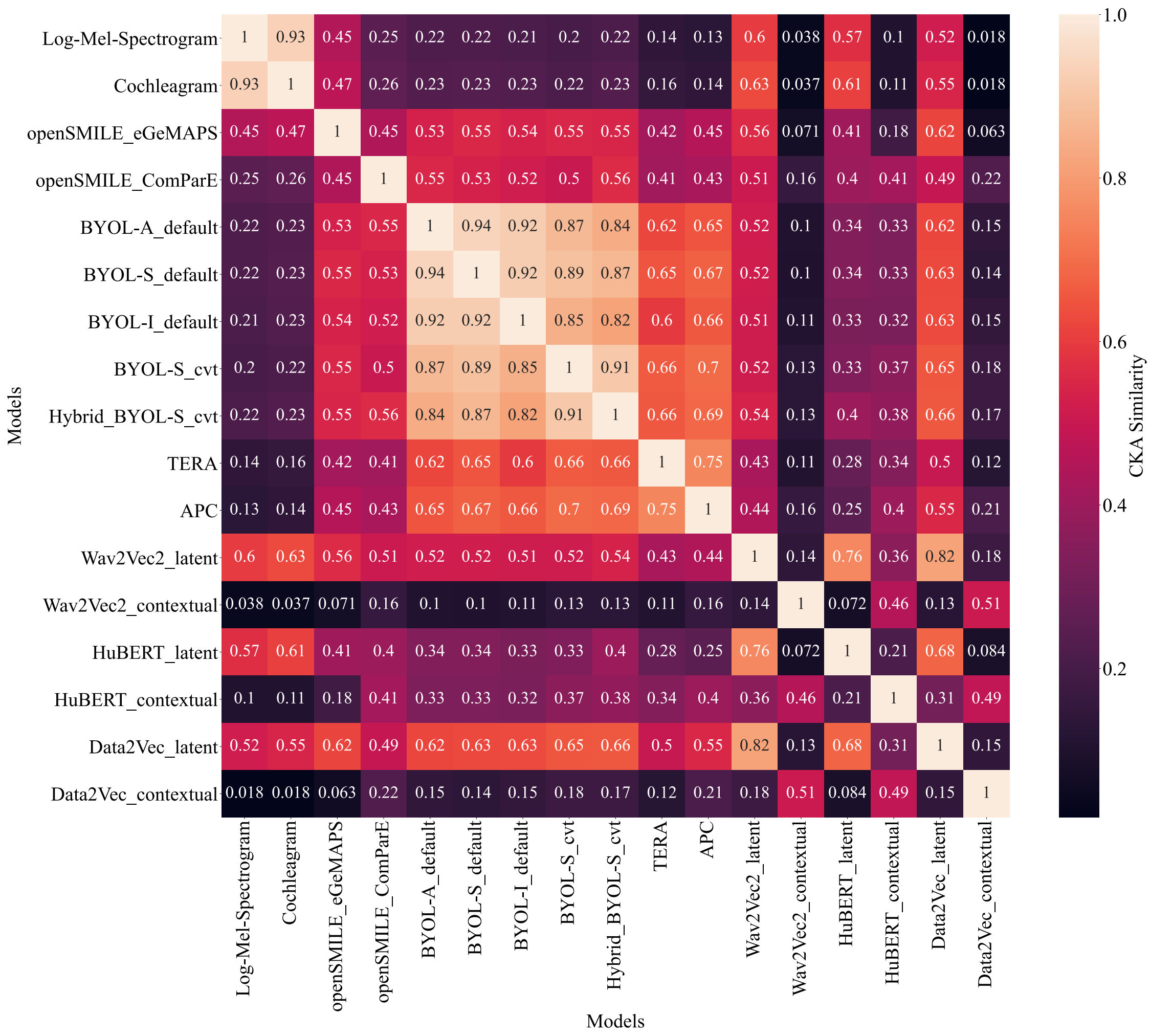

Figure 2.8: CKA analysis across all models' representations.

### 2.4.3 Discussion

As a control, we see that spectrogram and cochleagram representations share high CKA score. Moreover, BYOL models also share high score across each other regardless of variations in training data or encoder architecture. The representations from the final convolution layers (latent layers) of Wav2Vec2, HuBERT, and Data2Vec seem to share the highest CKA scores with spectrogram and cochleagram features. This similarity between latent layers and acoustic features suggests that these layers are encoding low-level acoustic features. Unlike the other SSMs, those three models are trained on raw 1D waveforms suggesting that earlier layers are likely to learn frequency representations similar to spectrograms and cochleagrams. This might indicate learning hierarchical information as a consequence of task optimization. Further experimentation is needed to investigate the similarities between hierarchical representations in humans and models. On the other hand, the contextual/final layers show very low CKA scores with almost all representations indicating that contextual layers might encode different aspects of speech compared to the rest. That might also explain the discrepancy in performance between the final layers of these models and the other candidate SSMs. Moreover, generative models such as TERA and APC feature high CKA scores with the BYOL-family models which might explain the similar performances in the benchmark. We also computed CKA for the TIMIT benchmark, see Figure 5.4.

## 2.5 Speech Experiments

### 2.5.1 Phonetic Information

One way to distort the phonetic structure in an utterance is to reverse the speech signal. As a result, the signal loses its phonemic temporal consistency while maintaining the frequency and spectral information. This allows us to study the impact of altering phonetic structure on models' performance.

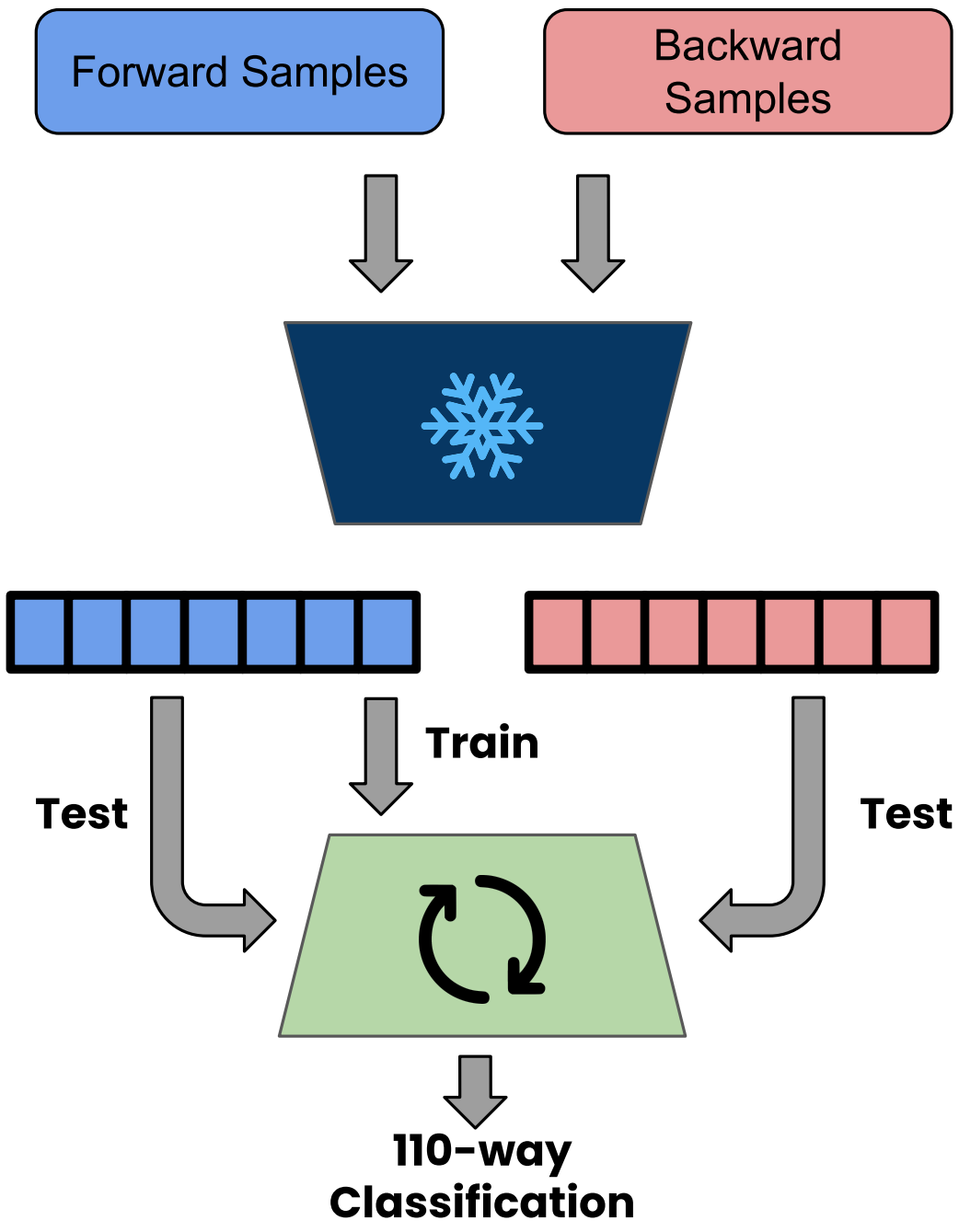

Figure 2.9: Pipeline for Backward Speech Experiment.

**Methods:** We created a parallel reversed version of CSTR VCTK dataset [112] using `pytorch` audio augmentation package [4]. For each model, we extracted the embeddings of both forward and backward utterances. Then, for each speaker, we split the generated embeddings into train and test sets (70% and 30%, respectively) to maintain equal ratio of samples across speakers. Both splits included forward embeddings and their corresponding backward ones. We trained a shallow linear layer to recognize speakers from forward embeddings only in the train set and tested its performance on both forward and backward embeddings in the test set, as illustrated in Figure 2.9. We implemented a grid search across different hyperparameters including number of nodes, learning rates, and activation functions. To better visualize the models' representations in a compressed lower-dimensional space, we used t-SNE [113] as a dimensionality reduction method. We selected the utterances of 10 speakers from the corpus (5F/5M) and reduced the dimensions of their representations.

[4]https://github.com/Spijkervet/torchaudio-augmentations

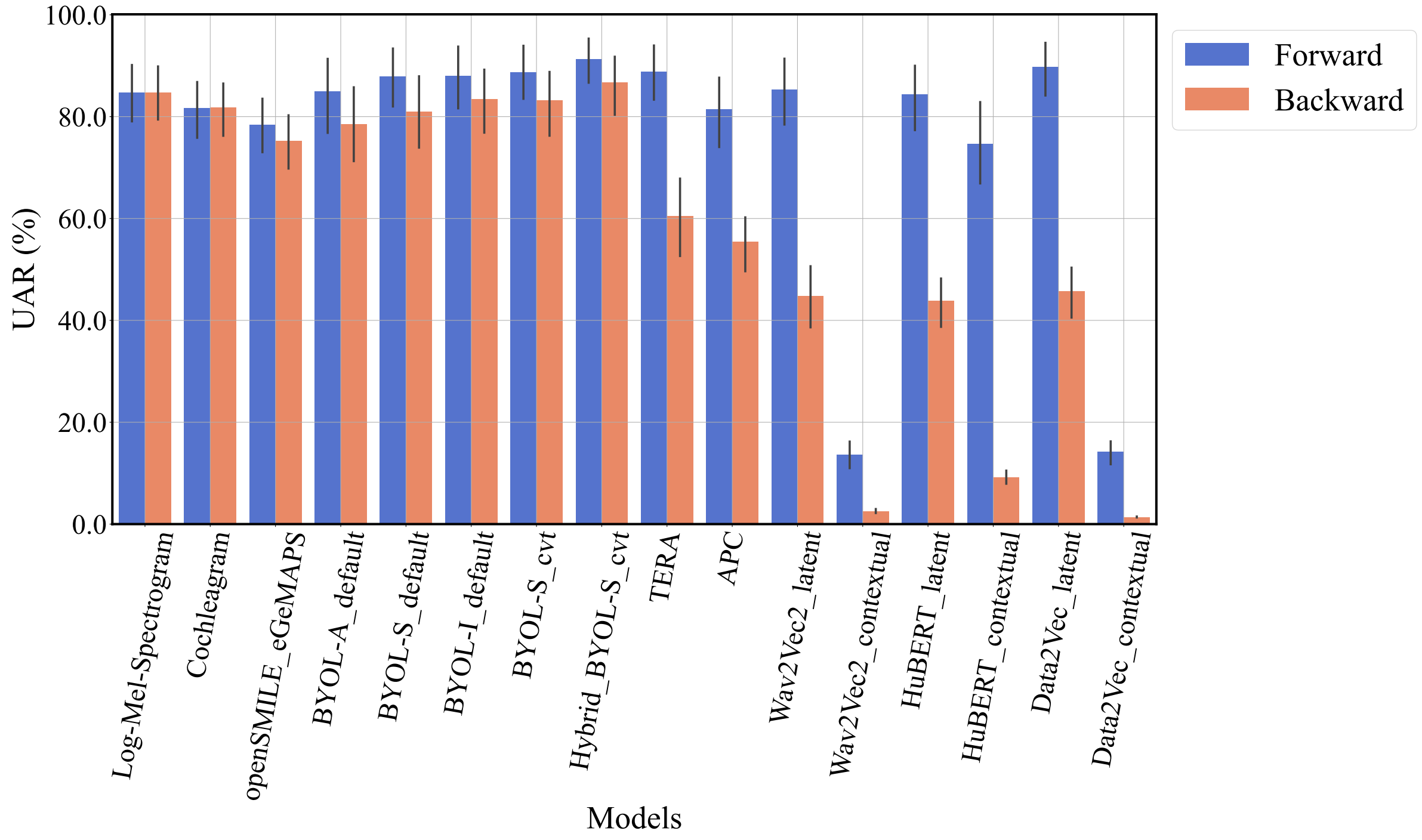

Figure 2.10: Performance of models on a ASpR task in case of forward and backward speech. The error bars indicate the standard deviation of the UAR across all hyperparameter sequences.

**Results:** We reported the UAR % on a test set for both categories; forward and backward speech, as shown in Figure 2.10. In Figure 2.11, we visualize a subset of candidate models and label different information related to speaker utterances such as: identity and sex of the speaker, type of stimulus (forward or backward), F0 and the number of syllables in each utterance.

**Discussion:** In Figure 2.10, the spectrogram or cochleagram representations serve as controls and, as expected, didn't show any difference as reverse samples simply mean flipping the features of the spectrogram. Also, we noticed no significant differences in performance across

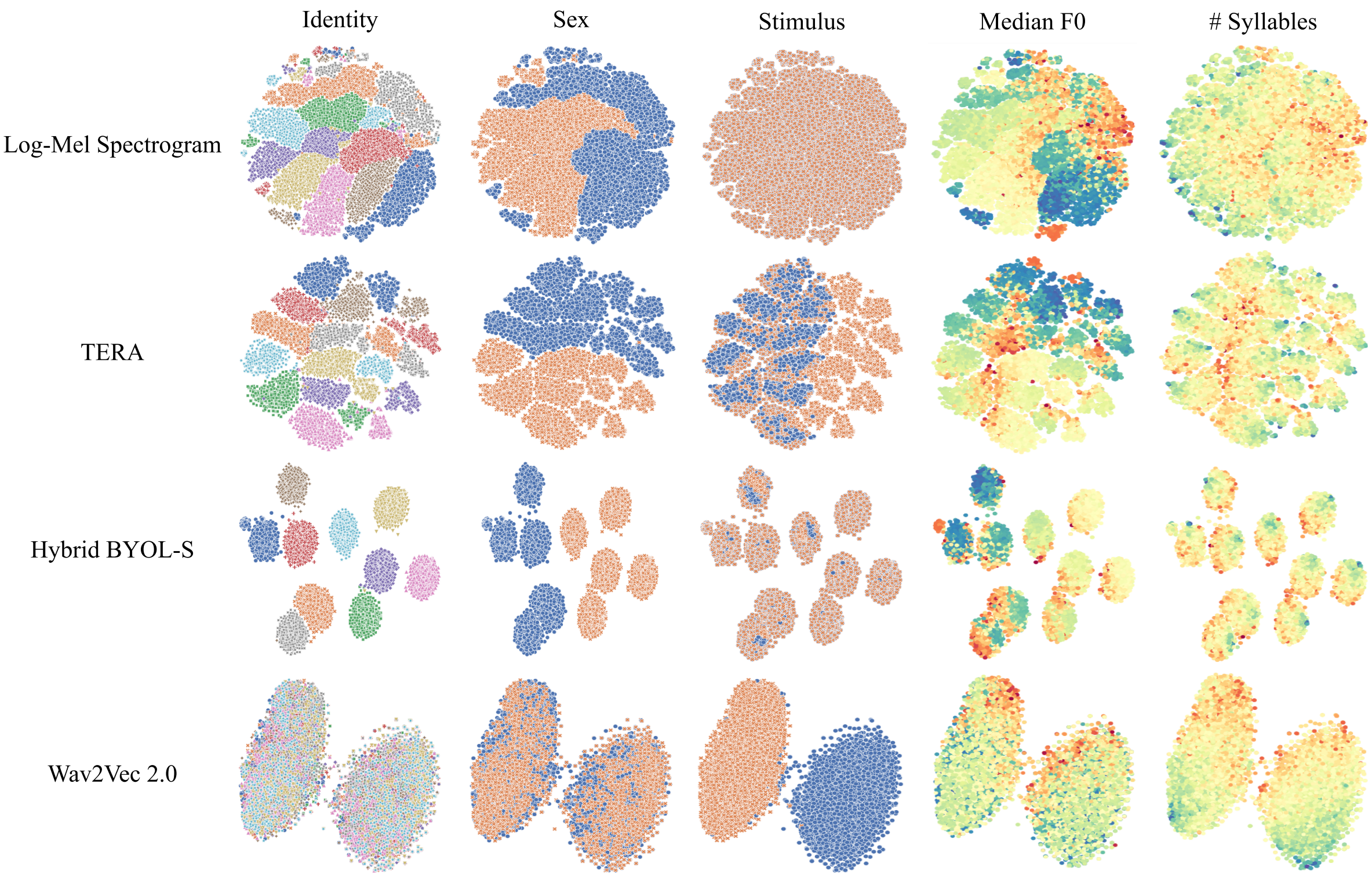

Figure 2.11: t-SNE plots for forward and backward speech experiment. The rows indicate the candidate representations. The columns indicate different labels. Identity: each color indicate a unique speaker (total 10 speakers), Sex: Male speakers in orange and Female speakers in blue, Stimulus: Forward samples in blue and Backward samples in orange, Median F0 and Number of Syllables: color scale from red to blue (low to high values), values are in log scale for better visualization.

BYOL-family models. However, there is a clear drop in performance across the rest of the models specifically Wav2Vec2, HuBERT, and Data2Vec for both latent and contextual layers. These results further support our earlier claim that the three mentioned models encode temporal ordering, and perhaps phonological constraints, of phonetic information in their representations which makes them vulnerable to such a variation.

Moreover, Figure 2.11 displays the differences between models' representations. For instance, most models exhibit relatively clear separations of speaker identities and sex suggesting that these models such as Hybrid BYOL-S and TERA might be capturing features with high precision for vocal identity. This might justify the high performance these models showed in the benchmark, in Figure 2.4. Conversely, Wav2Vec2 doesn't show any clear separation across identities or sex. Nevertheless, the two main clusters in Wav2Vec2 space are representing the forward and backward samples. Additionally, these clusters are showing some order related to number of syllables. These plots are additional evidence to highlight the highly-represented

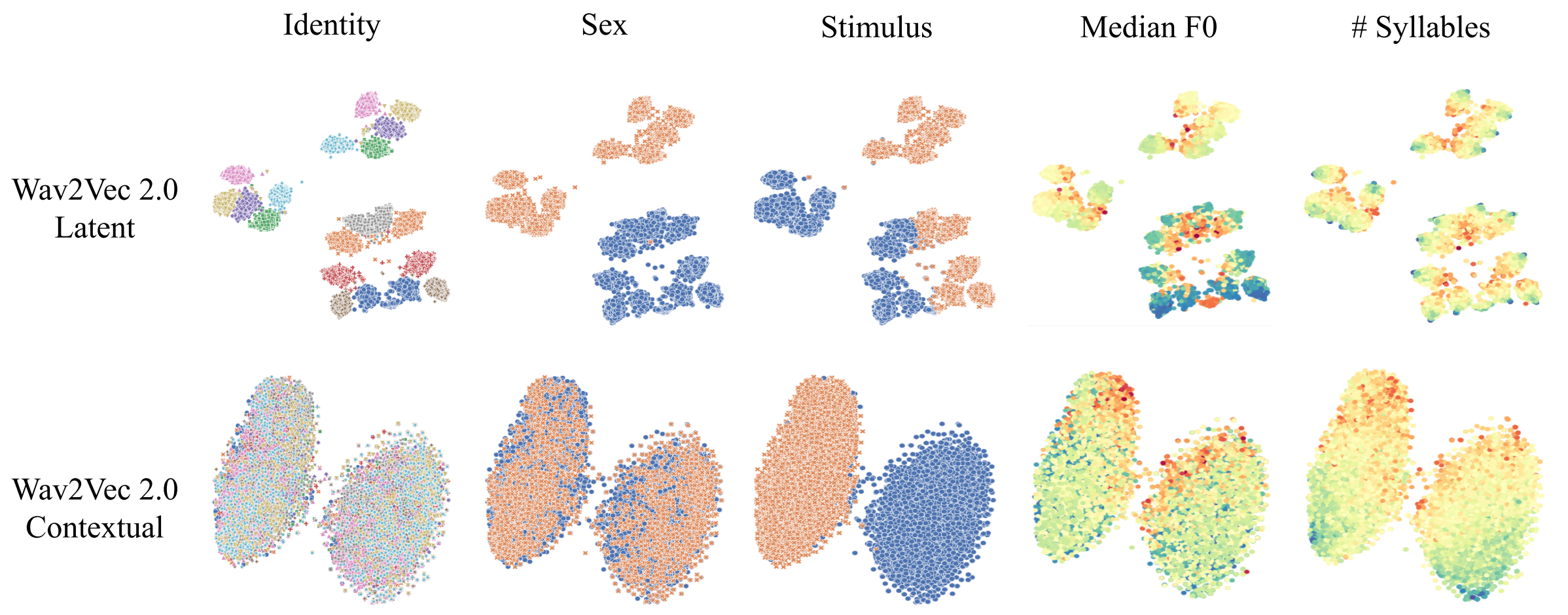

Figure 2.12: t-SNE plots for forward and backward speech experiment for latent and contextual layers of Wav2Vec2. The rows indicate the candidate representations. The columns indicate different labels. Identity: each color indicate a unique speaker (total 10 speakers), Sex: Male speakers in orange and Female speakers in blue, Stimulus: Forward samples in blue and Backward samples in orange, Median F0 and Number of Syllables: color scale from red to blue (low to high values), values are in log scale for better visualization.

phonetic information encoded in the final layers of Wav2Vec2.

To further investigate the encoded representations in Wav2Vec2, we show in Figure 2.12 the contrast between latent and contextual layers of the same model. It can be observed that the latent layer has captured speaker-related characteristics such as identity and sex, and is still maintain a clear separation for stimulus type. This Figure might explain the high performance of latent representations compared to contextual ones as well as the significant drop in both cases when predicting identity from backward samples.

Additionally, we have used other dimensionality reduction methods such as PCA, UMAP [114], and PaCMAP [115] to verify that the observed trends in clusters and representations are not algorithm-specific. Further information regarding the hyperparameter tuning methods used in this work and the results from other models are all shown in Appendix Section 5.2.

## 2.5.2 Frequency Information

The act of speaking requires orchestrating a number of anatomical structures with precise spatial and temporal coordination to generate a rich vocal signal. The source of this signal comes from the oscillations of the vocal folds (VFs) caused by the air from the lung passing through them. This oscillation creates a tone and we call this process *phonation*. The number of oscillation per second is defined as the fundamental frequency (F0) and we refer to its perceptual correlate as *pitch*. F0 is usually used as a predictor for sex since , on average, males tend to have lower F0 (lower pitch) compared to females. However, in this work, we are interested in examining the impact of altering this frequency component on identity processing. On phone conversations, we often over articulate in order to increase clarity for the listener. It is possible that we vary our pitch to convey additional information about our vocal identity or related vocal tract characteristics. Accordingly, it is hypothesized that sweeping harmonics might maximize speaker-related information more than steady pitch.

**Methods:** To test this hypothesis with the models, we used a dataset shared and described by [116]. This corpus comprises 2010 utterances from 15 German speakers (11F/4M). Each speaker contributed 80 recordings of sentences for training and 54 recordings of vowels for testing. The vowel utterances are divided into 6 pitch conditions (Steady-state: low, mid and high pitch & Sweeping: fall, fall-rise and rise) for each of 3 vowels (/a/, /i/ and /u/). Finally each condition and vowel is repeated 3 times.

We extracted the embeddings of all utterances using all candidate models. We trained a shallow layer on the embeddings of the sentences and tested the classifier to recognize speakers from the embeddings of the vowel utterances in all conditions. The implementation and the hyperparameter tuning was similar to the previous experiment.

**Results:** We report the UAR% for each model across the 6 pitch conditions in Figure 2.13, which shows that across almost all models the best pitch condition for speaker recognition is the falling followed by rising pitch. Also, we aggregated the sweeping conditions and the steady conditions to evaluate the overall performance in case of sweeping harmonics and steady pitch, as illustrated in Figure 2.14. We see from Figure 2.14 that sweeping harmonics are overall better predictors for speaker identity than steady ones.

**Discussion:** The best recognition condition with falling followed by rising pitch supports the hypothesis that adding more variability to the harmonics increases speaker related information. Conversely, the worst pitch condition on average is uttering vowels with high pitch. It is noticed that some models are performing almost at chance level (red line) and these are the models that we have shown, in previous experiments, as not encoding speaker information. Given the benefit of sweeping harmonics over steady ones align with a thesis on infant-directed speech (IDS). IDS shows higher acoustic variability compared to adult-directed speech (ADS). Recent work by [117] suggests that the origin of IDS might be an evolutionary outcome of signalling caregivers' vocal identity to their offspring. Our results also align with findings from behavioral experiments on humans [116] supporting our claim that such models are candidates to model speaker identity perception in humans.

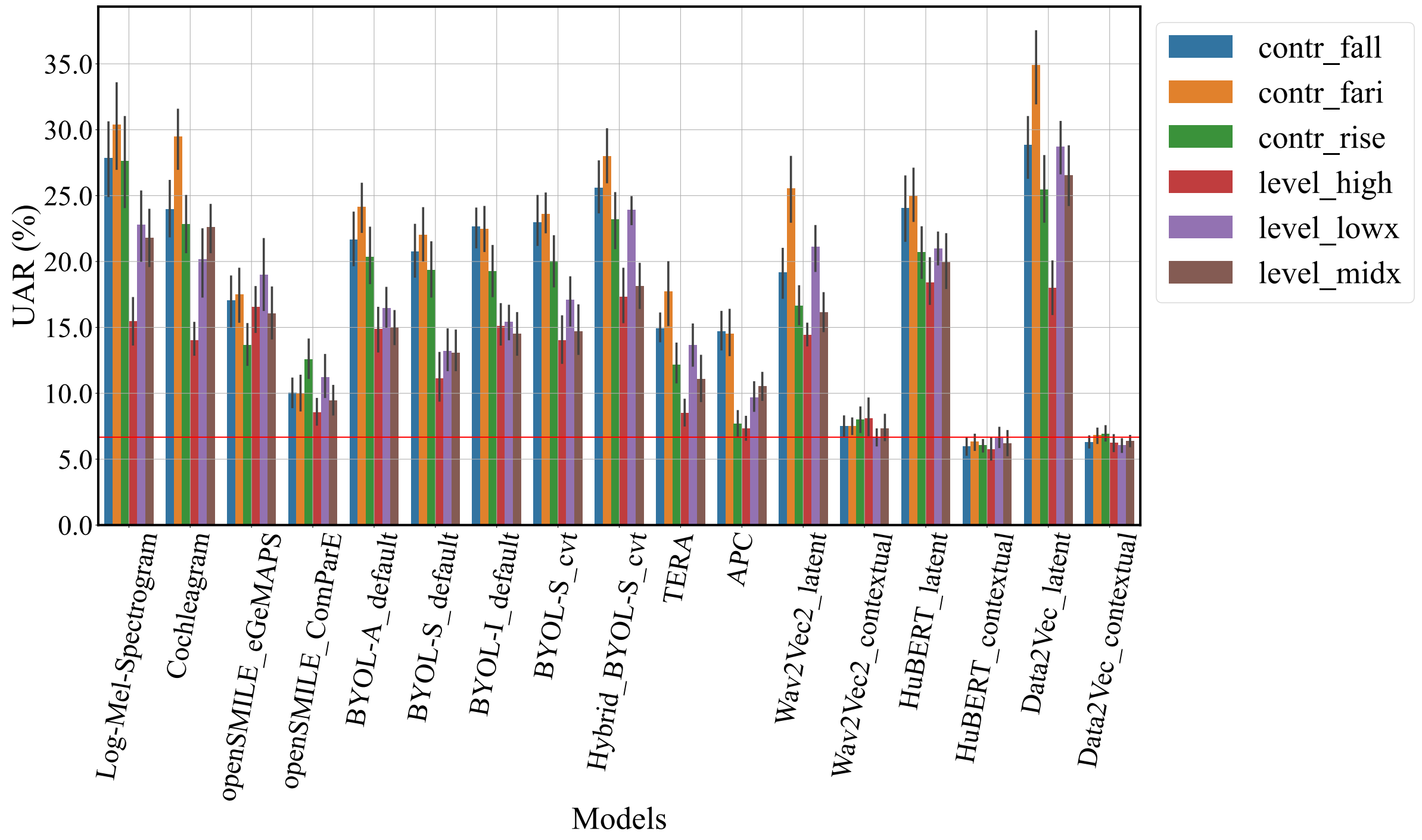

Figure 2.13: Performance of models on a ASpR task with the 6 pitch conditions. *contr_fall*: falling pitch contour, *contr_fari*: falling follwed by rising pitch, *contr_rise*: rising pitch contour, *level_high*: high steady pitch, *level_midx*: medium steady pitch, *level_lowx*: low steady pitch. The error bars indicate the standard deviation of the UAR across all hyperparameters sequences.The red line indicates chance level performance.

### 2.5.3 Linguistic Information

A phenomenon known as Language-familiarity effect suggests that listeners tend to identify voices more accurately when the language spoken is their native language compared to other languages [39, 118]. Thus, in this section, we are curious if such phenomenon could be observed in models.

**Methods:** For this experiment, we used SpiCE corpus [119]. It is a bilingual speech dataset that comprises 30-minute interviews with 10 bilingual speakers. The speakers speak both English and Cantonese. We pre-processed the data by removing the interviewers voice from the utterances using Audacity. Also, we cut all the pauses using `Librosa` Python package and setting a threshold of 45dB (i.e. below that is considered silent). All the extracted segment from the latter step were concatenated to form utterances containing only the speaker's voice. Finally, we chunked each speaker's utterance into 3-sec clips. Accordingly, for each speaker,

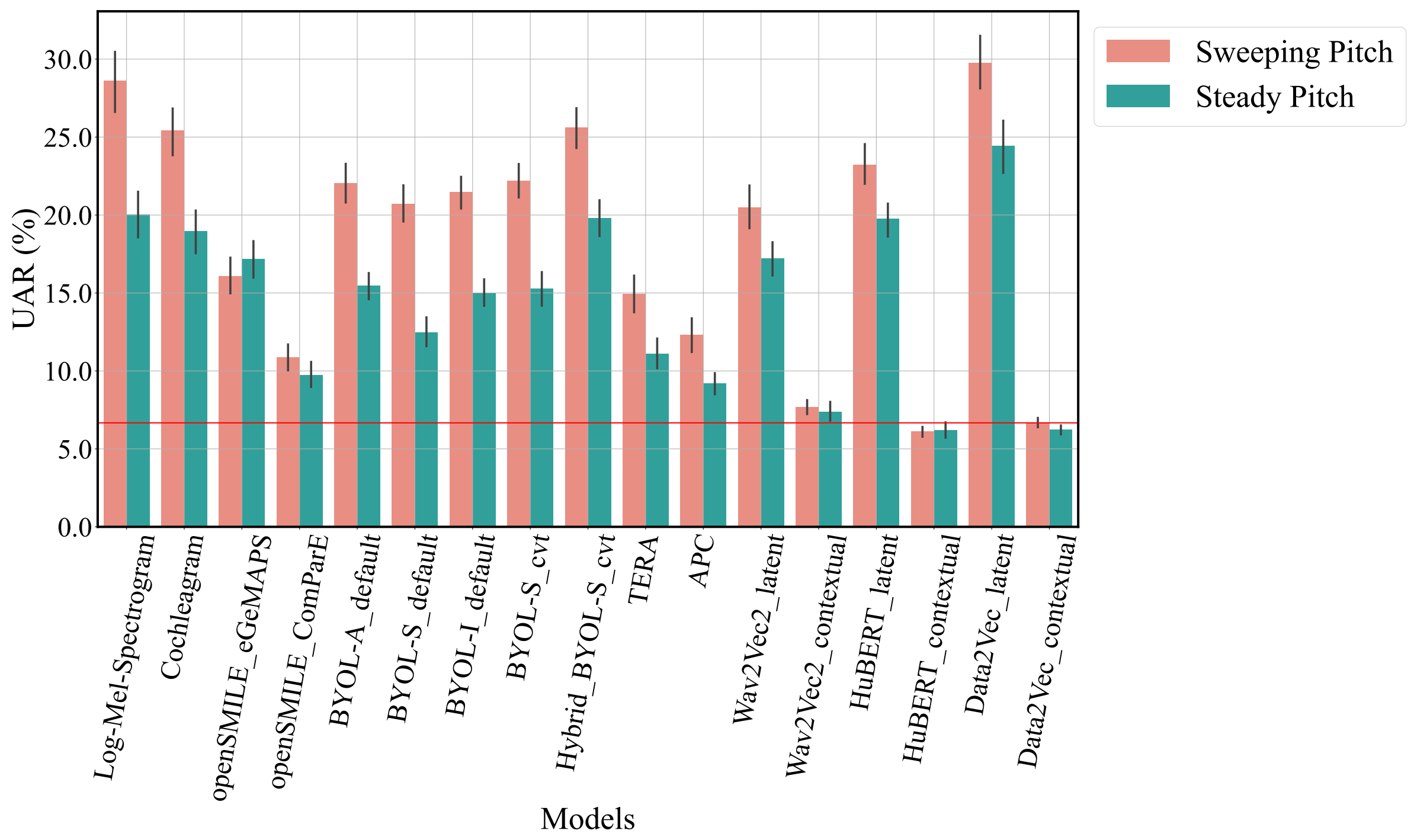

Figure 2.14: Performance of models on a ASpR task in case of sweeping and steady pitch. The error bars indicate the standard deviation of the UAR across all hyperparameters sequences. The red line indicates chance level performance.

we have 3-sec utterances in both languages, yielding $5,238$ utterances total for all speakers.

To examine the language-familiarity in models, we trained a shallow layer on embeddings from all models on one language and tested the performance of the classifier to recognize speakers from recordings in the other language. The implementation and the hyperparameter tuning was similar to the previous experiments.

**Results:** In Figure 2.15, we compare the performance when training a classifier on embeddings of English clips (left figure) and on embeddings of Cantonese ones (right figure).

**Discussion:** It is intuitive to find that the classifier would be able to recognize speakers better when the language uttered in the train and test set is the same. Indeed, we found this behavior illustrated in Figure 2.15. However, these findings support the claim that we need more than low-level acoustic features to learn speaker identity. There is a language comprehension component that contributes to vocal identity processing. Furthermore, we observe that all models show similar trends in performance with varying effect sizes between languages. Unlike previous experiments, it is interesting to note that Wav2Vec2, HuBERT, and Data2Vec final layers exhibit the highest effect size between languages reflecting the importance of joint encoding of linguistic and phonetic characteristics in these layers.

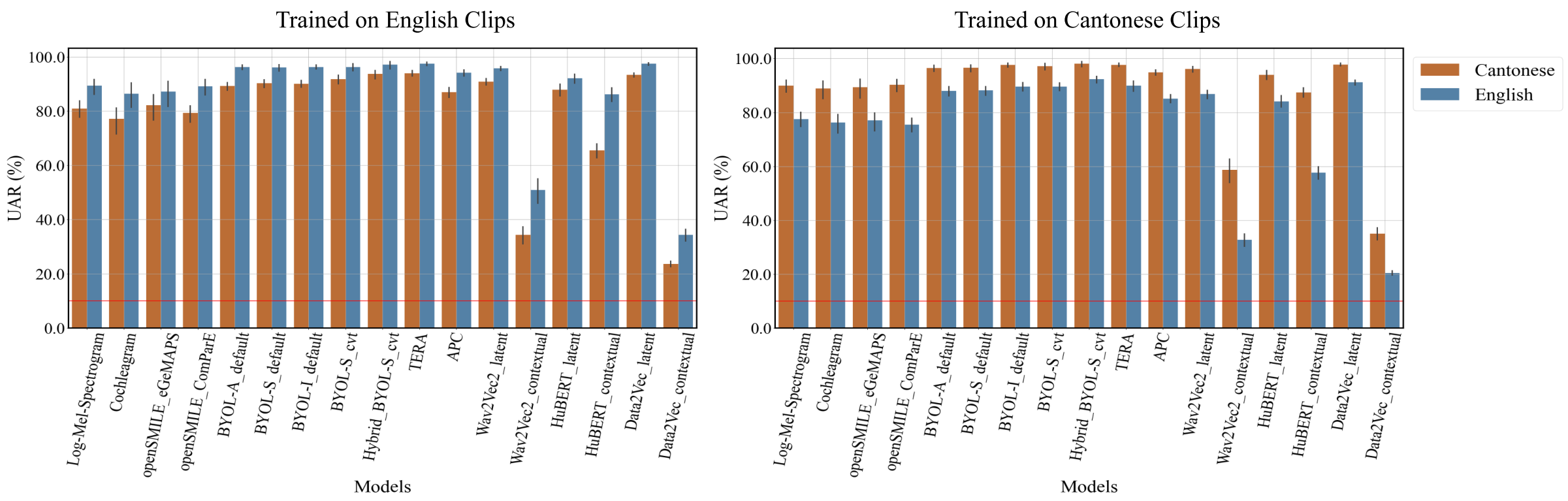

Figure 2.15: Performance of models on a ASpR task when trained on English clips (left figure) and Cantonese clips (right figure). The error bars indicate the standard deviation of the UAR across all hyperparameters sequences. The red line indicates chance level performance.

### 2.5.4 Paralinguistic Information

After exploring the effect of language on speaker identification, we were interested to also consider the paralinguistic aspects of the signal. The paralinguistic component covers any information inferred from the speech signal besides the uttered spoken words. This could be the emotional, physical, or mental state of the speaker. In this section, we studied one variable under the paralinguistic umbrella which is discrete emotions and the effect of altering emotions on speaker identification.

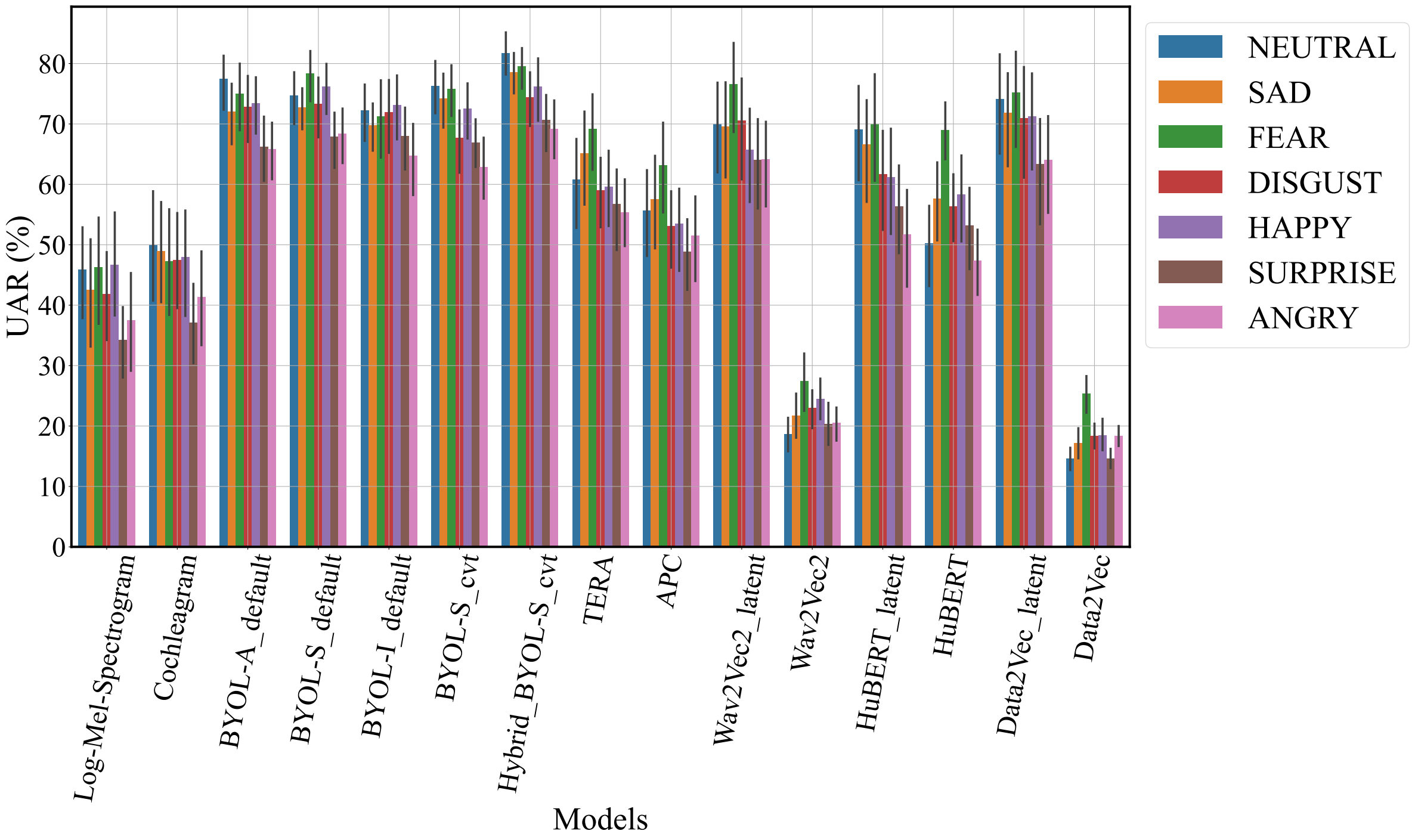

Figure 2.16: Performance of models on a ASpR task in case of categorical emotions. The error bars indicate the standard deviation of the UAR across all hyperparameters sequences.

**Methods:** In this experiment, we used SUBESCO corpus [120]. It is a speech emotion recognition (SER) dataset spoken in Bangla language. The corpus contains 7 hours of speech. It comprises $7,000$ utterances from 20 native speakers (10F/10M) expressing 7 different emotions (Neutral, Happy, Sad, Fear, Angry, Surprise, and Disgust). Each speaker produced $10$ sentences with 7 emotions and it was repeated $5$ times. Hence, $20 * 10 * 7 * 5 = 7,000$ utterances.

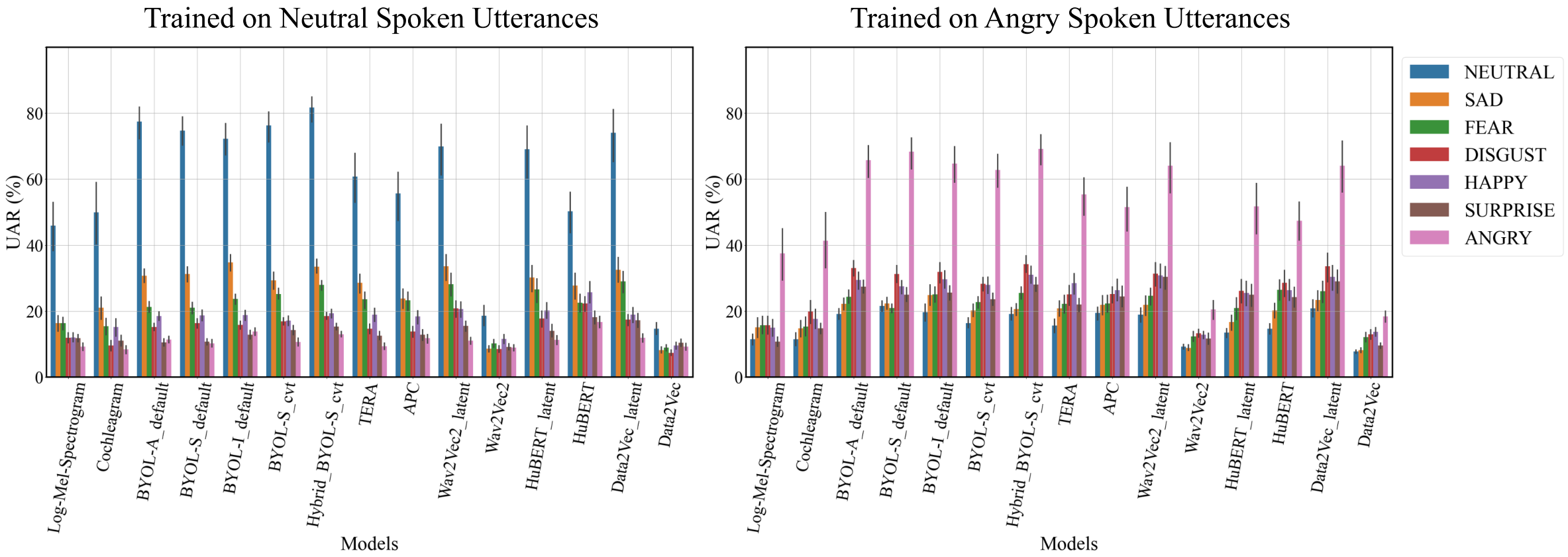

Figure 2.17: Performance of models on a ASpR task when trained on Neutral utterances (left figure) and Angry utterances (right figure). The error bars indicate the standard deviation of the UAR across all hyperparameters sequences.

First, we wanted to see which emotion improves identification performance across models. Accordingly, for each model, we trained a linear decoder on one emotion and tested on the same spoken emotion but unseen sentences. Then, we trained a linear decoder only on *Neutral* spoken utterances and tested on all other emotions. Same approach was adopted with *Angry* spoken emotions. That way, we could measure the variability in performance across different categorical emotions. Finally, we also explored the structure of the encoded representations in a lower-dimensional space as in Section 2.5.6.

**Results:** In Figure 2.16, the performance of each model to predict identity from a specific emotion category is shown. Then, we show in Figure 2.17 the ability of models to capture identity across different emotions while being fine-tuned on only one category; Neutral (left figure) and Angry (right figure). We visualized the encoded representations in Figure 2.18 for two models; HuBERT and Wav2Vec2 from two layers; latent and contextual.

**Discussion:** From Figure 2.16, we didn't notice any significant differences in performance across emotions except that all models showed worse performance when recognizing speakers from with Angry emotion. We might suspect at this point that there is no specific emotion category that maximizes identity information. Nevertheless, when we trained the decoder on one emotion category and tested on the rest, we have noticed, in some models, a gradual decline in performance. For instance, in case of training on Neutral samples, we noticed better performance on low arousal emotions (e.g. Sad and Fear) than high arousal emotions (e.g. Angry and Surprise). Conversely, when trained on high arousal emotion such as Angry, we observe the decline in the opposite direction. However, this trend is not witnessed in other models such as the final layer of Wav2Vec2 and Data2Vec. Finally, we have seen that in a lower dimensional space most models failed to encode clear clusters of speaker-related information such as identity and sex. That might indicate that extreme variations in emotions and expressions might impair identity processing for models and [41] have shown such effect in humans. However, it was interesting to add more evidence supporting our claim regarding the final layers of the three mentioned models are phonetically and linguistically rich. As shown in Figure 2.18, The final layers of these models created clear clusters of sentences uttered regardless of the speaker identity and the spoken expression which is not the case in the latent layers. This again supports the fact that these models might be learning hierarchical information spanning frequency to phonological representations.

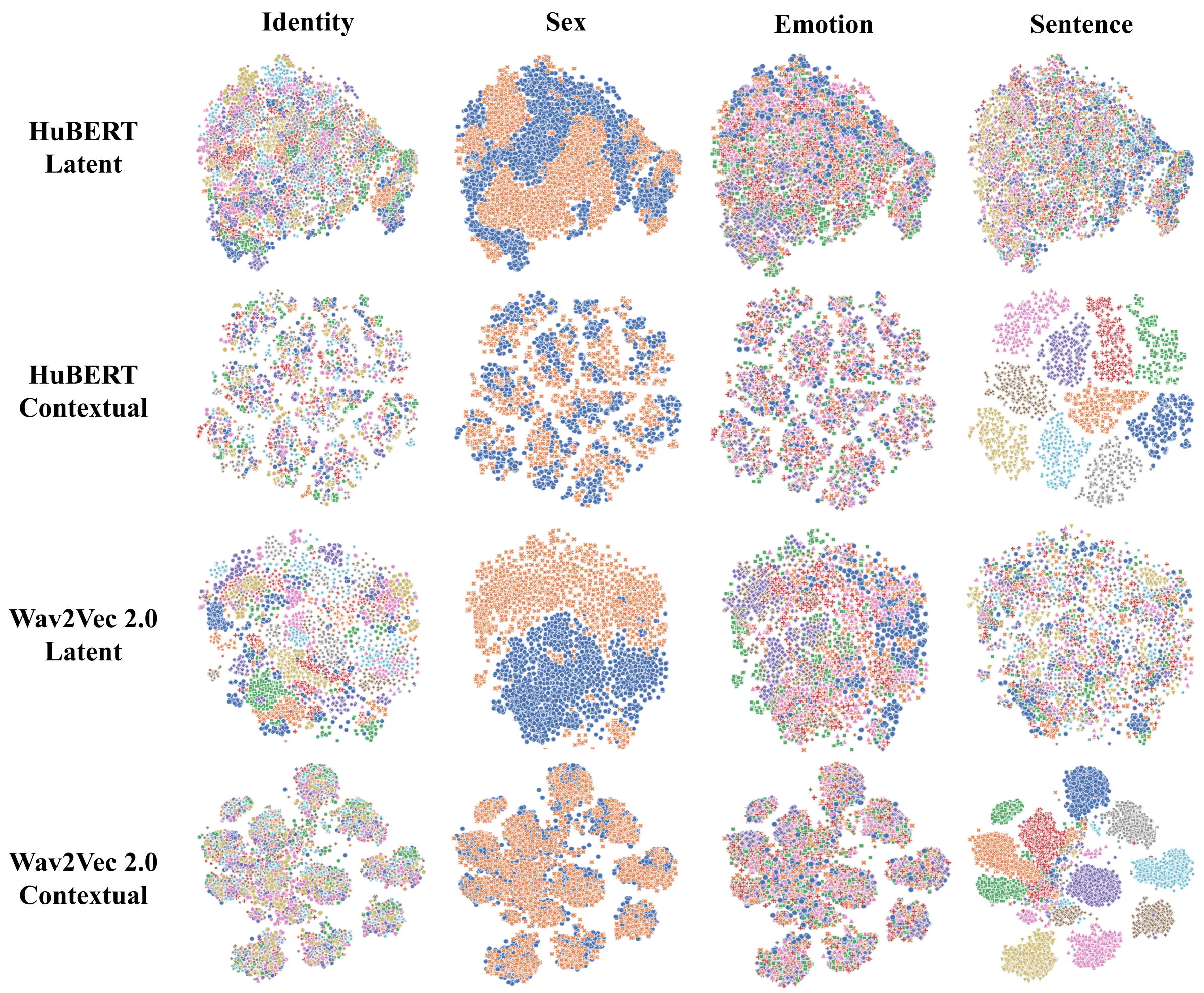

Figure 2.18: t-SNE plots for emotion speech experiment. The rows indicate the candidate representations. The columns indicate different labels. Identity: each color indicate a unique speaker (total 20 speakers), Sex: Male speakers in orange and Female speakers in blue, Emotion: 6 emotion categories, Sentence: the sentences uttered by speakers (total 10 sentences repeated in different emotions).

### 2.5.5 Speaking Style

In this section, we explore the differences between read/scripted speech and casual/spontaneous speech. Both speaking styles feature minimal vocal variations yet impactful when it comes to identity processing. It has been observed that speaking style could affect vocal identity processing in humans for unfamiliar voices. Previous studies [57–59] showed that in voice discrimination tasks humans tend to perform better when pairs of stimuli have the same speaking style (i.e. read-read or spontaneous-spontaneous). Also, they showed that humans perform better with read speech than spontaneous speech. Accordingly, we investigated the effect of speaking style on models' ability to recognize speakers.

**Methods:** We used a subset of the ALLSSTAR corpus [121] containing 26 speakers (14F/12 M). Each speaker was asked to read passages and sentences from stories (read/scripted utterances) and tell personal stories and answer random questions (spontaneous utterances). We chunked utterances to 3-sec clips.

In a similar vein to previous experiments, we extracted the clip embeddings from all models and trained two shallow classifiers. One was trained and tested on read clip embeddings and the other was trained and tested on spontaneous ones. That way we can compare which speaking style is better for recognizing speakers. We balanced the number of utterances used in both categories.

**Results:** In Figure 2.19, we report the UAR% of the test sets in both speaking styles, for all models.

**Discussion:** From Figure 2.19, one can notice similar trends across most models of slightly higher performance in case of read speech compared to spontaneous speech. However, this is not the case for Wav2Vec2 and Data2Vec but they feature overall very poor performance to recognize speakers.

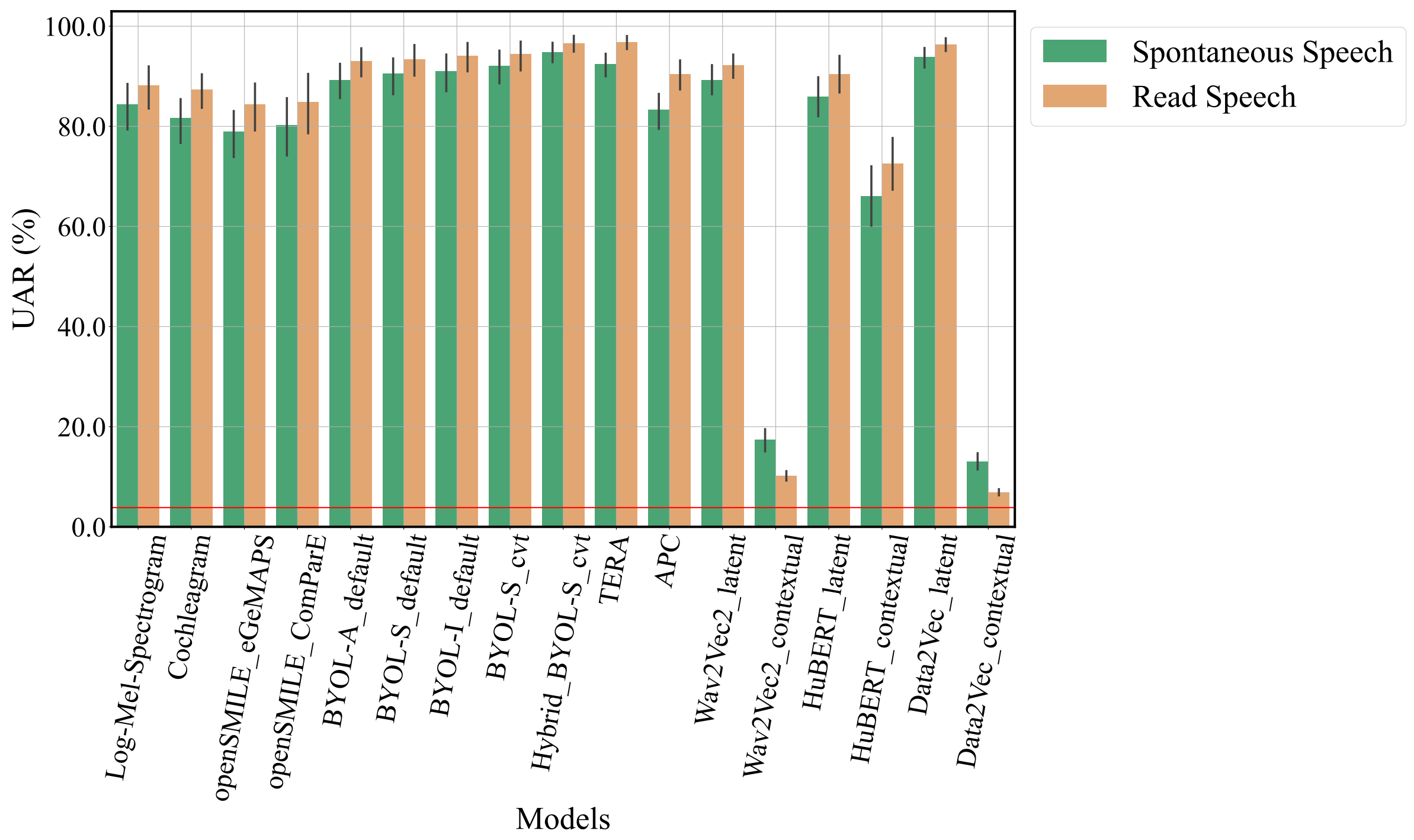

Figure 2.19: Performance of models on the ASpR task for read and spontaneous speech. The error bars indicate the standard deviation of the UAR across all hyperparameters sequences. The red line indicates chance level performance.

### 2.5.6 Audio Background

In this final experiment we investigate the effect of background noise on models' performance and the ability of models to capture identity features that are transferable in a noisy environment.

**Methods:** [122] created a noisy version of CSTR VCTK corpus [112] with babbling noise in the background. We used the utterances with clear and babbling backgrounds of 10 speakers (5F/5M) and extracted their embeddings for each model. Similarly, we trained a classifier on clear background utterances and tested it on clear and babbling background utterances.

**Results:** Figure 2.20 shows the UAR% for each model in both background environments, clear and babbling. Low-level acoustics such as spectrograms and cochleagrams are, relatively speaking, the least affected by the background noise. BYOL-S with CvT as an encoder shows better performance on babbling noise compared to BYOL-S with a CNN encoder.

**Discussion:** As expected, all models were impacted by the background noise, yet with varying effect sizes. The better performance of BYOL-S with CvT highlights the contribution of encoder architecture to the robustness of the learned representations. Also, low-level acoustics such as spectrograms and cochleagrams are, relatively speaking, the least affected by the background noise.

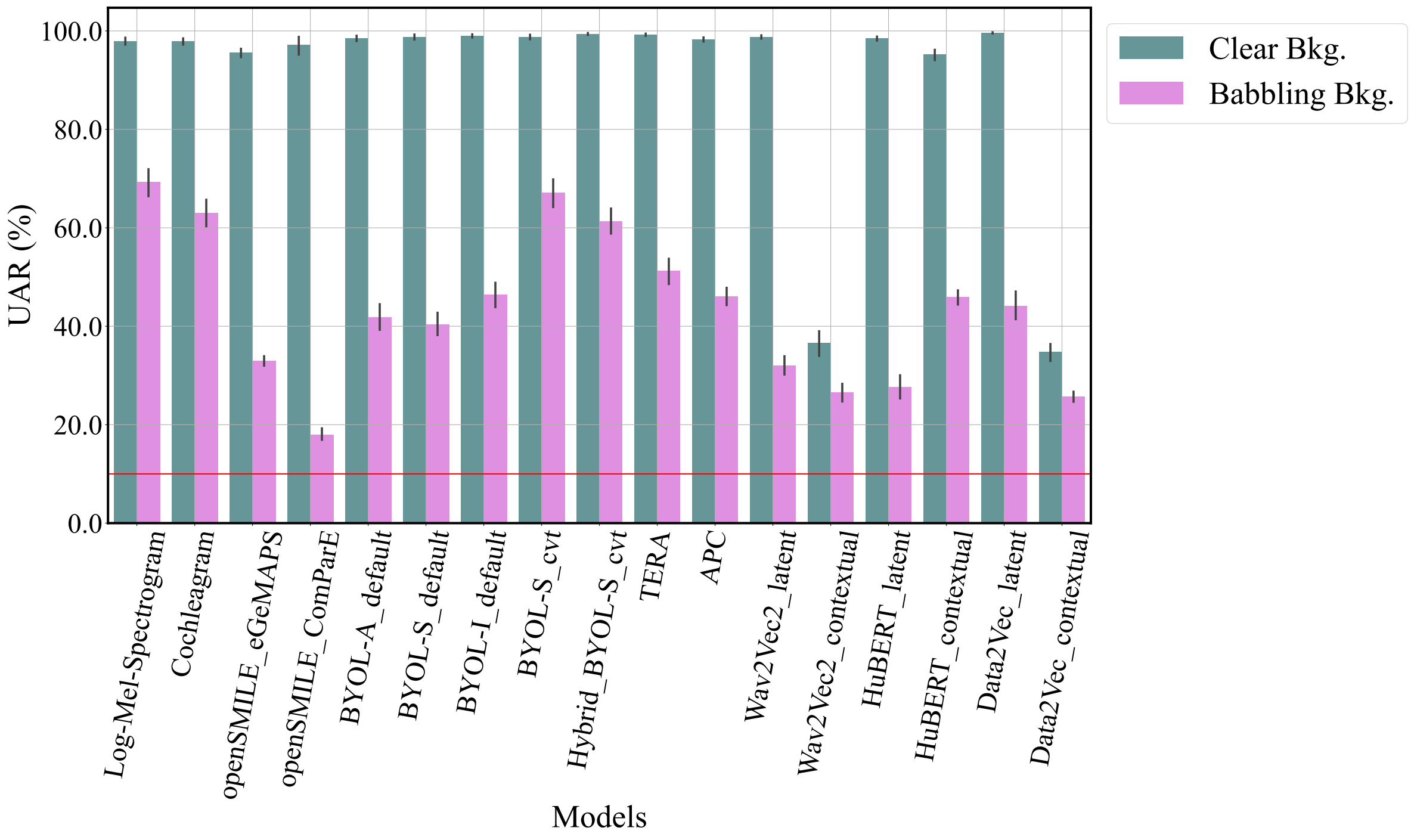

Figure 2.20: Performance of models on a ASpR task in case of clear and babbling backgrounds. The error bars indicate the standard deviation of the UAR across all hyperparameters sequences.The red line indicates chance level performance.

## 2.6 Conclusion

In this chapter, we showed that representational spaces from SSMs embed significant information about speaker identity and can be related to multiple aspects of human perception of voice-based identity. The results demonstrate the predictive power of the representations across and within models on large-scale ASpR tasks. We see that some models showed significantly higher performance compared to acoustic features. The poor performance of the final representational layer of some powerful SSMs led us to discover that such models maximize speaker identity information in earlier layers of the network, thus offering a novel mechanism to better understand and use representations from different layers of these models. Using a set of experiments to examine the effect of perturbing specific and salient acoustic, phonemic, prosodic, and linguistic characteristics of speech on speaker recognition performance, we demonstrated that many of these models' representational spaces can retain significant information about identity, and that both dynamical variation in speaking and similarity of language offers additional benefits to speaker recognition. Both of these outcomes match current experiments on human behavior. These findings collectively provide significant insight into possible perturbations to reduce identifiability, and the potential for such model performance as analogs of human behavior.

# Chapter 3

# Voice Encoding in Humans and SSMs

## 3.1 Introduction

Following the assessment of the goodness of multiple SSMs to capture speaker identity traits under several intrinsic and extrinsic variations, we sought to study the models' encoding spaces as analogous to the perceptual space of humans. As we discussed in Section 1.5, previous work in literature has reported cerebral [95] and behavioral [93] correlates to acoustical distances, mainly using euclidean distance, as a proxy for studying the similarities between an acoustic space and a perceptual space. However, in this chapter, we would like to challenge this notion of distances since we hypothesize that computing linear distances such as euclidean might fail to capture the non-linearities or other lower dimensional manifolds of the perceptual space. Furthermore, we argue that using a non-linear learnable decision model (decoder) instead of a linear distance metric might better explain human perceptual judgements. As we are conscious of the distinctiveness of both systems in terms of learning dynamics and temporal precision, we still hypothesize that both spaces might exhibit similar judgements on a behavioral level. This hypothesis is substantiated by the similarities in perceptual trends between models and humans demonstrated in the previous chapter. Accordingly, we believe that these similarities might reflect compatible coding, perhaps hierarchical coding as well, of vocal characteristics. In the previous chapter, we studied the efficacy of models in the context of speaker recognition, whereas in this chapter, we focus on a speaker discrimination task. Thus, we can evaluate the ability of models to tell people *apart* versus tell people *together*. Here are the research questions we are tackling in this chapter:

- Is there a correlation between linear distances computed in the embeddings space and the

perceptual space of humans?

- Does learnable decision models explain human behavior better than linear distance metrics?
- What are the commonalities and differences between the representational spaces of the models and humans?

Accordingly, we divided this chapter into three main studies that aimed to juxtapose encoding spaces of models and humans to evaluate their similarities as well as examine the use of distance metrics in this context. The first study involves analyzing the positioning of voice pairs in each respective space and investigating the appropriateness of different distance metrics. Testing this approach will encompass computing different distance metrics between pairs of speakers' embeddings extracted from a particular model. The distances, in this context, indicates the proximity of speakers within the model's encoding space. Then, provide these pairs as stimuli to test humans' ability to discriminate between speakers. Accordingly, we can simply study the correlation between embeddings' distances and human performance. A positive correlation would indicate that the more distant a pair of speaker embeddings are the easier it is for humans to tell them apart and conversely more challenging for smaller distances, indicating similar representational spaces. This approach is illustrated in Figure 3.1 which has been adopted in previous work using acoustic features/spaces instead of data-driven spaces [91, 94, 95] (see Section 1.5 for more details).

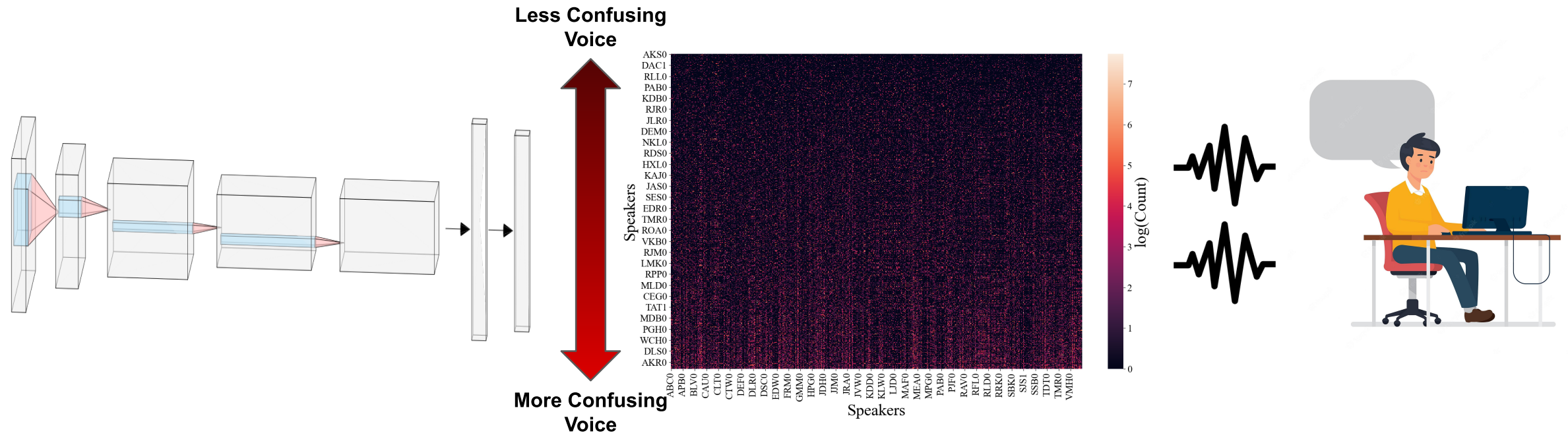

Figure 3.1: Compare distances between voice embeddings from a model to human performance.

The second study is built on some of the findings discussed in the first one. Here, we compare the distances across utterances in a model's space with a non-linear transformation of this space developed for speaker recognition. We train the same model on a speaker recognition task and we evaluate the correlations between the model's ability to recognize/confuse speakers with the distances across these speakers. Thus, in the second study, we compare two spaces of the same model, an encoder space and a decoder one.

The third study is aimed to train a learnable decision model on top of the encoder space to discriminate between speakers, then test the performance of both models and humans on the same discrimination task, as shown in Figure 3.2.

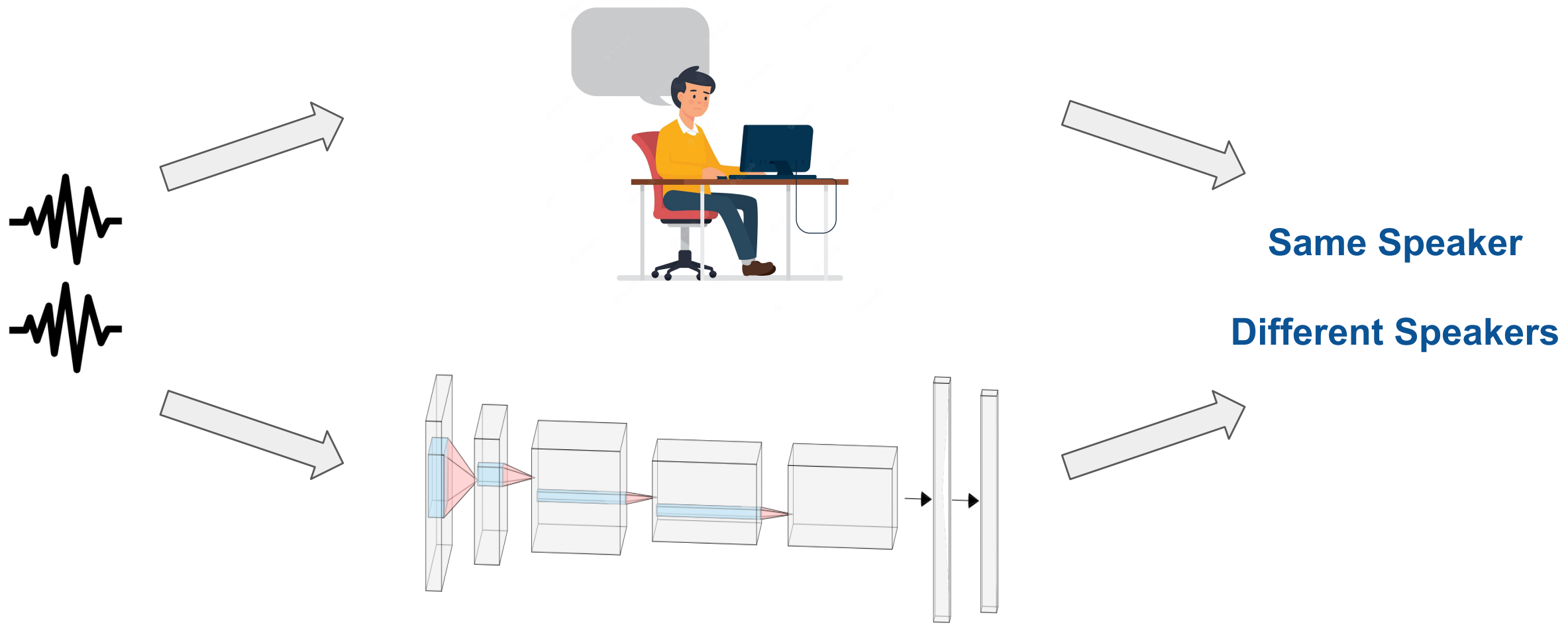

Figure 3.2: Compare Behavioral judgements of both models and humans.

## 3.2 Study 1: Investigating Distance to Behavior

In the first study, we extracted the embeddings of speaker utterances from different datasets using the best-performing model on VC1 benchmark which is an intermediate layer in HuBERT model (transformer layer 8). Then, we computed multiple distance metrics across pairs of embeddings to sample some of them as stimuli for a behavioral experiment conducted on humans. In the upcoming subsections, we explain in detail the selection of stimuli, the behavioral experiment design, and the results of this study.

### 3.2.1 Stimuli Selection

The aim of this subsection is to prepare the stimuli that will be used in the behavioral experiment. To do so, we performed some analyses on multiple datasets to evaluate how a model can *position* a speaker in its encoding space. We compiled 3-sec speech utterances from three corpora; LibriSpeech [123], VCTK [112], and TIMIT [110]. Due to the paucity of 3-sec clips in LibriSpeech corpus, we chunked the utterances from the test set of the corpus into 3-sec long clips, yielding $10,050$ utterances. For VCTK and TIMIT, we compiled $4,384$ and $655$ utterances that were 3-sec long. Thus, we selected a total of $15,089$ samples from $599$ speakers to be analyzed and preprocessed.

### Pre-processing and Feature Extraction

The pre-processing included three main steps. First, we normalized the loudness of all utterances using `pyloudnorm` python package[1]. Then, resampling the utterances to 16kHz using the resampling function in `librosa` python package [124][2]. Finally, convert the audio arrays into `torch` tensors.

After pre-processing the samples, we fed the audio tensors to a pretrained HuBERT model to extract the embeddings from the layer of interest (transformer layer 8). Subsequently, we explored the hierarchical relations between all speakers' embeddings by visualizing the hierarchical clusters using `clustermap`, as shown in Figure 3.15.

It looks like the speakers clustered together are usually from the same dataset (the color code is for dataset). This is also reflected in the similar patteren inthe embeddings values. Each dataset has its unique value pattern as shown in the heatmap texture. Thus, it seems that the model is also capturing dataset-specific information, perhaps microphone setup, and that's interesting since the three selected datasets have relatively clean and anechoic backgrounds yet the model could observe some differences. We further explore these dataset differences by computing different distance and correlation metrics.

### Computing Distance Metrics

The distance and correlation metrics we considered are euclidean distance, cosine distance, hausdorff distance, and spearman correlation, the equation for each method can be found in Appendix 5.2. We computed these four metrics across all possible pairs of embeddings. Then, we averaged the distances for each speaker to evaluate how far one speaker from the rest. We show the results of computing euclidean distance while the rest of the metrics are documented in Appendix (see Figures 5.7, 5.9, and 5.11). The results are shown in the form of a heatmap for all $599$ speakers and their distances between each other as shown in Figure 3.3.

[1] https://github.com/csteinmetz1/pyloudnorm
[2] https://librosa.org/doc/latest/index.html

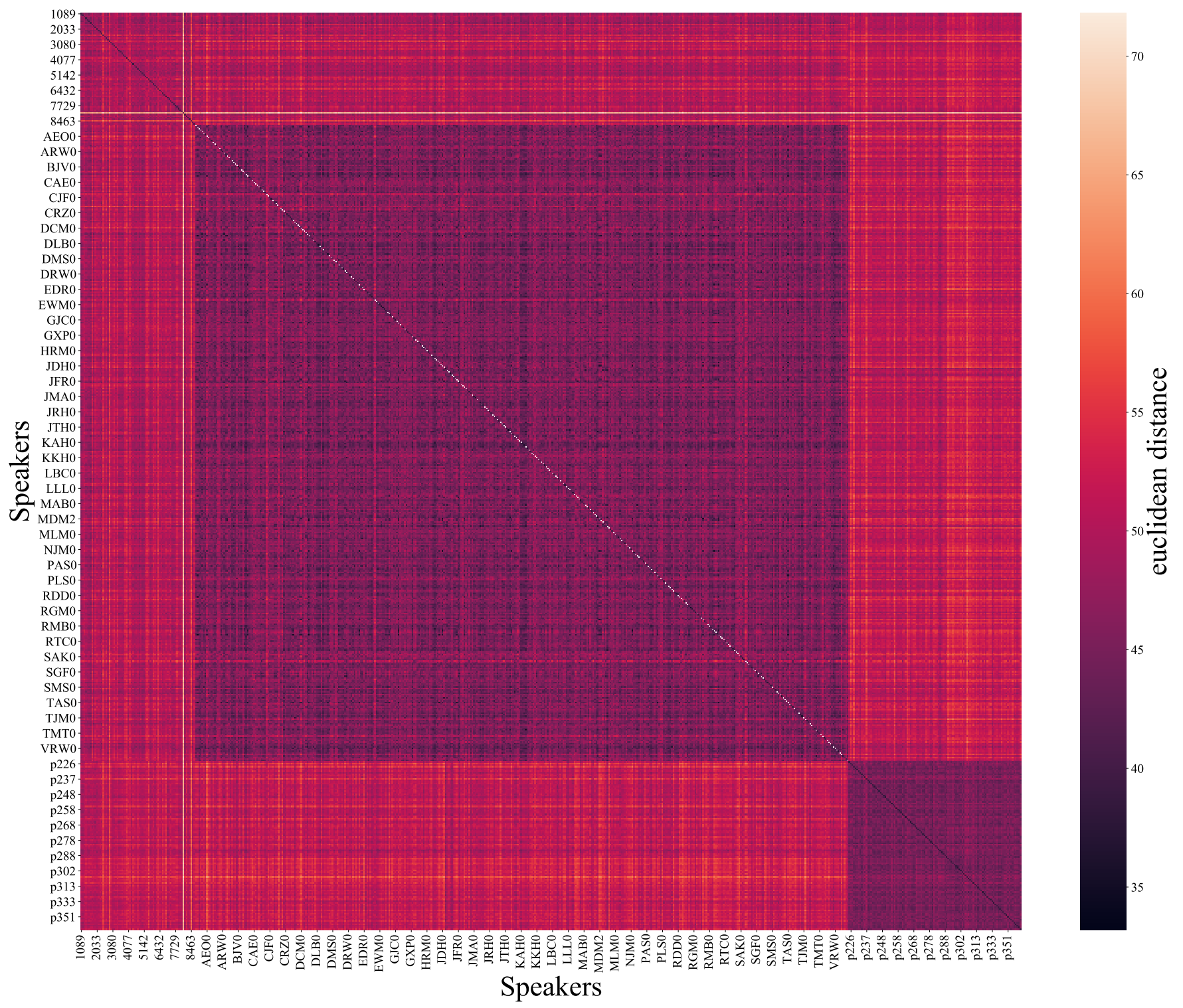

Figure 3.3: Heatmap showing pairwise euclidean distances between $599$ speakers.

Figure 3.3 confirms the dataset-related effect. As shown above, speakers from the same dataset have smaller distances between their embeddings (e.g. big dark box is TIMIT and the small one is VCTK). However, it is not quite evident with Librispeech.

Furthermore, we discarded the distances computed between same sex (e.g. to avoid trivial/easy trials in the experiment), different speakers uttering same sentence (to avoid similarities due to context), and same speaker uttering same sentence (i.e. same file, hence distance=0). After filtering these pairs, we plotted the distribution of standardized distances that were computed between pairs of different speakers against the ones computed between pairs of same speaker samples, as shown in Figure 3.4. The histograms for the other metrics are shown in Appendix (see Figures 5.8, 5.10, and 5.12).

We sampled pairs from the overlapping region for every metric assuming that these pairs might be challenging for humans. Then, we selected the common pairs across them resulting in $1,122$ pairs of speakers ($670$ same speaker pair and $452$ different speakers pair). We computed a score for each pair. That score was an average of all metrics and defined as follows:

$$Score = \frac{Euclidean + Cosine + Hausdorff + (1 - Spearman)}{4} \tag{3.1}$$

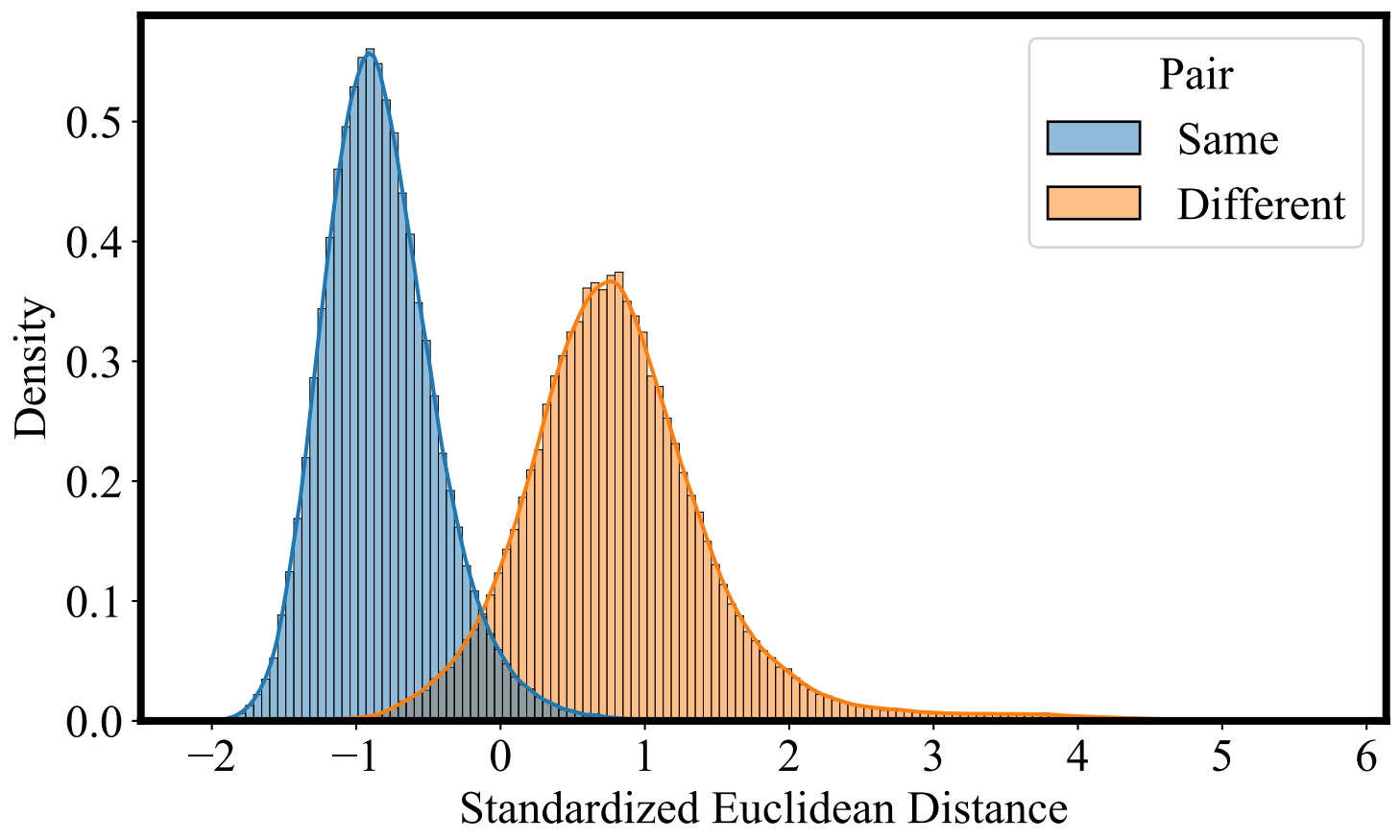

Figure 3.4: Distribution of standardized euclidean distances for *same* and *different* pairs.

That score basically indicates the proximity of two samples within one pair, the larger the score the further the samples from each other based on the selected metrics. With this, we selected the largest 50 *same* pairs and the smallest 50 *different* pairs, as illustrated in Figure . We manually listened to all 100 pairs and removed one pair because the content was semantically related. This pair was replaced with the subsequent pair in the score ranking. We adopted this strict criteria for stimuli selection in order to make the discrimination task challenging for humans. Lastly, to avoid subjects picking up microphone or background settings when listening to the stimuli, we added background white noise in only one stimulus of a pair within a trial. We balanced the order of which sample will be noisy across all trials. Accordingly, we have 50 trials with the first stimuli being noisy and the second one is clean while the other 50 trials have the noisy stimuli in the second one.

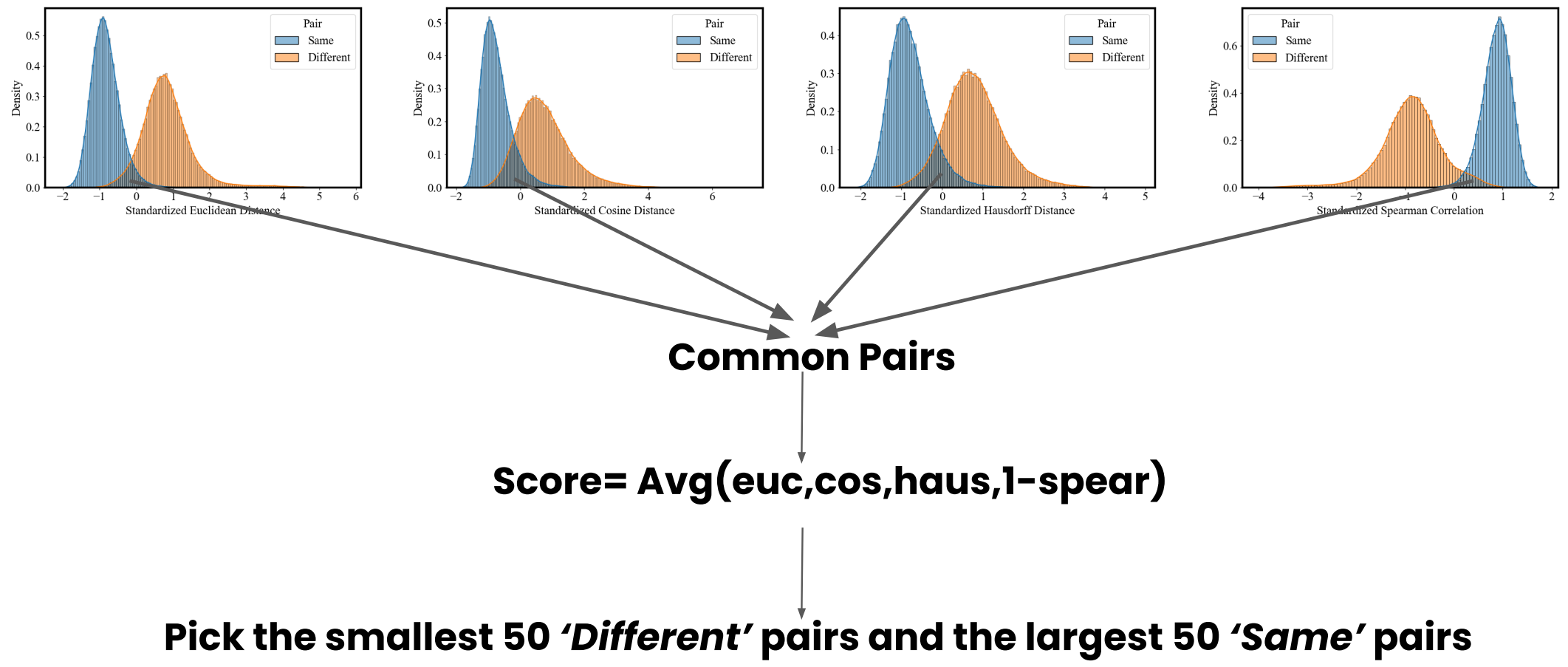

Figure 3.5: Steps for stimuli selection using multiple metrics.

It is worth noting that the final pairs of utterances selected for the behavioral experiment turned out to be all from the VCTK corpus. This might again relate to the dataset-specific distances we noticed in all metrics computed.

### 3.2.2 Procedure

The study was implemented using the online platform[3] GORILLA [125]. We designed a voice discrimination task in which subjects listen to a pair of 3-sec voices and report if the voices come from the same speaker or not. We also asked subjects to rate their confidence regarding their response from $1-7$ with 7 being most confident.

The experiment comprised 100 trials selected as discussed in Section 3.2.1. The trials were balanced per condition (*Same* and *Different*) and order of noisy and clean backgrounds (*Noisy-Clean* and *Clean-Noisy* trials). All trials were randomized for each subject. The experiment allowed the subject a 1 min break after every 25 trials.

Knowing that this is an online experiment, we added screening tests to ensure that the subjects have the required audio settings for this task. First, a browser autoplay test is conducted to check if the audios can be played smoothly during the experiment. To guarantee the data quality received, we asked subjects to be wearing headphones throughout the experiment, which was verified using a headphone screening test. The test comprises 6 trials in which subjects will be presented with three white noise clips with gaps between them. One of the noise clips has a hidden faint tone within and the task is to report which clip contains the tone [126], which can only be done reliably when wearing headphones. The criterion for passing the screening test is to answer all 6 trials successfully. Participants were given two chances in this test before they were screened out. All 6 trials were presented in a random order. A pilot was conducted with 7 participants to test the efficacy of the experiment.

### 3.2.3 Participants Recruitment

We recruited participants for this experiment via the Prolific platform[4]. A total of 120 participants completed the study. We specified that the subjects should be located in the USA and have a minimum approval rate of 90% in previous experiments. After completing the study, we manually checked the submissions of each participant to respond with either approved or rejected. Our rejection criteria included:

[3] https://gorilla.sc/

[4] https://www.prolific.co

- Failing the screening tests (Browser Autoplay Check and Headphone Check).
- Exiting the experiment early.
- Timing out (taking more than $40$ mins to finish the experiment).
- Giving intentionally low-effort responses.

For the first and second items we didn't need a manual check the subject was automatically rejected from the experiment. However, before rejecting subjects, either manually or automatically, we suggested that they return their token instead to avoid harming their approval rate record. Nevertheless, we immediately rejected subjects that intentionally gave a low-effort response since they had already consumed a token from our Gorilla account. The way we judged if the subject performed poorly intentionally is by computing the $d'$ for each subject as discussed in Appendix 5.2. In the end, we approved a total of $105$ submissions. We only paid approved subjects for their participation. We paid them based on the amount suggested by Prolific which was $12$\$/hour. We estimated our experiment to take 30min, thus, we paid approved subjects $6$\$ each. This resulted in a final dataset of $105$ subjects ($38$ female and $67$ male) with average age of $36.5 \pm 11.8$.

### 3.2.4 Results

In summary, the $105$ participants performed the task with an average accuracy of $78.21\% \pm 6.56$ and with an average sensitivity (d') of $1.68 \pm 0.48$. The average reaction time to respond was $2.59 \pm 1.27$ sec. In the following sections, we further investigate the performance of participants. For statistical analysis, we reported Cohen's d to elucidate the effect size across different conditions.

#### Telling People *together* vs. *apart*

In this analysis, we studied the performance of participants to discriminate between different speakers or identify same speaker in different instances. As shown in Figure 3.6, participants performed significantly better to *different* trials compared to *same* ones (Figure a). Also, participants responded correctly when they were more confident of their answer (Figure b).

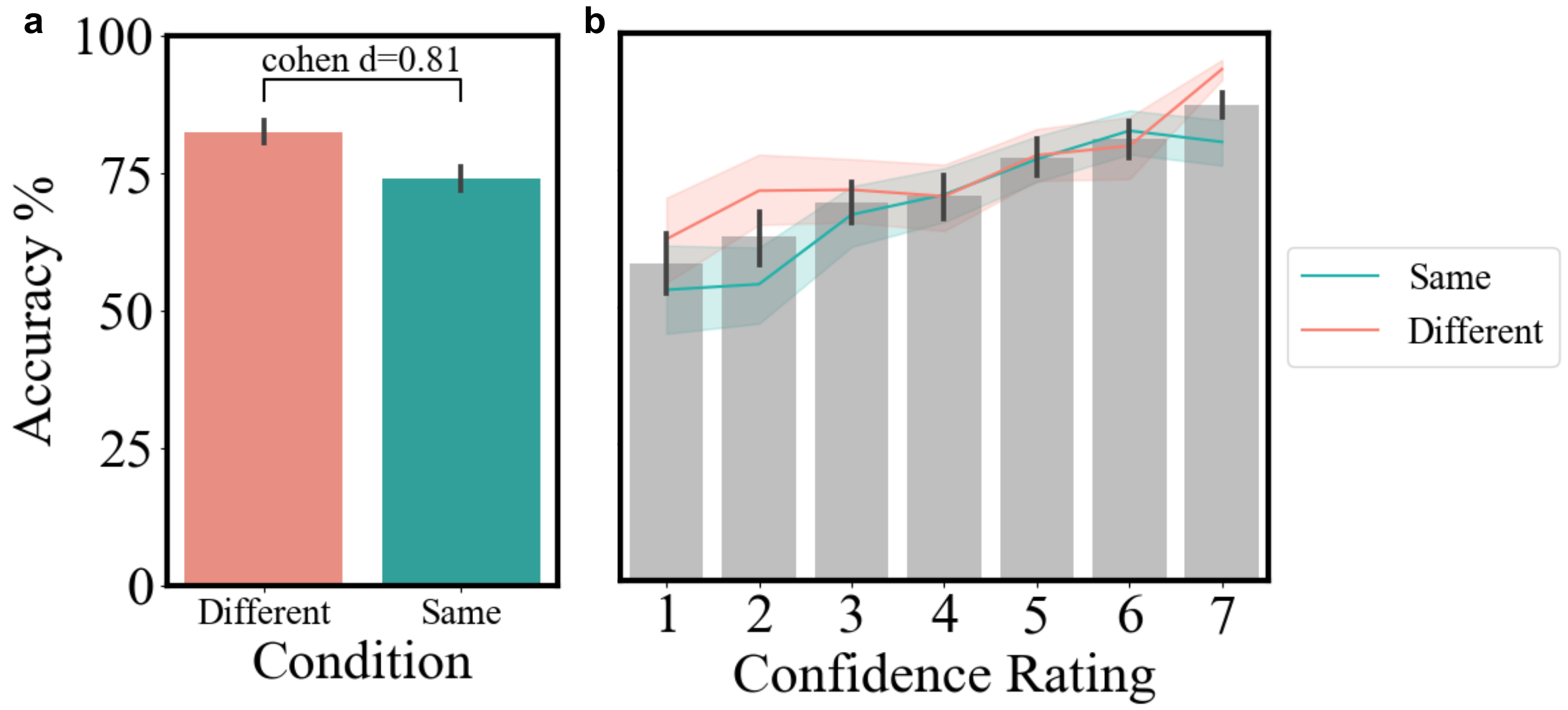

Figure 3.6: (a) Performance Accuracy between two conditions (same and different). (b) Performance across different confidence rating.

**Does the order of background setting affect performance?**

Subsequently, we reported the performance of participants when presented with clean background stimuli followed by a noisy one (C-N) compared to the opposite (N-C). Figure 3.7.a shows no significant differences between both orders. However, Figure 3.7.b shows an increase in effect size between Same and Different trials when participants are presented with noisy stimuli first which is illustrated further across different confidence ratings (Figure 3.7.c).

**Model's Distances and Human Performance**

Lastly, we aimed to answer the original question of this chapter if distances in the encoding space of models correlate with human performance. Thus, for each trial, we reported its average performance across all participants against distance and correlation metrics. Accordingly, Figure 3.8 shows the relation between multiple metrics with human performance for each trial/pair and we reported pearson correlation and p values for each sub-figure. We also showed the correlation for *Same* and *Different* pairs separately, see Figure 3.9.

### 3.2.5 Discussion

In this study, we asked the question if distances between speakers' embeddings in data-driven space explain or correlate with human perceptual space. Thus, we compiled utterances from different corpora to extract speaker emebddings using HuBERT best-performing layer in VC1 benchmark (see Section 2.2). Consequently, we noticed that speakers from same dataset were clustered together featuring similar embedding values and patterns, as illustrated in Figure 3.15. That observation might indicate that the model is capturing dataset-specific features, perhaps microphone setup. We further confirmed this observation by computing pairwise distances between all speakers as in Figure 3.3 showing very low distances between speakers belonging to the same dataset (see dark squares along the diagonal). Interestingly, The dataset-specific trend was not noticed between speakers from the LibriSpeech corpus. Knowing that LibriSpeech is a collection of utterances from different audio books which means different recording studios, hence, no dataset-specific features appear in the heatmap. Diversifying the pre-training data with different audio environments and augmentations might improve the model's invariance to dataset-specific configurations.

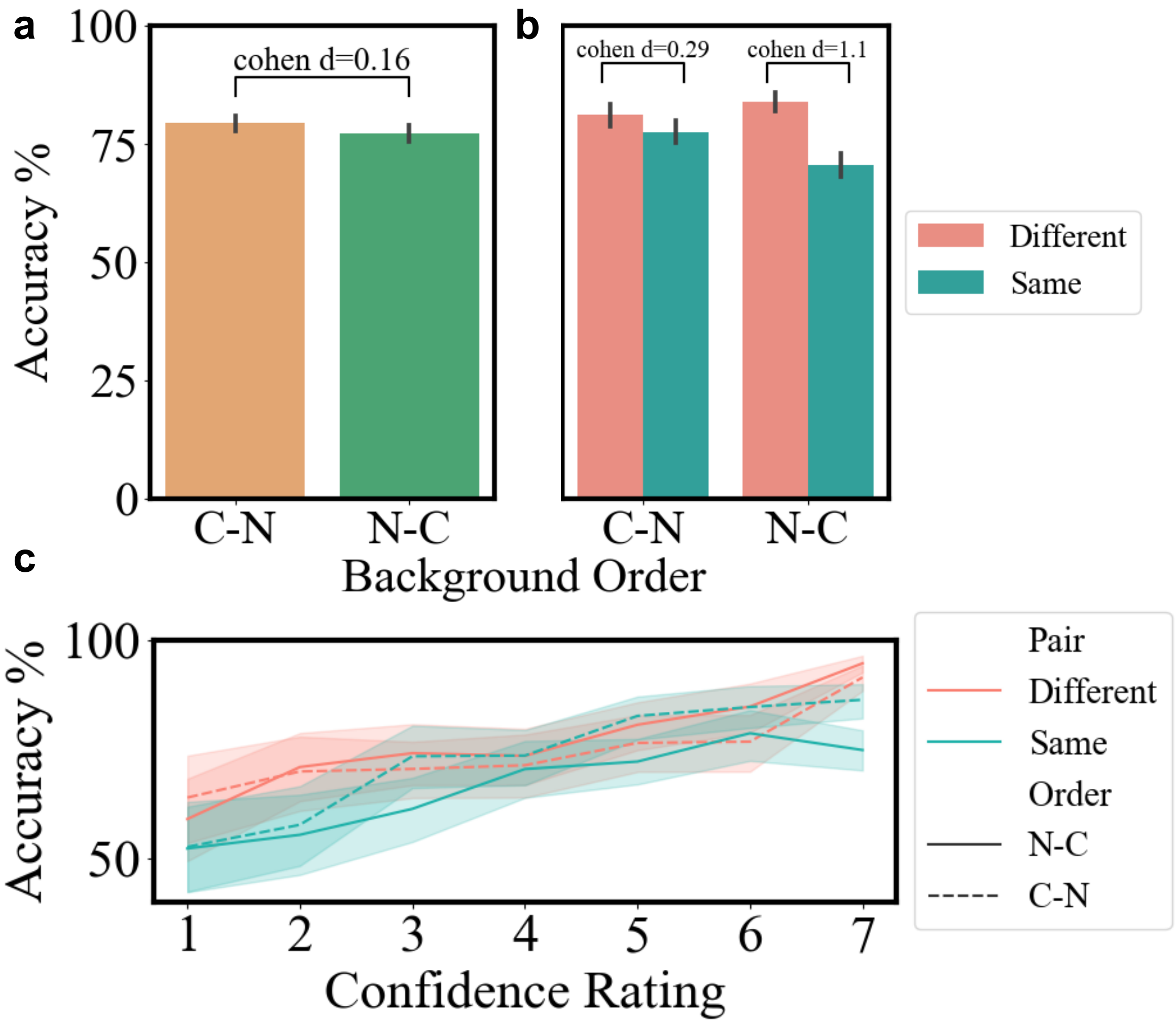

Figure 3.7: (a) Performance Accuracy between two background orders (C-N and N-C). (b) Performance on Same and Different trials in both background orders. (c) Performance across different confidence rating in all conditions and orders.

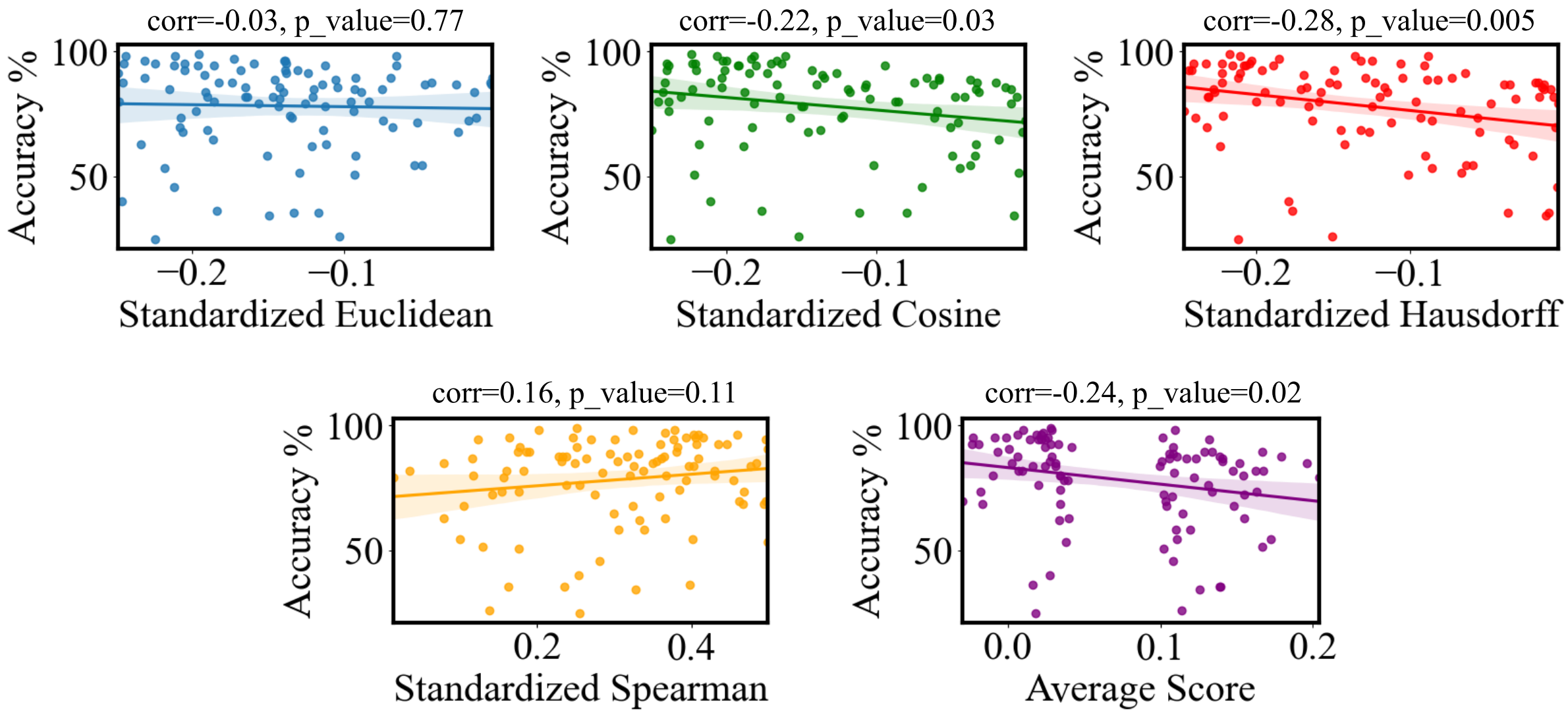

Figure 3.8: Average Accuracy is reported against multiple metrics (a) Euclidean Distance, (b) Cosine Distance, (c) Hausdorff Distance, (d) Spearman Correlation, and (e) Average Score of all metrics. Pearson correlation and p Values are reported for every sub-figure.

After compiling the pairs of speakers as trials, we conducted an online behavioral experiment where participants performed a speaker discrimination task. We found that participants were significantly better at telling people apart than together, as in Figure 3.6.a, which is aligned with previous results in literature [40, 41, 58, 59]. Additionally, we noticed that the differences between Same and Different trials are amplified when participants reported that they were either very confident or the least confident in their answers (see Figure 3.6.b). It is reasonable to believe that in order to identify the voices to be from the same person, more compelling evidence is required compared to just discriminating between different speakers, where any featural mismatch would be sufficient. Afshan and colleagues suggested that we might rely on acoustic variability when telling people apart while in case of telling people together we rely on speaker-specific idiosyncrasies instead [59]. Our stimulus duration was chosen to minimize the possibility of such idiosyncracies.

We checked if the order of stimuli with different backgrounds will affect performance. As expected, no significant differences between both orders was observed in Figure 3.7.a. However, it was noticed that the effect size between same and different trials was exacerbated in case of N-C order (see Figure 3.7.b). This indicates that when participants listen to the noisy clip first their ability to tell people together worsen. One possible interpretation might that in a noisy environment it might be difficult to capture speaker representations that transfer to different background setting, hence, easier to respond different instead of same. That being said, further

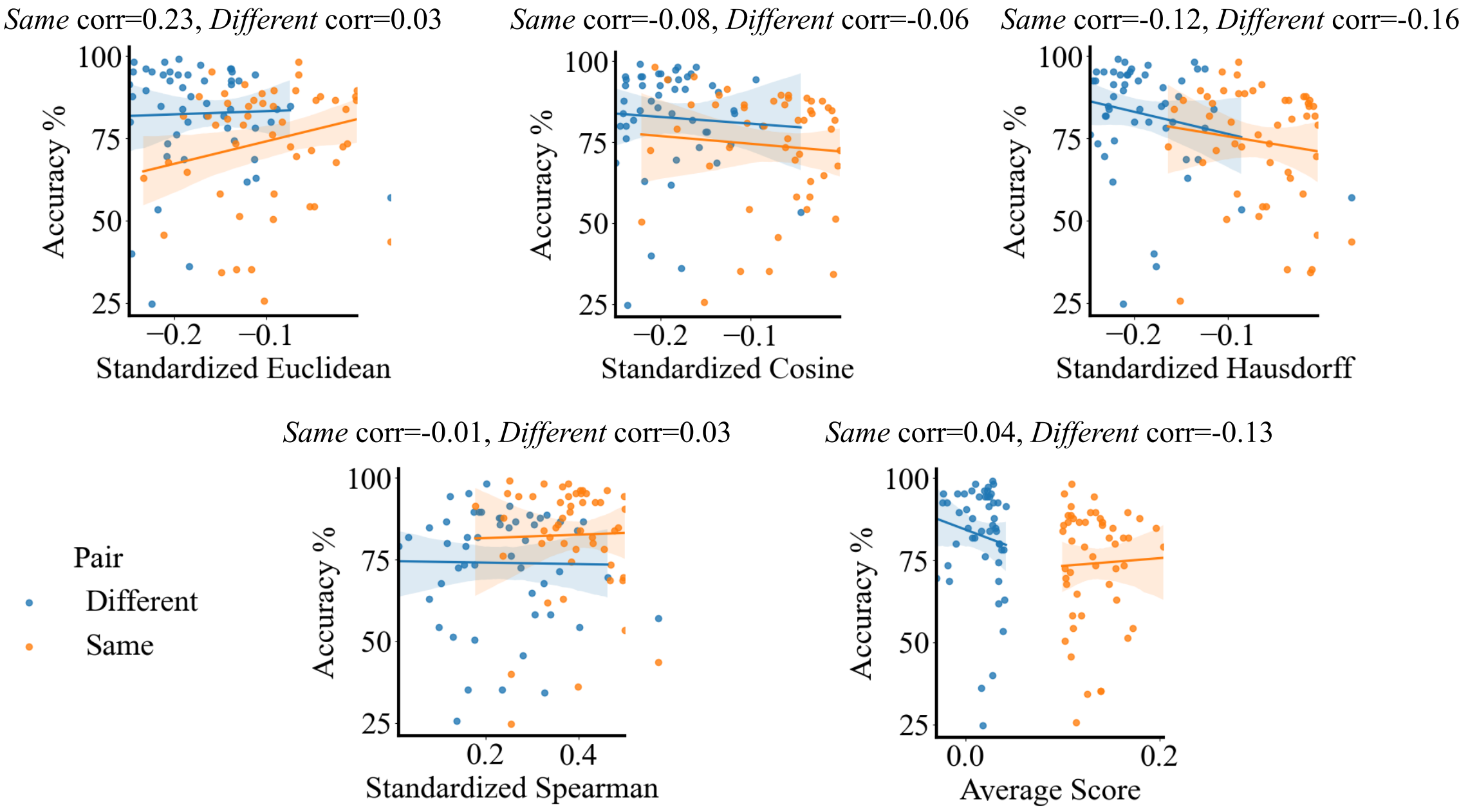

Figure 3.9: Average Accuracy is reported against multiple metrics (a) Euclidean Distance, (b) Cosine Distance, (c) Hausdorff Distance, (d) Spearman Correlation, and (e) Average Score of all metrics. Pearson correlation is reported for every sub-figure. No significant correlation, p Value > 0.1 in all cases.

analysis and experiments might be needed to study such behavior.

Lastly, we evaluated the relation between different distance and correlation metrics and human performance. As shown in Figure 3.8, no significant correlation has been observed for all metrics except for cosine and hausdorff but with low correlation. Also, we found no significant correlations when separating Same and Different trials, see Figure 3.9. This suggests that we might need to revisit the appropriateness of the distance metrics in the context of speaker proximity. These linear distances might not capture the complexity of the representational spaces of either models or humans. For instance, there is a hypothesis for faces implying the existence of a smooth non-linear manifold along which faces representations lie [127]. In a similar vein to face manifolds, one might suggest that voices might be constrained in a latent space and transitioning, or morphing, from one voice to another might require moving across a manifold instead of linear displacement. Further work is needed to confirm the voice manifold hypothesis.

## 3.3 Study 2: Distance and Behavior in Models

From the previous study, we showed that distance metrics does not correlate with human performance on a speaker discrimination task. In this study, we investigate if this finding is also valid between distances in an encoding space and model's behavior on a speaker recognition task. We used TIMIT [110] corpus to train the model on a ASpR. We selected this dataset it comprises large number of speakers (630 speakers) with balanced number of utterances per speaker (10 utterances) yielding a total of 6300 utterances. All utterances were downsampled to 16kHz and the loudness was normalized to $-23$dB.

### 3.3.1 Computing Distance Metrics

Similarly to the first study, we extracted the embeddings of all utterances using the best-performing layer in HuBERT (on TIMIT benchmark, see Appendix 5.2). It is worth noting that the optimal layer in HuBERT on TIMIT benchmark is transformer layer 6 while on VC1 was the subsequent one, layer 7. Cosine, Hausdorff, and Spearman were computed across all 630 speakers. We did not compute euclidean distance given the very low correlation demonstrated in the previous study 3.8.

Figure 3.10 shows the cosine distance matrix for all speakers. Similar matrices are shown in Appendix for the other metrics (hausdorff and spearman), Figure 5.13 and 5.14.

### 3.3.2 ASpR Training Procedure

Another way to evaluate the positioning of the encoded voices in a model's space is by evaluating which speakers the model gets confused between. One might hypothesize that speakers' positioning in the original encoding space is similar to the one in a decoder space (speaker recognition space), meaning the closer the speakers in a space the more difficult for the model to recognize them. Accordingly, we computed a confusion matrix to show the number of times the model misidentified one speaker with another. This was implemented by training a shallow layer decoder on recognizing speakers using the extracted embeddings. We bootstrapped that task for $10,000$ trials where in each trial we randomly sampled 7 utterances from each speaker for training and keep 3 for testing, see Figure 3.11 for illustration. We saved the predictions for each trial in addition to the ground truth. Then, counted the times the model misidentified each speaker with another speaker and show the results as a confusion matrix. For better visualization, we sorted the vertical axis of the confusion matrix from speakers with less number of misidentifications (top) to speakers with high number of misidentifications (bottom), as shown in Figure 3.12.

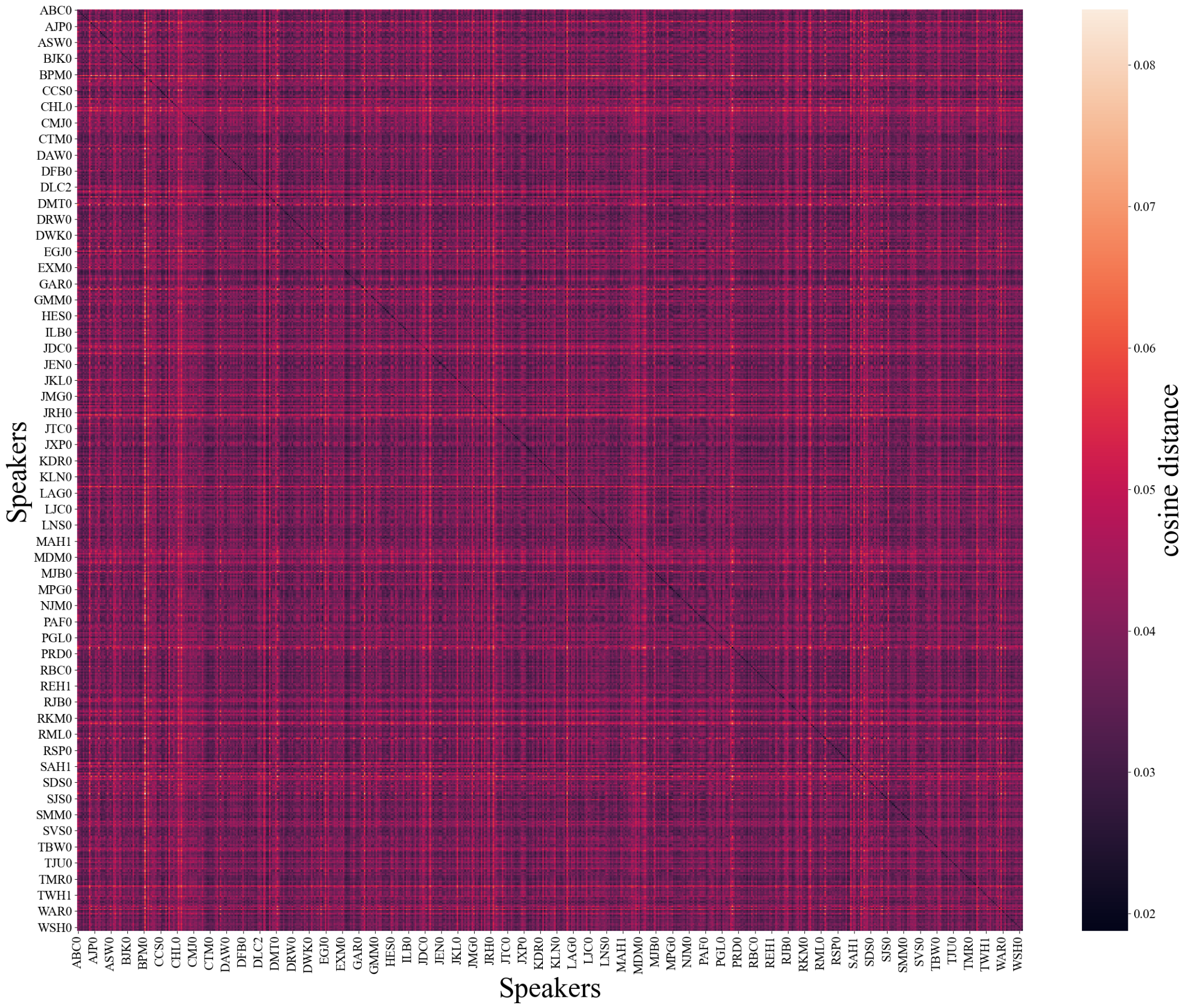

Figure 3.10: Cosine Distance Matrix across speakers.

### 3.3.3 Encoding Distances versus Decoder Performance

It is worth mentioning that the TIMIT corpus comprises utterances with variable durations. The duration average of clips is $3.1 \pm 0.8$ seconds. Accordingly, it is important to examine if clip duration is a confounding variable for model's performance. Thus, we plotted the relation across clip duration, the number of misidentifications for each speaker, as a proxy for model's behavior, and the distance and correlation metrics, as a proxy for speaker proximity. The results are shown in Figure 3.16.

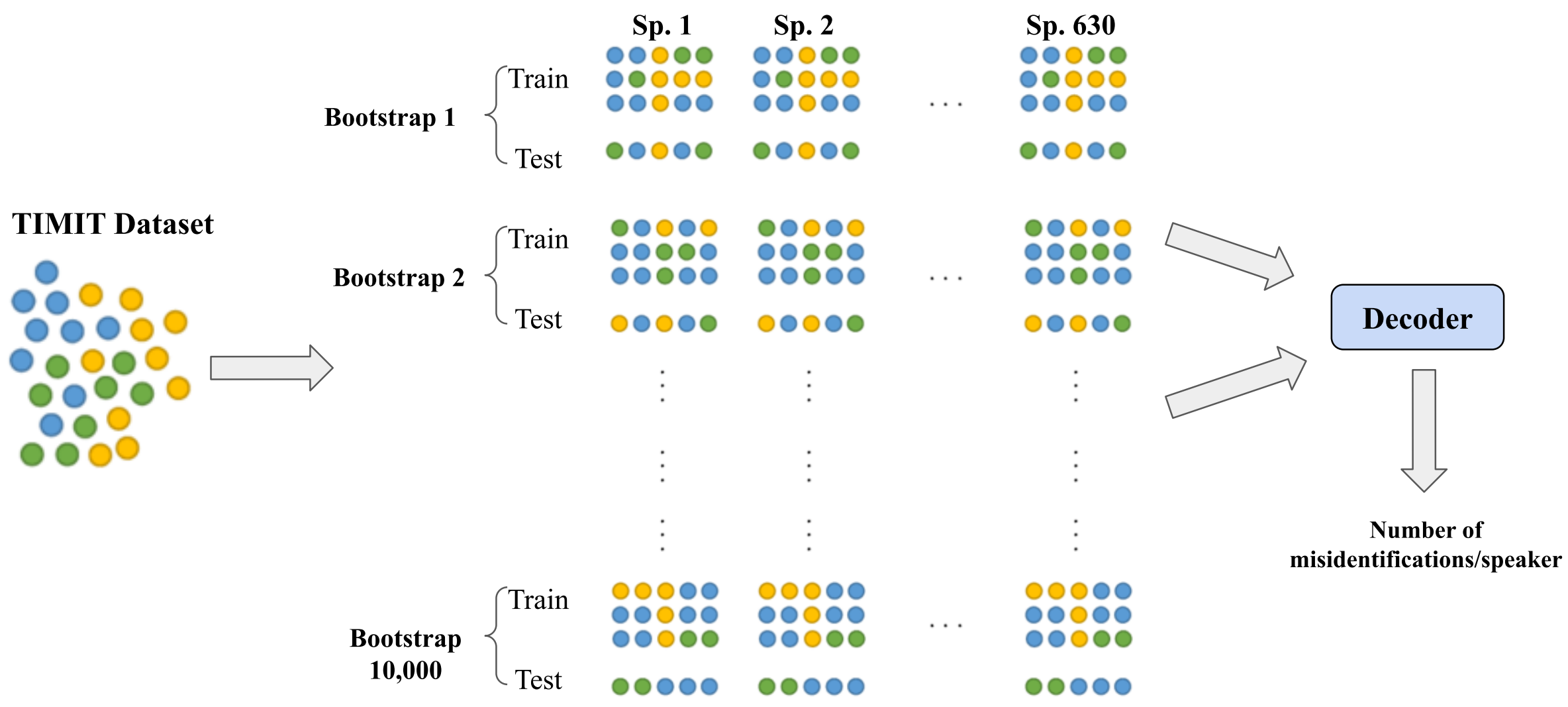

Figure 3.11: ASpR training using Bootstrapping for 10,000 trials. The decoder is a shallow linear layer. *Sp.*: Speaker.

### 3.3.4 Discussion

The duration of utterances showed no significant correlation with the number of misidentifications (pearson correlation=$-0.05$, $p_value = 0.22$). This indicates that the duration of clips did not affect the model's performance to recognize speakers in this case. As expected all distance and correlation metrics show high correlation amongst each other. Furthermore, it is interesting to see no significant correlation between number of misidentifications and cosine distances (pearson correlation=$-0.02$, $p_value = 0.65$), hausdorff distances (pearson correlation=$-0.02$, $p_value = 0.67$), and spearman correlation (pearson correlation=0.03, $p_value = 0.47$). This suggests that the decoder trained for ASpR task learned a different non-linear decision model to recognize speakers. These results supports the argument in the first study that linear distance metrics are not sufficient to model decision hyperplanes in humans (study 1) or models (study 2).

## 3.4 Study 3: Speaker Discrimination in Models and Humans

In the third study we replaced the linear distances with a learnable decision model. We trained a decoder on speaker embeddings extracted from multiple SSMs on an automatic speaker discrimination (ASpD) task. Consequently, the decoder will learn to transform the encoder (SSM) embeddings into a decoding space where a more intricate non-linear decision metric is learned and utilized instead of fixed linear distances. Furthermore, we evaluated the performance of these decoders against human performance from the behavioral experiment conducted in the first study (see Section 3.2).

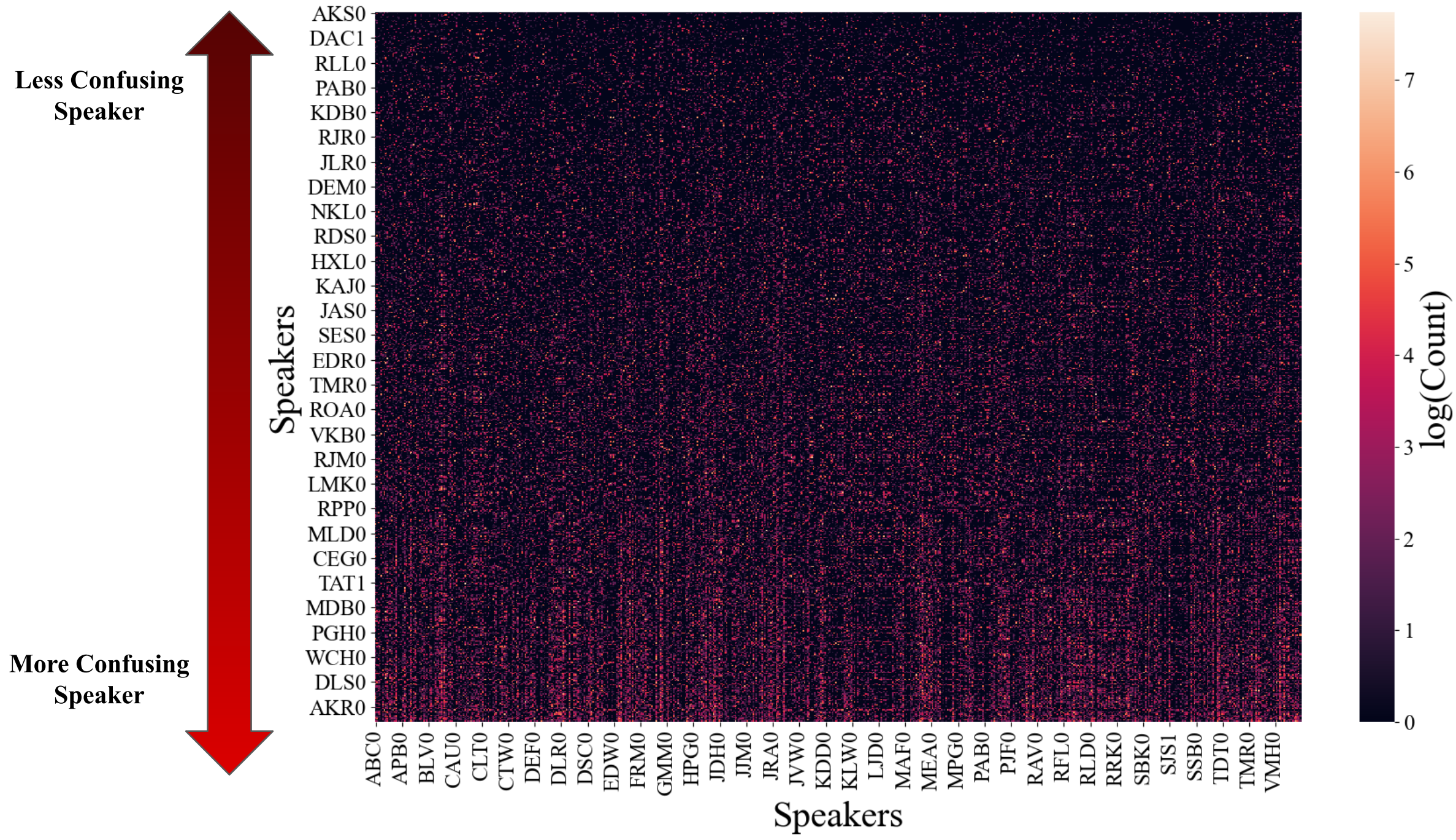

Figure 3.12: ASpR Confusion Matrix, sorted based on the total number of times the model misidentified a speaker. The logarithm of the number of misidentifications is used for better visualization

### 3.4.1 ASpD Training Paradigm for Models

The training included randomly selecting a 3-sec long speaker utterance from a corpus pool, then selecting another 3-sec utterance sharing the same speaker voice but uttering different sentence. Moreover, a third utterance is selected to be from a different speaker but sharing same sex yielding triplets of utterances. The triplets are then fed to the pre-trained encoder (SSM) to extract speaker embeddings and eventually concatenate the embeddings coming from the same speaker as well as the embeddings from different speakers. Consequently, we end up with two inputs to the decoder; one from the same speaker and the other one is from different speakers. The aim of the training is to feed these two inputs to the decoder to learn to discriminate between speakers (telling people *together* vs. *apart*) in a supervised manner (binary classification). It is worth noting that the extracted speaker embeddings are temporally pooled using *mean+max* pooling. The steps to train models on an ASpD task are illustrated in Figure 3.13.

In this study, we used LibriSpeech corpus [123], the decoder was trained on the 960 hours in the training set of the corpus (2338 speakers) and validated on the dev set (73 speakers). We chunked the utterances from LibriSpeech into 3-sec long clips and duplicated the dataset by adding background white noise to all clips. Given the enormous amount of pairs that could be generated from these clips ( 72 Billion pairs), we constrained the training to 1 million pairs per epoch balanced between both conditions (Same and Different) and 100K pairs for the validation set. To that end, the clips were sampled randomly leading to four different background setting orders (N-C, C-N, N-N, or C-C) while in the behavioral experiment we implemented two options for participants (N-C or C-N), see Section 3.2.

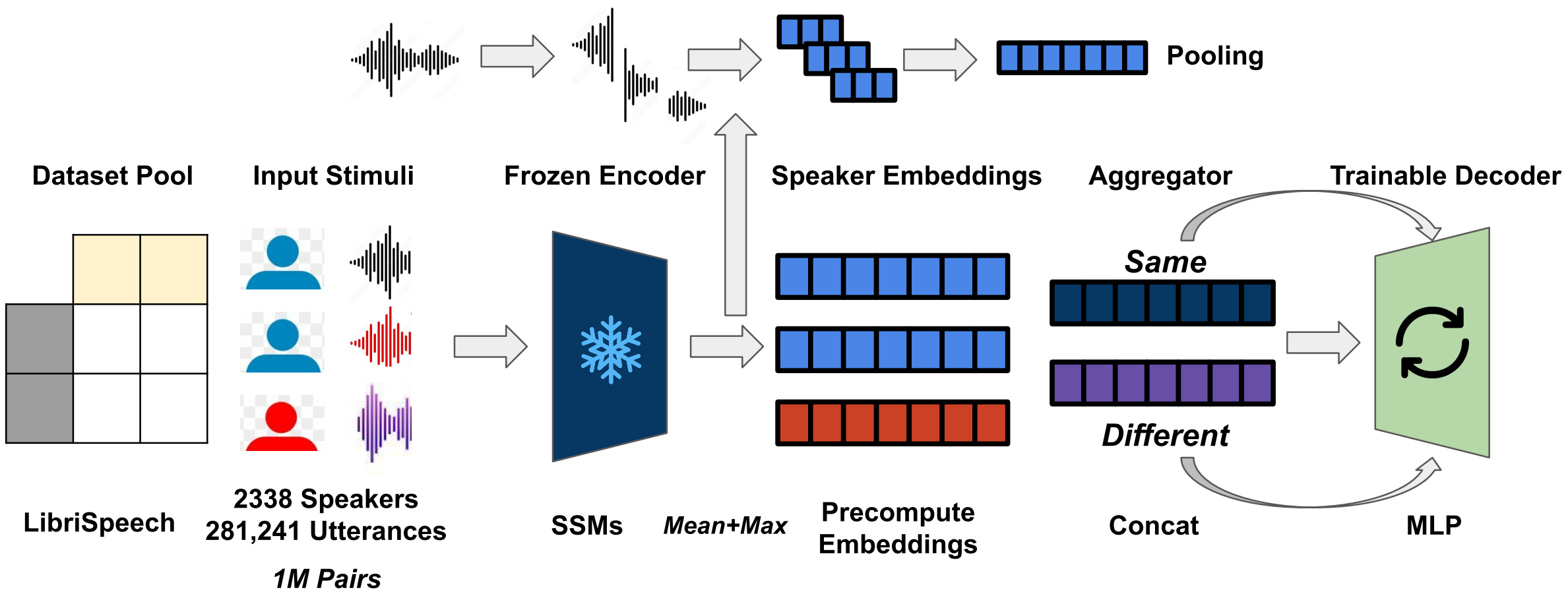

Figure 3.13: ASpD Training Pipeline for SSMs.

We selected two SSMs from the top-performing models in the speech experiments (see Section 2.5) as a proof-of-concept. Thus, we chose TERA [102] and Hybrid BYOL-S/CvT [103] as candidate models for this study. The decoder is a multi-layer perceptron (MLP) followed by a sigmoid layer. The decoder is trained to optimize a Binary Cross Entropy loss using `adam` optimizer. The loss and accuracy for the training and validation sets were reported. In this training paradigm, a set of hyperparameters were selected to be tuned on the validation set. To reduce the large space of hyperparameters, we divided the hyperparameter tuning into two training runs. First, we trained the decoder with the three encoders and tuned a set of hyperparameters that included batch size (64, 128, and 256), learning rate (1e-4 and 1e-3), and number of MLP layers ranging from one layer to four layers. We used Ray library[5] for hyperparameter tuning. The whole training was implemented using `pytorch lightning`[6] for 100 epochs with patience of 10 epochs in case of no improvement in the validation loss. For the two encoders, we found that using only one layer with batch size of 64 and learning rate of 1e-4 yielded the highest validation accuracy for TERA (82.5%) and comparable one for Hybrid BYOL-S (86.3%) compared to the rest of the hyperparameter sequences. See Table 3.1 and 5.3 for detailed results. Second, we ran the second round of training but this time with the optimal hyperparameters and only tuning the number of nodes in the single-layer decoder (512, 1024, 2048, and 4096). The trained decoders are then tested on the stimuli used in the behavioral experiment to compare models' responses with humans.

[5]https://docs.ray.io/en/latest/tune/index.html
[6]https://pytorch-lightning.readthedocs.io/en/stable/

Table 3.1: Hyperparameter Tuning with **TERA** model on an ASpD task

| Tunable Hyperparameters | | | Validation Accuracy |
|---|---|---|---|
| **Learning Rate** | **Batch Size** | **Decoder (MLP) Layers** | % |
| 0.0001 | 64 | [4096] | 82.5 |
| | | [4096,256] | 79.6 |
| | | [4096,256,128] | 81.1 |
| | | [4096,256,128,64] | 80.2 |
| | 128 | [4096] | 82.1 |
| | | [4096,256] | 81.0 |
| | | [4096,256,128] | 80.7 |
| | | [4096,256,128,64] | 80.6 |
| | 256 | [4096] | 82.2 |
| | | [4096,256] | 79.9 |
| | | [4096,256,128] | 79.7 |
| | | [4096,256,128,64] | 79.0 |
| 0.001 | 64 | [4096] | 78.0 |
| | | [4096,256] | 79.0 |
| | | [4096,256,128] | 80.9 |
| | | [4096,256,128,64] | 79.0.2 |
| | 128 | [4096] | 79.0 |
| | | [4096,256] | 79.4 |
| | | [4096,256,128] | 79.9 |
| | | [4096,256,128,64] | 79.9 |
| | 256 | [4096] | 73.2 |
| | | [4096,256] | 80.0 |
| | | [4096,256,128] | 79.2 |
| | | [4096,256,128,64] | 80.5 |

### 3.4.2 Performance of Models

In Figure 3.14, we show the performance of trained decoders with multiple encoders on the stimuli tested on humans in the behavioral experiment against human performance.

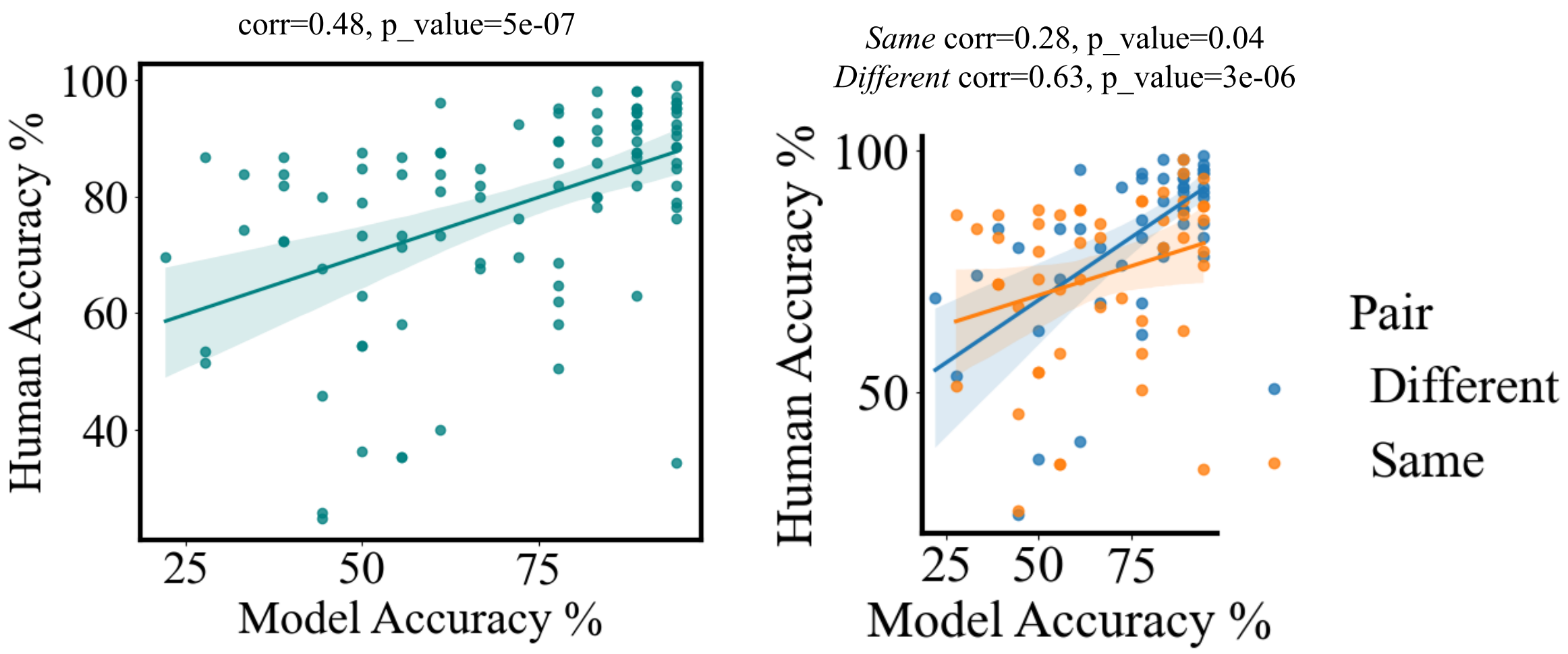

Figure 3.14: Correlation between humans and models performance across all trials and separate trials (Same and Different). Spearman's correlation and corresponding p values are reported.

### 3.4.3 Discussion

From this preliminary results, we showed that with simple decoder architecture, performance on a discrimination task is significantly correlated with human performance on across trials. Additionally, we computed the correlation between the performance of models and humans on *Same* and *Different* trials separately yielding higher correlation in telling people apart than together. Thus, with further experimentation on different decoder architecture and SSMs it is feasible to reach human-like performance on this task. However, the main challenge is to study if these models share similar responses and judgements as humans. Then, we can highlight the similarities and differences between representational spaces of models and humans.

## 3.5 Conclusion

In this chapter, we investigated different means of comparison between the encoding spaces of models and humans on a behavioral level. We showed that linear distances can not capture the complexity of the non-linear decision models either in humans or models. Alternatively, we proposed that using learnable decoders might approximate to the perceptual space relatively better compared to distances. Our results are presented as a proof-of-concept substantiating the benefits of leveraging machine learning algorithms to simulate the representational perceptual space of humans.

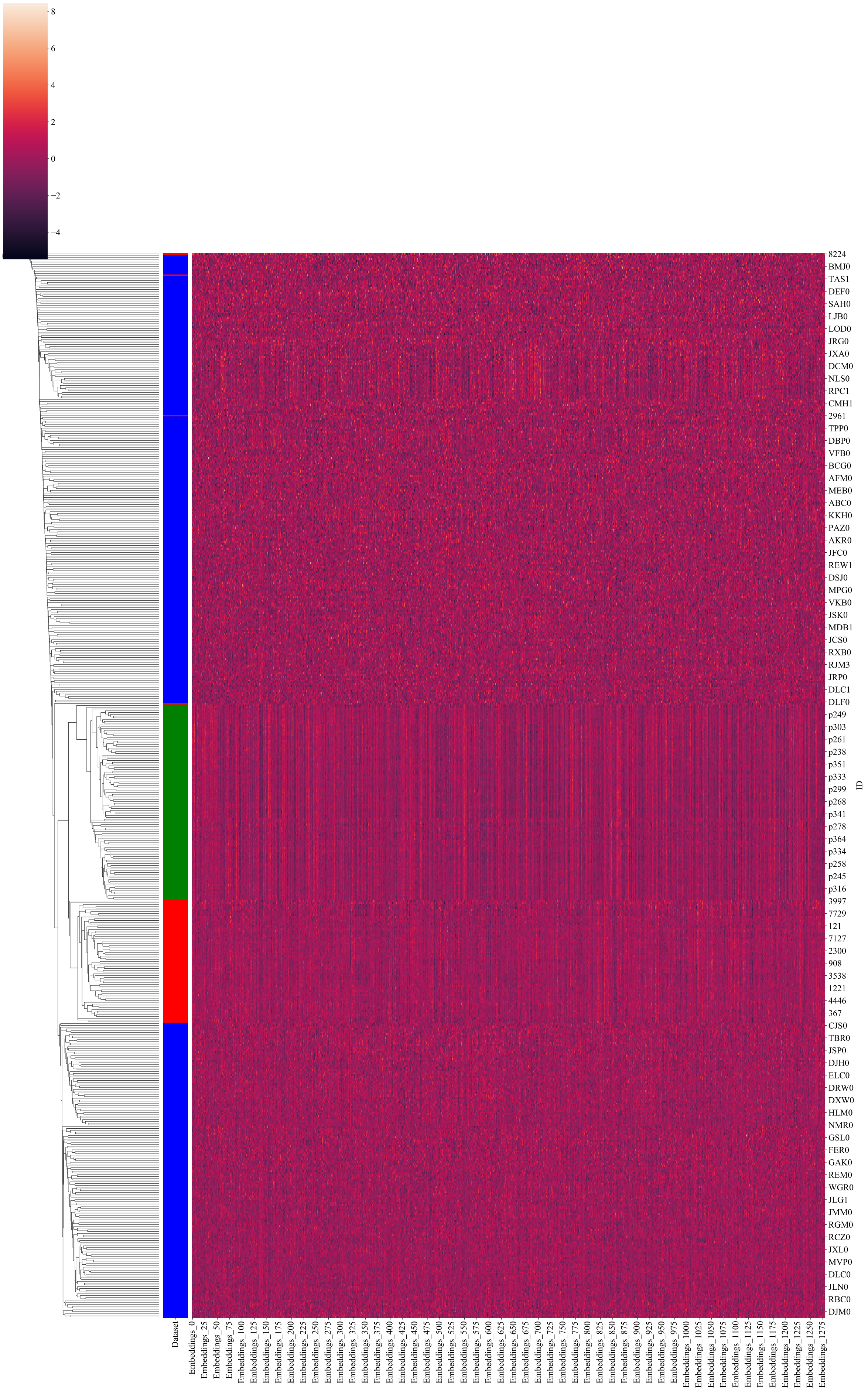

Figure 3.15: Clutermap of the utterances embeddings extracted from HuBERT best layer.

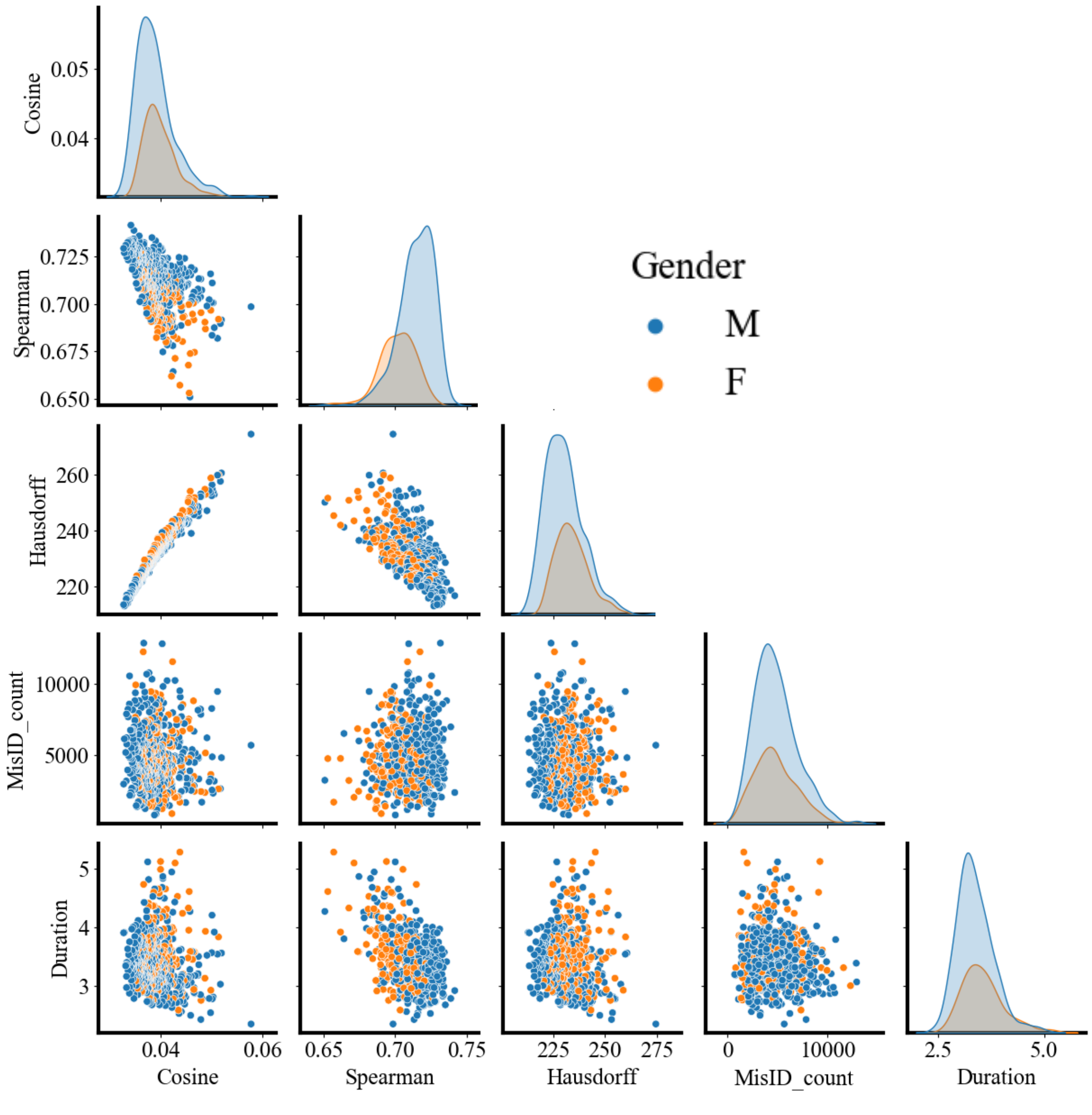

Figure 3.16: Pairplot showing correlation across all computed variables, clip duration, number of misidentifications, cosine distance, hausdorff distance, spearman correlation. The color code indicate gender, $M$:Male, $F$:Female.

# Chapter 4

# From Models to Brains

## 4.1 Introduction

In the previous chapters, we studied SSMs as candidate models for examining human perception. We identified similar perceptual trends and behavioral judgments in both recognition and discrimination tasks. Taking a step further, in this chapter, we aim to investigate the correspondence between models' encoding spaces and human neural representational space. Previous studies have suggested that some audio ANNs can predict brain cortical responses in auditory regions better than acoustic features such as spectrotemporal filters [128–137]. These studies examined the ability of an array of ANNs with different architectures, task objectives, and inputs to predict fMRI responses to a variety of natural sounds (e.g. speech, music, and environmental sounds). However, to the best of our knowledge, no work has been done yet to study brain-model correspondence in the context of speaker identity. Therefore, in this chapter, we tackle the following questions:

- Where is the information content of speech models best represented in the brain?
- Can we recognize speaker identities from brain activity?

One way of studying this problem is using naturalistic imaging paradigms. This approach came into prominence recently to lessen the gap between real life ecological stimuli and experimental setups [138, 139]. Such paradigms involve naturalistic stimuli, such as movies and stories, evoking dynamic changes in brain activity. Eventually, we are interested in modeling these dynamic sensory representations using SSMs' embeddings.

For decades, researchers developed different methods to study the information represented in the brain activity space. The most popular approaches are encoding and decoding models [140]. We use encoding models to study the correspondence of a feature space to the brain activity space (e.g. voxel-based space) [141]. Encoding models predict brain activity that is evoked due to a stimuli. Therefore, these type of models allow us to quantify the association of a particular input stimuli to brain regions [140]. In this work, we feed the naturalistic audio stimuli to the SSMs for embeddings extraction. Then, we fit a linear model on the embeddings space to predict voxel responses and report the performance on different SSMs. This method is referred to as *Voxel-wise Modeling*. An illustration of the framework for voxel-wise modeling on visual stimuli is shown in Figure 4.1.

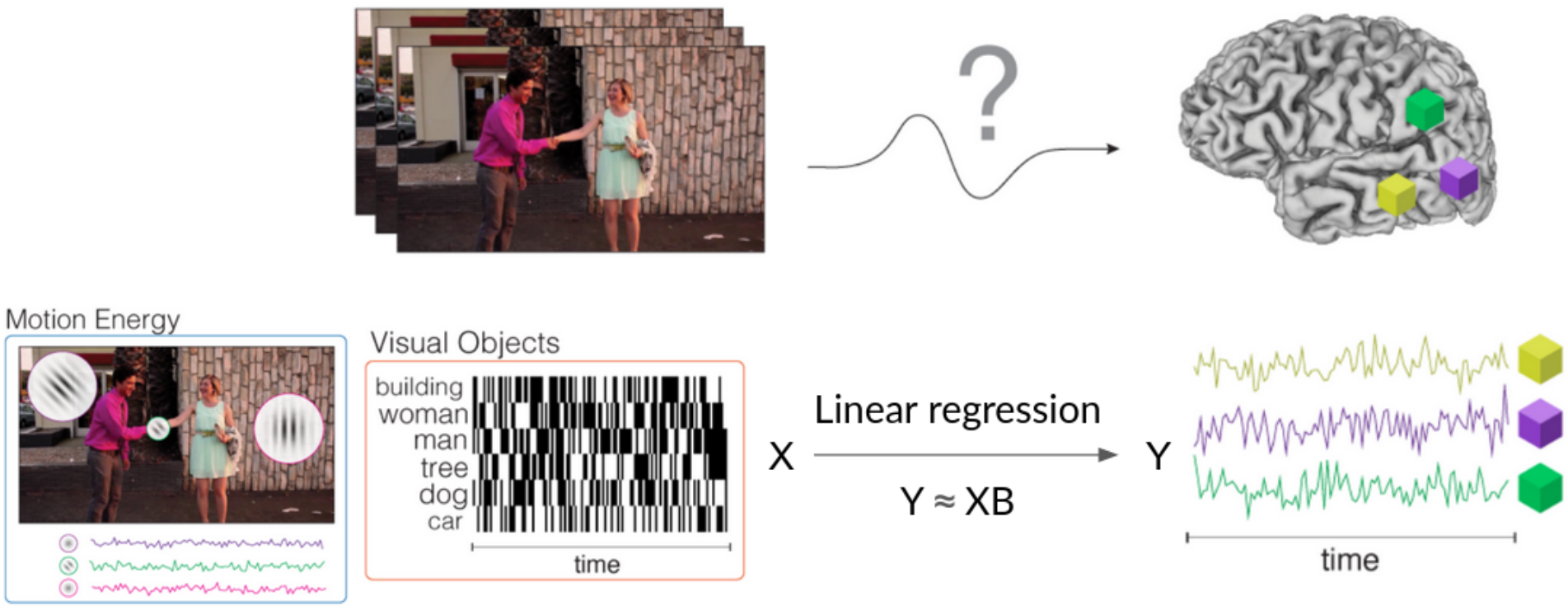

Figure 4.1: Framework for Voxel-wise Modeling (This figure is taken from [1]).

On the other hand, decoding models are useful for evaluating if the information in a specific brain region is associated with a certain behavior [142]. In simpler terms, we map the brain activity space to the feature space via building a decoding model (e.g. linear classifier). In our case, we used a decoding model to test the ability of the recorded activity to separate the narrator's voice from other sounds in the movie (e.g. other voices, music, environmental sounds, silence,..etc.). This approach is studied as a first step towards understanding identity networks in the brain.

## 4.2 Methods

### 4.2.1 Naturalistic fMRI Dataset

In this analysis, we selected StudyForrest dataset[2], a large naturalistic imaging data comprising several tasks and multiple data modalities such as fMRI scans, structural brain scans, physiological recordings and behavioral data from subjects either listening to or watching the movie *Forrest Gump* (see Supplementary Figure 5.15). The particular sub dataset of interest in this chapter is of subjects listening to a German audio version of the movie being presented while being imaged by an fMRI scanner [143]. This dataset is publicly available on the OpenNeuro data archive with ID `ds000113`[3].

This dataset comprises 7 Tesla fMRI recordings from 20 German participants (8F/12M) while listening to the German audio-description of the movie *Forrest Gump*. Other data types that were collected include T1w, T2w, DTI, angiography, and susceptibility-weighted images. In this task we focused on the functional imaging data and associated T1w data.

The audio movie is divided into eight segments with average duration of 15 min per segment. The movie is presented chronologically across eight runs/segments in two MRI sessions per participant [143]. For each run, 4D volumetric images (160x160x36) were stored in a NIfTI format with a time of repetition (TR) of 2s. The authors provided a version of the dataset to correct for severe geometric distortions. We discarded one subject (`sub-10`) from our analysis, for whom the distortion corrected NIfTI files were missing.

### 4.2.2 fMRIPrep and Preprocessing

After installing the dataset through `datalad`[4], which followed the BIDS standard [5], we preprocessed the NIfTI files using *fMRIPrep* software package [144]. fMRIprep is preprocessing tool for neuroimaging datasets. The pipeline comprises several processing steps (e.g. co-registration, normalization, unwarping, noise component extraction, segmentation, skull-stripping, etc.), see Appendix.

Three main outputs were generated from fMRIPrep. First, a visual `HTML` report per subject showing a summary of different steps of the pipeline intended for visual quality assessment.

[2]https://www.studyforrest.org/
[3]https://openneuro.org/datasets/ds000113
[4]https://www.datalad.org/
[5]https://bids.neuroimaging.io/

Second, derivative (preprocessed data) files. Multiple types of derivatives are generated, however, for our data analysis, we considered the functional derivatives containing preprocessed BOLD images and FreeSurfer [145] derivatives which comprise surface reconstruction data and grayordinates files, which are a 2D representation of the cortical gray matter and can be referred to as *surface voxels*, saved in a CIFTI format. Lastly, fMRIPrep calculates an array of confounding variables (nuisance regressors), these are artifact-based variables that contribute to the measured signal affecting the true brain-induced signal (BOLD). Accordingly, we select a set of confounding variables and regress out their influence from the recorded signal. This step is referred to as *Confound Regression*. For further details on the analysis carried out by fMRIPrep, see fMRIPrep generated boilerplate in Appendix 5.2.

After executing fMRIPrep on all subjects, first, we smoothed the resulting CIFTI files (surface data) using `connectome workbench`[6] with 4mm full width at half maximum (FWHM). Then, we ran confound regression on the smoothed images to remove the effect of multiple regressors:

- 6 motion parameters (trans_x, trans_y, trans_z, rot_x, rot_y, rot_z)
- All DCT-basis regressors (high-pass filtering)
- 5 aCompCor components
- Cerebral Spinal Fluid (CSF) global signal
- Framewise displacement (estimated bulk-head motion)

Lastly, we standardized the grayordinate values using z-score. This pipeline was executed on all runs for every subject. Figure 4.2.2 shows the surface images for one subject before and after preprocessing. It is worth noting that functional images were acquired with partial brain coverage where the field-of-view was centered on Sylvian fissure covering many brain regions associated with hearing, speech, and language [143]. Hence, the dark spots on the top and bottom of the surface images in Figure 4.2.2 indicate brain areas not measured during scanning.

Furthermore, we preprocessed the audio stimuli used in the experiment as well to extract their embeddings using SSMs. The 8 audio stimuli files were converted into `wav` files and downsampled to 16kHz to be compatible with models' training settings. Subsequently, the stimuli were chunked into 2-sec long clips aligning with the TR samples acquired. Finally, we extracted the audio embeddings of these clips from all layers of the three SSMs (HuBERT, Wav2Vec2, and Data2Vec).

[6]https://www.humanconnectome.org/software/workbench-command/-cifti-smoothing

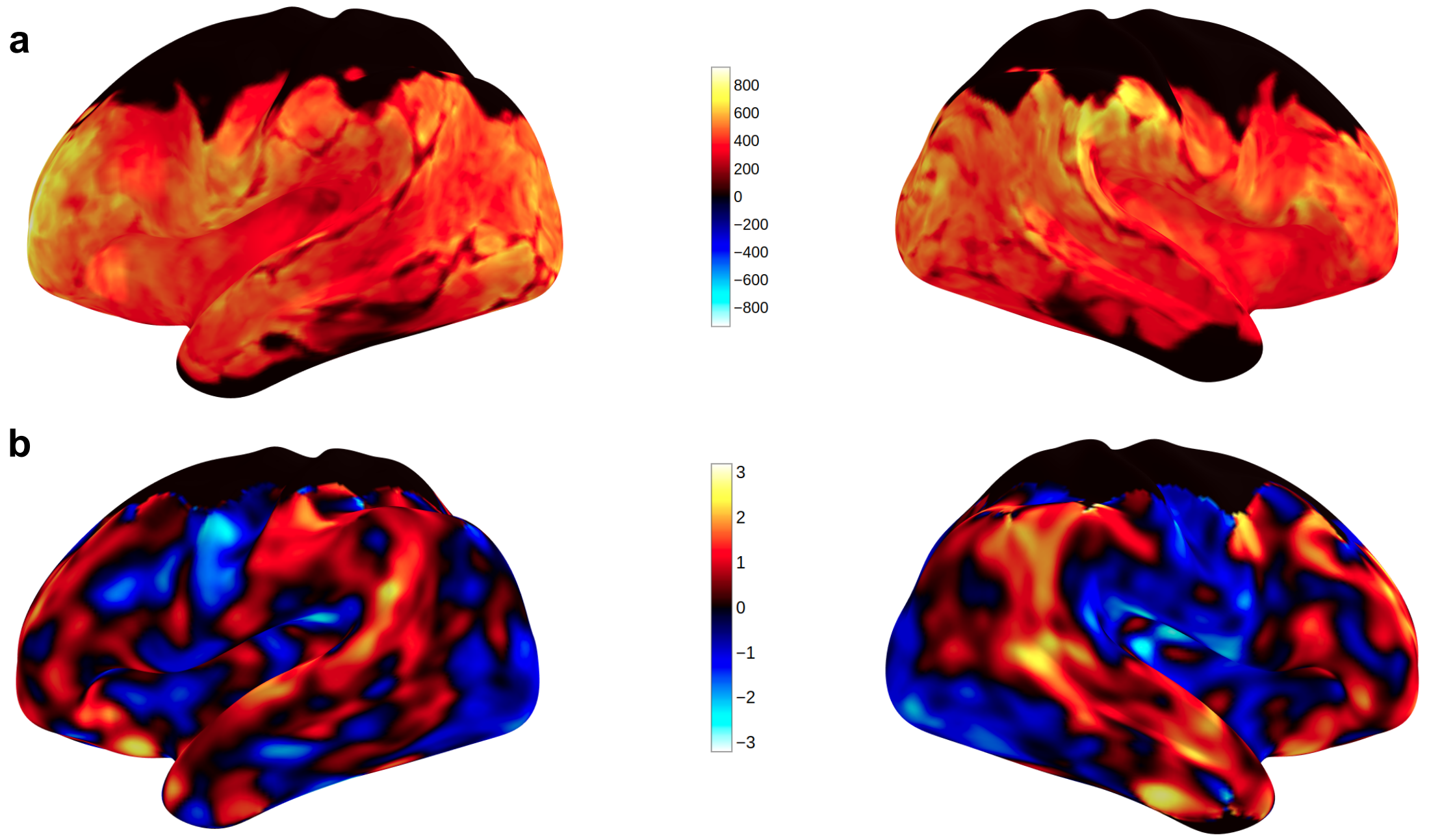

Figure 4.2: Cortical images for one subject in one run at a specific TR (a) Raw Surface Image. (b) Pre-processed Surface Image.

### 4.2.3 Brain Parcellation

Using brain parcellation methods helps associate the information in a feature space to a specific brain region. These brain regions are usually derived from resting-state functional connectivity (RSFC) studies which measure the synchrony in the fMRI signal across different regions to identify brain networks [146]. We chose the recent homotopic Yan2023 method [147] to parcellate the brain for two main reasons. First, this method provides homotopic parcels that are beneficial for lateralization studies. Second, it derives brain parcels using a homotopic variant of the gradient-weighted Markov Random Field (gw-MRF) algorithm which combines two common methods for parcellation, local gradient and global similarity yielding more homogeneous parcels than generated using both methods [148]. In this work, we extracted brain parcellations with the resolution of 400-area parcels. These parcels are matched to 17 cortical networks estimated using a multi-session hierarchical Bayesian model (MS-HBM) [149], see

Figure 4.3. The parcellation files were downloaded from this repo[7].

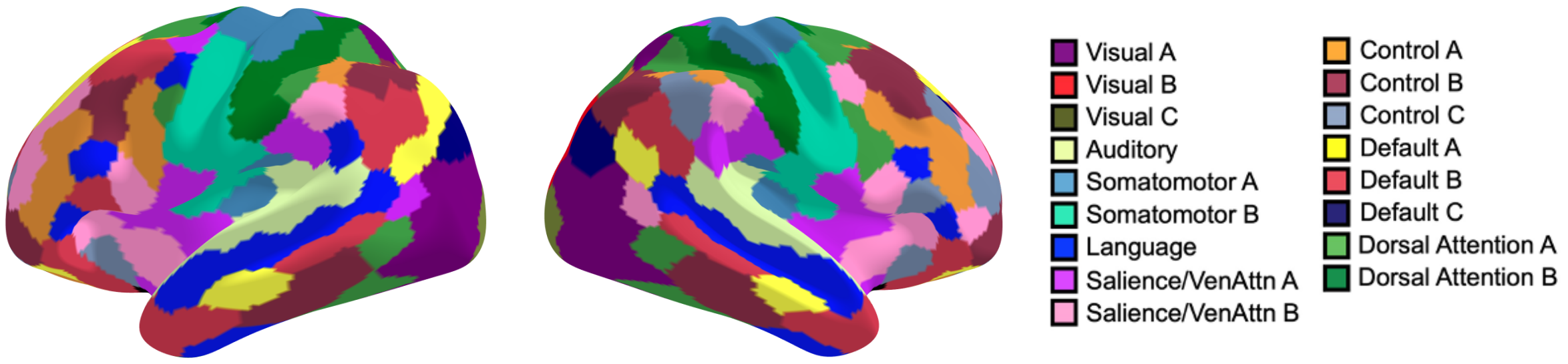

Figure 4.3: Yan2023 400-area parcellation matched to Kong2019 17 Networks.

### 4.2.4 Voxel-wise Encoding Modeling

After preparing the surface data and the stimuli embeddings, we built a regularized linear regression model (ridge regression) to predict the grayordinate values using the models' embeddings. First we split the data into train and test, we combined 7 runs as a training set and 1 run as a testing set. We chose the $7^{th}$ run/segment to hold out since it has the largest number of samples/TR (542 samples). Then, we defined the cross-validation (CV) scheme. We used the leave-one-run-out approach where for each CV step the model is trained on 6 runs and validated on 1. The embeddings were standardized before training and we also added a delayer to account for the slow response in the BOLD signal. Two parameters were fine-tuned to yield optimal results on the validation set, we tested with different delayer values as well as multiple values for the regularization term (alpha). Finally we computed pearson correlation and the coefficient of determination ($R^2$) between the predicted values and ground truth. We used `KernelRidgeCV`, a function in the `himalaya`[8] Python package that supports multiple-target linear models and implements a kernel trick for faster performance. See Figure 4.4 for illustration of the process. We ran the analysis separately for each layer in the three selected models per subject data. We discarded two additional subjects due to inconsistencies in the number of expected TRs, leaving us with total of 17 subject.

[7] https://github.com/ThomasYeoLab/CBIG/tree/master/stable_projects/brain_parcellation/Yan2023_homotopic

[8] https://github.com/gallantlab/himalaya

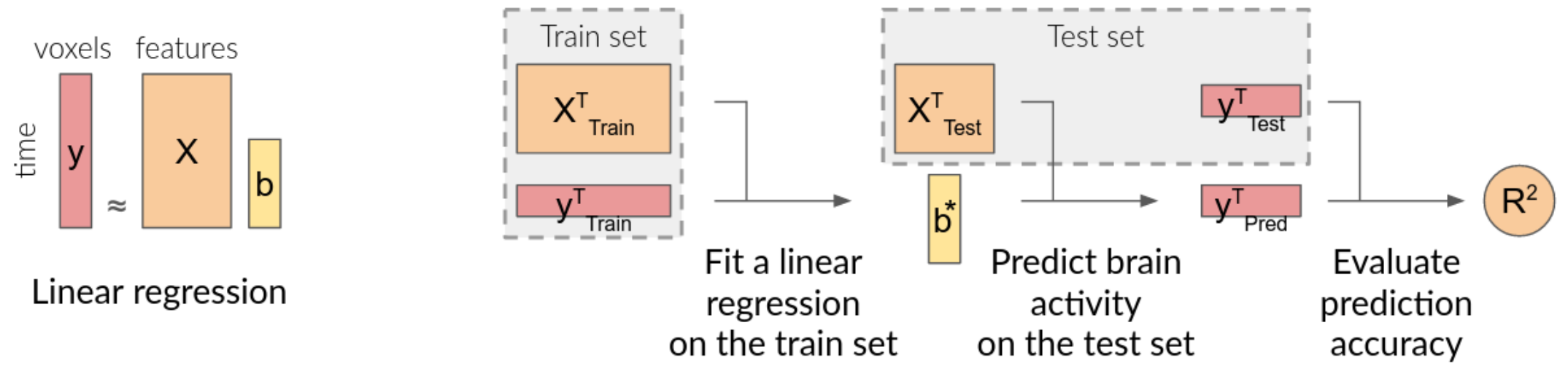

Figure 4.4: Steps for Training an Encoding Model (This figure is taken from [9]).

### 4.2.5 Decoding fMRI Signal

In this experiment, we aimed to evaluate the ability of brain activities to classify the narrator's voice against any other sounds or voices. To do so, first, we segmented the narrator's voice from the whole audio movie using `pyannote-audio` model [150]. Then, the segments with the narrator's voice were aligned with the TR samples recorded to label all samples with binary values (i.e. 1 if narrator is speaking and 0 otherwise) creating our binary target vector. The data splitting and the cross-validation scheme were similar to Section 4.2.4. To account for the delay in the BOLD signal, the label vector was shifted with different delays (1, 2, and 3 TR shifts). Also, we tried convolving the label vector with glover HRF function then thresholding the signal based on the median of the signal. Finally, we fit the grayordinate samples using support vector machine classifier (SVC) and optimize performance during CV by fine-tuning the regularization hyperparameter. Then, we evaluated the classifier performance by reporting the balanced accuracy metric to compensate for the imbalanced labels in the data.

## 4.3 Results

As mentioned in the Methods, we computed two parameters to evaluate the predictions of each layer within the models. In Figure 4.5, we show the average pearson correlation between the predicted values and ground truth. The correlation is computed for each layer across all models (n=3) and subjects (n=17).

To better visualize the mapping of the models' emebddings to the cortical surface. We plotted the $R^2$ scores of the median best-performing layer for each model across all subjects on the brain surface as shown in Figure 4.6. The layers plotted in this figure are transformer layer 12, transformer layer 7, and convolution layer 6 for HuBERT, Wav2Vec2, and Data2Vec, respectively. We used a threshold to only plot values above the 90th percentile of the $R^2$ scores.

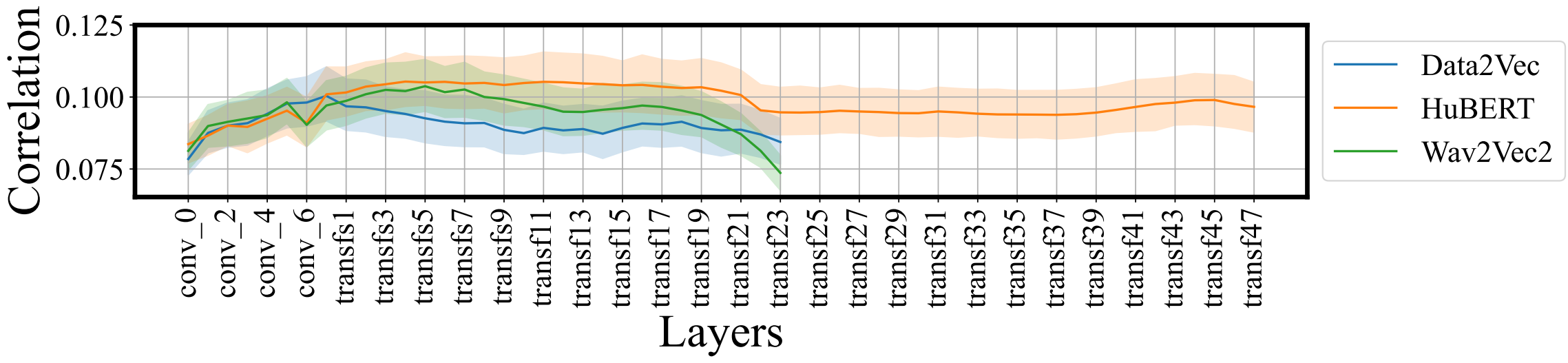

Figure 4.5: Layer-wise Analysis for Wav2Vec2, HuBERT, and Data2Vec on predicting grayordinate values. Average Pearson Correlation is reported for each layer across models on 17 subjects.

Similarly, we juxtaposed the three different layers we studied in Section 2.3. For each model, the $R^2$ scores of the latent, best-performing (on VC1 benchmark see Section 2.2), and contextual/final layers are plotted on the cortical surface averaged across all subjects, as shown in Figure 4.7.

## 4.4 Discussion

In this chapter, we explored the robustness of some SSMs to predict brain responses as a result of auditory stimuli using voxel-wise modeling. We leveraged naturalistic imaging paradigms, audio movies in this context, due to the considerable number of speakers acting with large voice variability. The audio movie included predominantly speech as well as other sounds (e.g. music, environmental sounds). Thus, in our experiments, we tested general auditory encoding as a step towards evaluating speaker identity encoding for future work.

In Figure 4.5, we see the average correlation between the predicted values and the actual grayordinate values for each layer across all models. The three models show very low correlation scores. This is expected because the encoder was trained to predict brain responses from the whole cortex including non-stimuli-induced brain regions. These regions might feature noisy fluctuations during the auditory stimuli harming the predictive power of the speech models. However, these results are similar to the work done by Vaidya et al.[129]. They studied speech encoding in SSMs and showed that these models resulted very low correlation when predicting the whole cortex. Nevertheless, predicting specific ROIs, in this case auditory regions, yielded higher correlation values. Accordingly, it might be difficult from this analysis to evaluate trends across layers similar to 2.5.

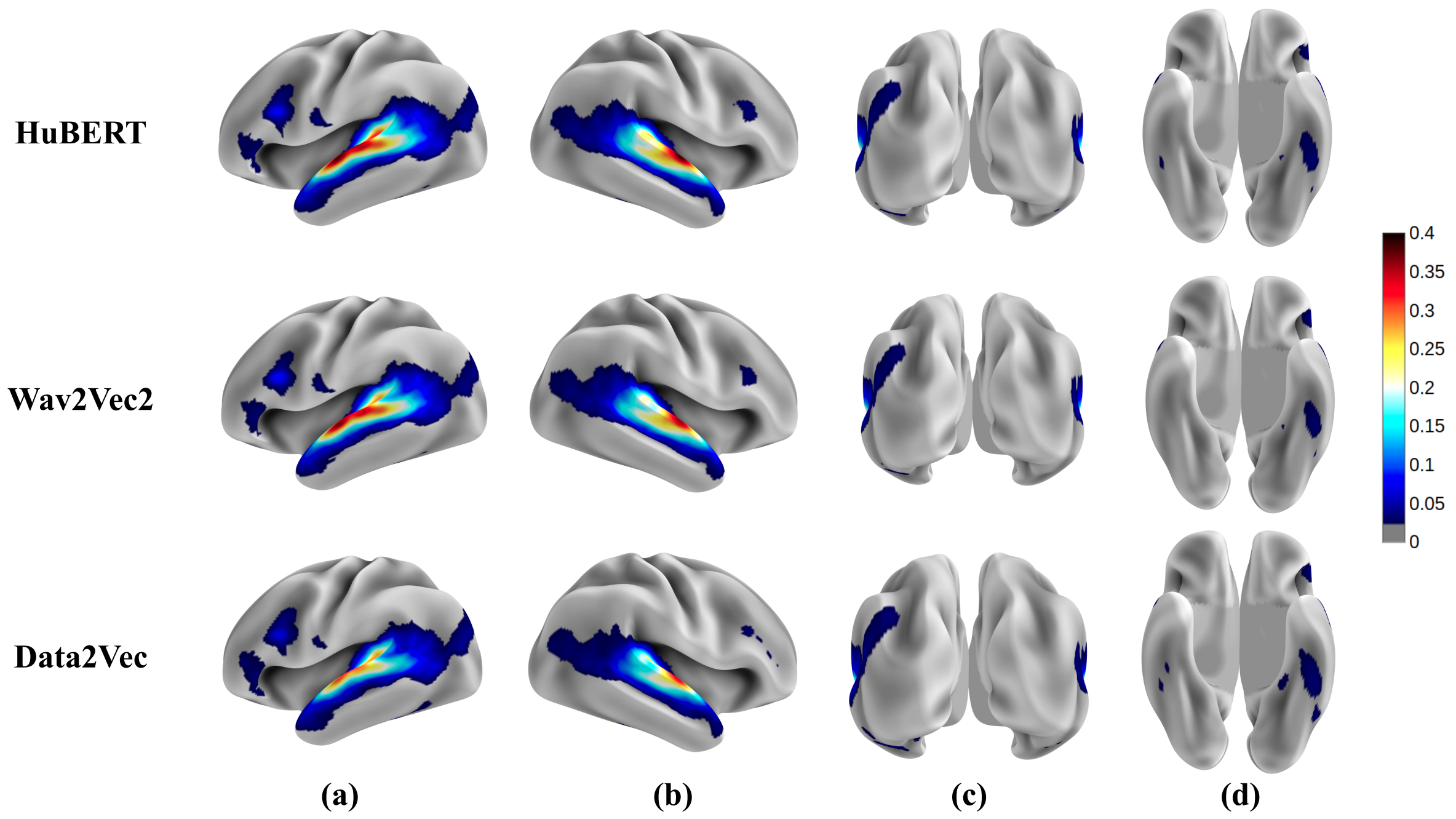

Figure 4.6: Surface maps of the median best layer/stage across subjects for each Model. (a) Right Hemisphere Lateral view (b) Left Hemisphere Lateral view (c) Posterior view (d) Inferior view.

To that end, we selected the best-predicting layers in each model (i.e. yielding the highest correlation value) and plotted the $R^2$ scores on the cortical surface as shown in Figure 4.6. That way we can identify and visualize the brain regions best-predicted by these models. Similar patterns in $R^2$ scores are observed across the three models. As expected, these models show high predictive power in auditory regions along the STG/S, specifically on the left hemisphere. Interestingly, we notice some mapping around Broca's area in the Inferior Frontal Cortex (IFC). This region is known to be associated with language and speech production [151] which might indicate similar encoded representations.

Lastly, we revisited the three different model stages studied in Section 2.3. In similar vein to Figure 4.6, we showed the $R^2$ mappings on the cortical surface for the latent (final convolution layer), best-performing (on VC1 benchmark), and contextual (final layer) layers for the three models. Similar mappings are observed across the three layers. However, the best layer for the three models is showing slightly higher $R^2$ values in the auditory regions compared to the other layers, indicating better predictions to these regions using the embeddings from this layer. This might validate our previous arguments that intermediate layers in these models capture more robust auditory representations in contrast to other layers.

Further work is needed to disentangle speaker identity processing from general auditory coding. Future work might involve chunking the audio movie into speakers'/actors' clips to study the ability of models' embeddings to capture brain responses due to speaker stimuli only. Furthermore, using brain responses to predict speakers' identity might allow us identify brain networks that encode identity representations.

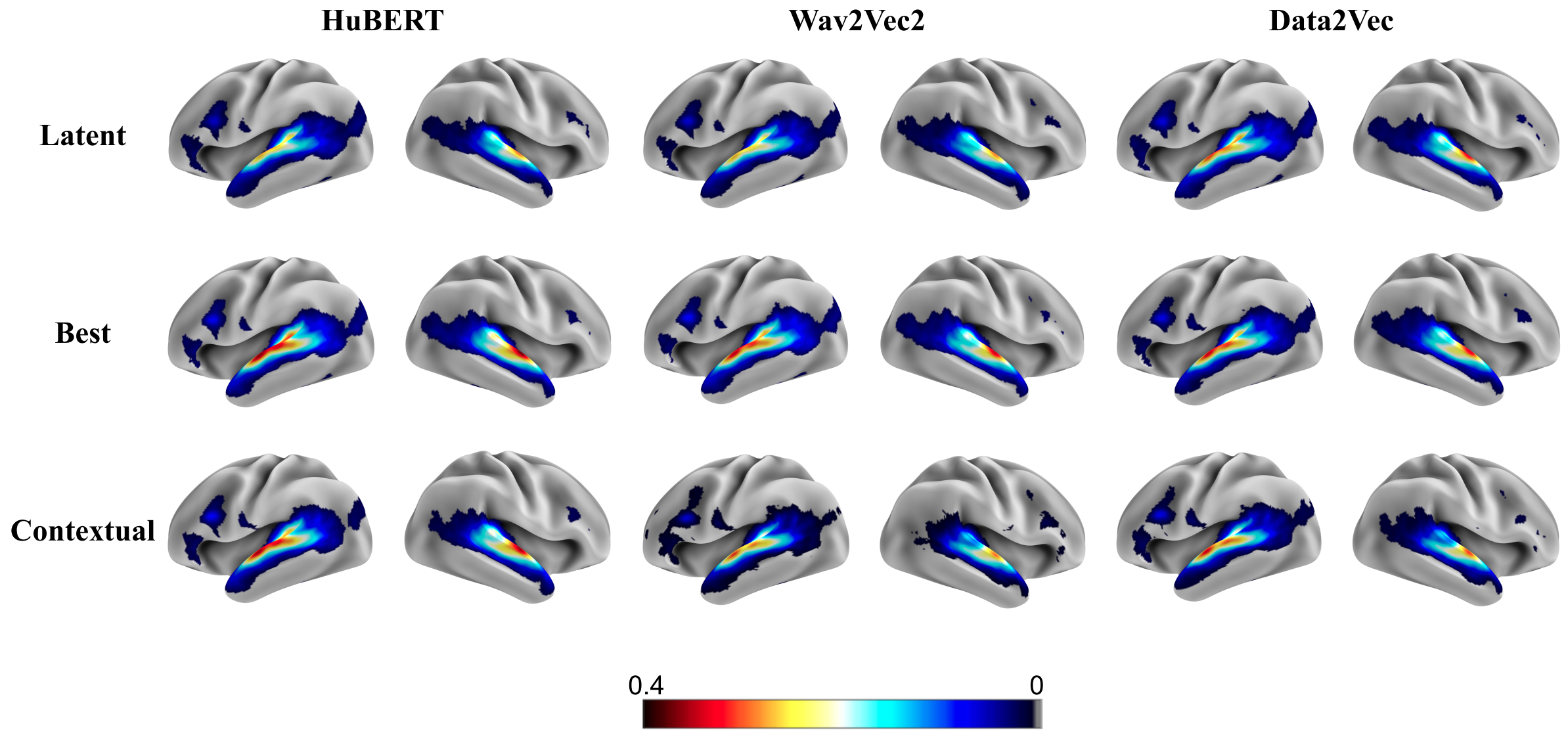

Figure 4.7: Surface maps of latent, best, and contextual layers for each Model averaged across all subjects.

# Chapter 5

# Conclusion and Future Directions

## 5.1 Summary of Research

In this thesis, we studied the problem speaker identity perception through the lens of SSMs, which are complex data-driven systems. While these SSMs were trained on implicit tasks without any explicit supervisory signals related to speaker identification, these models demonstrated high predictive power across multiple downstream tasks [98]. In the first chapter of this work, we used a number of SSMs, drawn from different categories of such models (generative, contrastive, and predictive), and evaluated their appropriateness as candidate models for speaker identity coding. We showed that these models capture more robust speaker representations than acoustic or handcrafted ones. Additionally, we identified different types of encoding stages in some models. Subsequently, we established some equivariances and invariances of the models through a set of experiments that altered different properties of speech including acoustic, spectral, phonemic, linguistic, and environmental components. The experiments revealed information about the content, specificity, and disruptibility of the underlying representational spaces of these models as they relate to speaker identification.

In the second chapter, we examined the potential of linear distance metrics as a proxy for voice proximity in human perceptual space. We showed through a behavioral experiment that these metrics cannot capture either the decision space of models when trained on a ASpR or humans performing a speaker discrimination task. In the behavioral experiment, we reported that humans are significantly better at telling people apart than together meaning that the decision space might exhibit some biases to consider. Additionally, we trained models on ASpD task and found that model can generalize and perform well on unseen speakers indicating that these learnable decoders might be better candidates to study voice proximity instead of linear distance and correlation metrics.

Lastly, in the third chapter, we showed that some models can predict brain responses in auditory regions along the STG/S as well as parts of the IFC. Additionally, we found that intermediate representations that should better performance on ASpR task, also better predict auditory encoding compare to latent and contextual representations.

These findings, taken together, advocate for using these SSMs in speaker identification models and to investigate speaker identity coding in human perceptual and cognitive tasks.

## 5.2 Future Directions

This project is a step towards incorporating more complex model systems to study speaker identity perception. While these systems, SSMs, demonstrate promising performance on identity-related tasks, they still exhibit some pathologies that require further improvements. For instance, the temporal dependence on context in naturalistic audio is usually undermined by the combined use of evaluating 3-sec frames of audio and pooling and aggregation methods in the DL models. Even though these representations perform reasonably well on downstream tasks, considering the temporal complexity of the signal might allow us to capture features that encode prosodic aspects of speech that tend to be averaged out due to temporal pooling. Further improvements can occur in the model architecture development stage, using architectures that account for temporal information (e.g. sequence modeling). Another line of improvement related to temporal processing can be in the downstream analysis phase. It is important to investigate if speaker recognition is carried out in a purely holistic template matching task or there is a temporal-weighting component (i.e. the more we listen to a speaker the easier the task becomes). Designing ASpR downstream tasks to give higher weights to earlier samples might provide some insights regarding the problem of how many time samples we need to recognize a voice. Most behavioral studies tend to use 3-sec utterances for either recognition or discrimination tasks. Also, the Vall-E model [70] claims to synthesize speaker's voice using only a 3-sec enrollment utterance. It looks like 3-sec is a magical number for voice studies, however, the impact of voice duration to identity processing remains unclear.

Furthermore, we might need to shed some light on the contribution of task objectives of encoding spaces. In this work, we showed that the final layer of HuBERT, Wav2Vec2, and Data2Vec encoded more linguistic and phonemic features, while not capturing much of speaker-related features. We hypothesized that such behavior might be result of the task objective during training. Since these models were trained to recognize speech, the training may have steered the model towards extracting speech-related components while discarding any variance due to speaker or environmental information. We also validated our hypothesis through a set of speech experiments that compared the representations in different layers or stages of the model. Thus, it might be worth specifically exploring behavior of such supervised models when the task objective itself is identity learning. Leveraging recent advances in contrastive learning algorithms might be beneficial for studying speaker discrimination. Also, multi-task learning has shown promising results in the field of audition [128, 130]. Thus, training models to recognize speakers (supervised learning) while discriminating with other voices (self-supervised learning) might lead to better precision of identity information. Leveraging the current advances in ANNs while constraining these models with perceptually-plausible inductive biases might allow us better understand perceptual phenomena in humans [152].

In the current project and in previous studies, models are showing human-level performance as well as similar perceptual trends. In future work, we might evaluate if similar behavioral responses and judgements might indicate similar hierarchical encoding strategies as well. In vision research, models optimized on an object recognition task showed hierarchical processing reminiscent of the ventral visual stream [153]. In a similar vein, one might hypothesize that two systems sharing similar outcomes might also share similar hierarchical transfer functions. Additional work may involve revisiting the notion of manifold learning and leveraging learnable manifold representations to revise our understanding of voice proximity.

Finally, future work may include extending the neuroimaging analysis to focus more specifically on speaker identity coding and differentiation in the brain. Building decoder models where brain responses predict or differentiate speaker identities can allow us to identify brain networks associated with identity. These experiments might be a great complement to the seminal lesion studies on *Phonagnosia* [28].

In conclusion, we believe that leveraging the synergy between computational systems such as ANNs and auditory perception will provide further insights towards rigorously and quantitatively modeling aspects of human perception.

# Appendix

## Candidate Models

We explored representations from different families of self-supervised models. Also, we included acoustic and low-level features as examples for handcrafted representations. All models incorporated in this paper are mentioned in Table 5.1 along with their embeddings dimensions, pretraining datasets and python package used for implementation.

Table 5.1: Candidate Models

| Model Name | Dimensions | Pretraining Dataset | Implementation |
|---|---|---|---|
| Handcrafted Models | | | |
| Log-Mel-Spectrogram | 128 | - | `torchaudio` |
| Cochleagram | 85 | - | `pycochleagram` |
| openSMILE/eGeMAPS | 88 | - | `openSMILE` |
| openSMILE/ComParE | 6373 | - | `openSMILE` |
| Generative Models | | | |
| APC | 512 | LibriSpeech | `s3prl` |
| TERA | 768 | LibriSpeech | `s3prl` |
| Contrastive Models | | | |
| BYOL-A/default | 2048 | AudioSet | `byola repo` |
| BYOL-I/default | 2048 | VoxCeleb 1&2 | `byola repo` |
| BYOL-S/default | 2048 | AudioSet (Speech subset) | `serab-byols` |
| BYOL-S/cvt | 2048 | AudioSet (Speech subset) | `serab-byols` |
| Hybrid BYOL-S/cvt | 2048 | AudioSet (Speech subset) | `serab-byols` |
| wav2vec 2.0 | 1024 | LibriSpeech | `hugging face` |
| Predictive Models | | | |
| HuBERT | 1280 | LibriSpeech | `hugging face` |
| data2vec | 1024 | LibriSpeech | `hugging face` |

### Handcrafted Representations

Handcrafted features (DSP-based) are features generated from domain-expertise transformations. These transformations help us unpack the audio temporal signal and extract information about its underlying composition. Handcrafted representations have either been used as low-level input to neural network models such as *Spectrograms* or competitive baselines for audio downstream tasks [154] such as *openSMILE* features [155]. That said, we extracted *Log-Mel-Spectrogram* (LMS), *Cochleagram* and *openSMILE* features. For spectrograms, we used `torchaudio` to process a 128-band LMS with frequency ranges from 5 to 20000 Hz, sampling rate of 16 kHz and hop size of 32 ms, yielding representations of dimensions $128 \times T$, where $T$, the number of spectrogram frames, depends on the utterance length. For cochleagrams, we used the `pycochleagram` [1] package to generate 85-band human-like cochleagram representations via implementing Equivalent Rectangular Bandwidth (ERB) filterbank that estimates the auditory filter bandwidths. Lastly, we used openSMILE (OS) [155], an open-source toolkit to extract acoustic features and low-level descriptors. This toolkit contains several feature sets for representing speech signal. Thus, we extracted two well-established feature sets, eGeMAPS (OS/eGeMPAS) feature set [156] comprising 88 features and ComParE (OS/ComParE) feature set with 6373 features [154].

## Data-driven Representations

To map out different families of self-supervised models, we followed the recent review about self-supervised models [157] in which they categorized state-of-the-art models, based on their pretext task, into three families, *Generative*, *Contrastive*, and *Predictive* models. Generative models are models that were trained to generate or reconstruct data from limited input information. An array of approaches fall under this category, however, we have used two representatives from this family in our experiments. One model is *Autoregressive Prediction Coding* (APC) [158] which learns to predict the spectrum of a future frame based on past frames. Another generative model is TERA [102] (Transformer Encoder Representations from Alteration). This model learns to reconstruct original frames from corrupted input. For contrastive category, we included wav2vec 2.0 [104] as an example for conflating contrastive and masking learning. Recently, a novel contrastive model was proposed by [106] called *Bootstrap Your Own Latent for Audio* (BYOL-A). This model learns general-purpose representations from comparing augmented views of a single audio input. It is originally based on the BYOL model proposed by [159] for image representations. Lately, we have witnessed new derivations of BYOL-A that showed effective results in several downstream tasks. For instance, training the BYOL paradigm on speech dataset only to create BYOL for speech (BYOL-S) yielded competitive results on emotion recognition benchmarks called SERAB [107] and general tasks in the HEAR benchmark [98]. [109] created several variants of BYOL-A. They trained different encoders on the BYOL-A paradigm in addition to the default encoder architecture that comprised a simple convolution neural network (CNN) [106]. The best performing encoder tested on BYOL-S was a convolution vision transformer (CvT) [160]. Furthermore, a latest variant of BYOL-S was proposed by [103] that learns handcrafted (openSMILE/ComParE) and data-driven features simultaneously generating robust speech representations. This model is denoted as Hybrid BYOL-S. That being said, in this paper we used the original BYOL-A with the default CNN architecture (BYOL-A/default) in addition to BYOL-S with CNN and CvT encoders and Hybrid BYOL-S with CvT as examples of contrastive self-supervised models. We also pretrained BYOL-A on speaker utterances from VoxCeleb corpus [99] denoted as BYOL-I (BYOL for Identity). For candidate predictive models, we used HuBERT [101], a model that is trained to predict the pre-determined assigned k-means clusters given masked speech units. In a similar vein, data2vec model [105] is trained to predict contextualized representation given a masked input. This model was tested on three modalities, speech, images and text.

[1] https://github.com/mcdermottLab/pycochleagram

# TIMIT ASpR Benchmark

We benchmarked the performance of different handcrafted and data-driven features against a speaker recognition task using TIMIT dataset. TIMIT [110] is a read speech corpus that comprises 630 speakers uttering ten phonetically rich sentences with total of 6300 utterances. The dataset has speakers of eight different dialects of American English. We used this corpus as a ASpR downstream task. In a similar vein to Section 2.2, we extracted all handcrafted and data-driven features from this dataset and reported the models' performances. We trained a classifier on seven randomly-selected utterances per speaker and kept three utterances for testing.

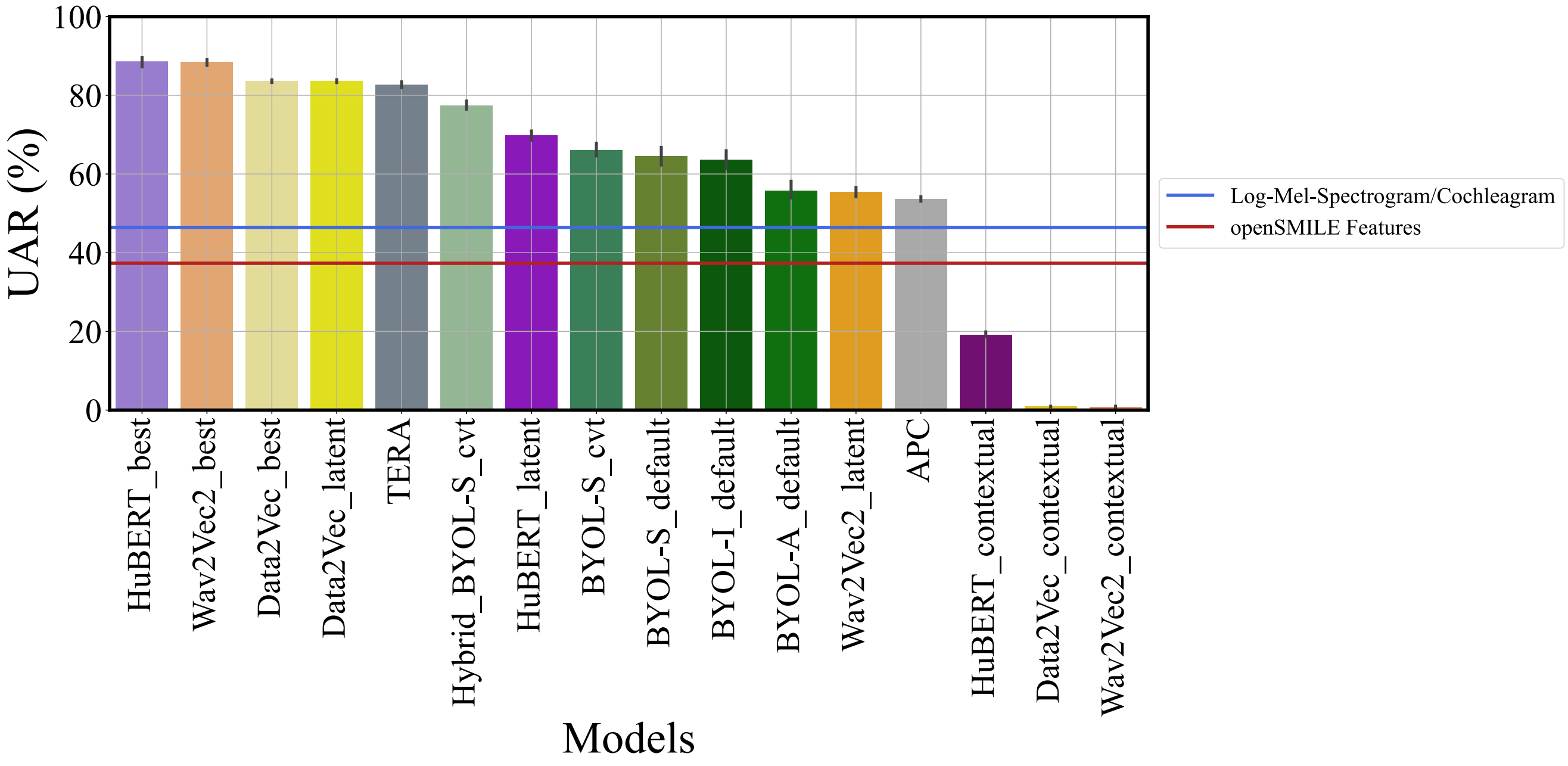

Figure 5.1: ASpR performance benchmark on TIMIT dataset using single linear layer. Models are sorted descendingly (from left to right) based on their mean UAR %, the error bars shows the standard deviation in performace across three repeated runs. The red horizontal line indicates average performance of openSMILE features (both sets). The blue horizontal line indicates the average performance ofspectro/cochlea-grams. The bar plot colors indicate the category of the models. BYOL- models are in shades of green, generative models are in shades of grey, Wav2Vec layers are in shades of orange, HuBERT layers are in shades of purple, and Data2Vec layers are in shades of yellow.

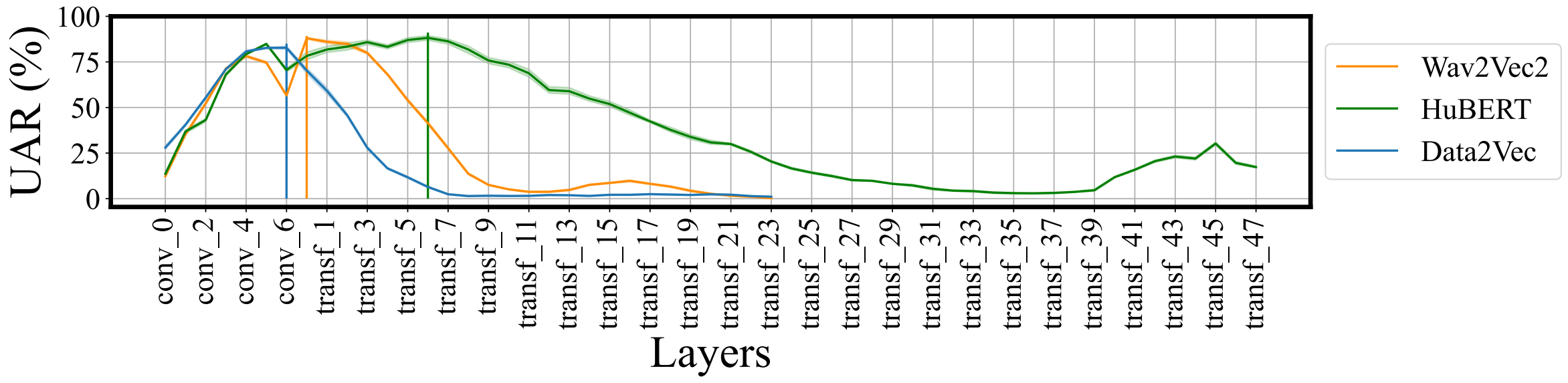

Figure 5.2: Layer-wise analysis for Wav2Vec2, HuBERT and Data2vec on TIMIT corpus. The vertical lines highlight the best-performing layer for each model. conv: convolution layer, transf: transformer layer.

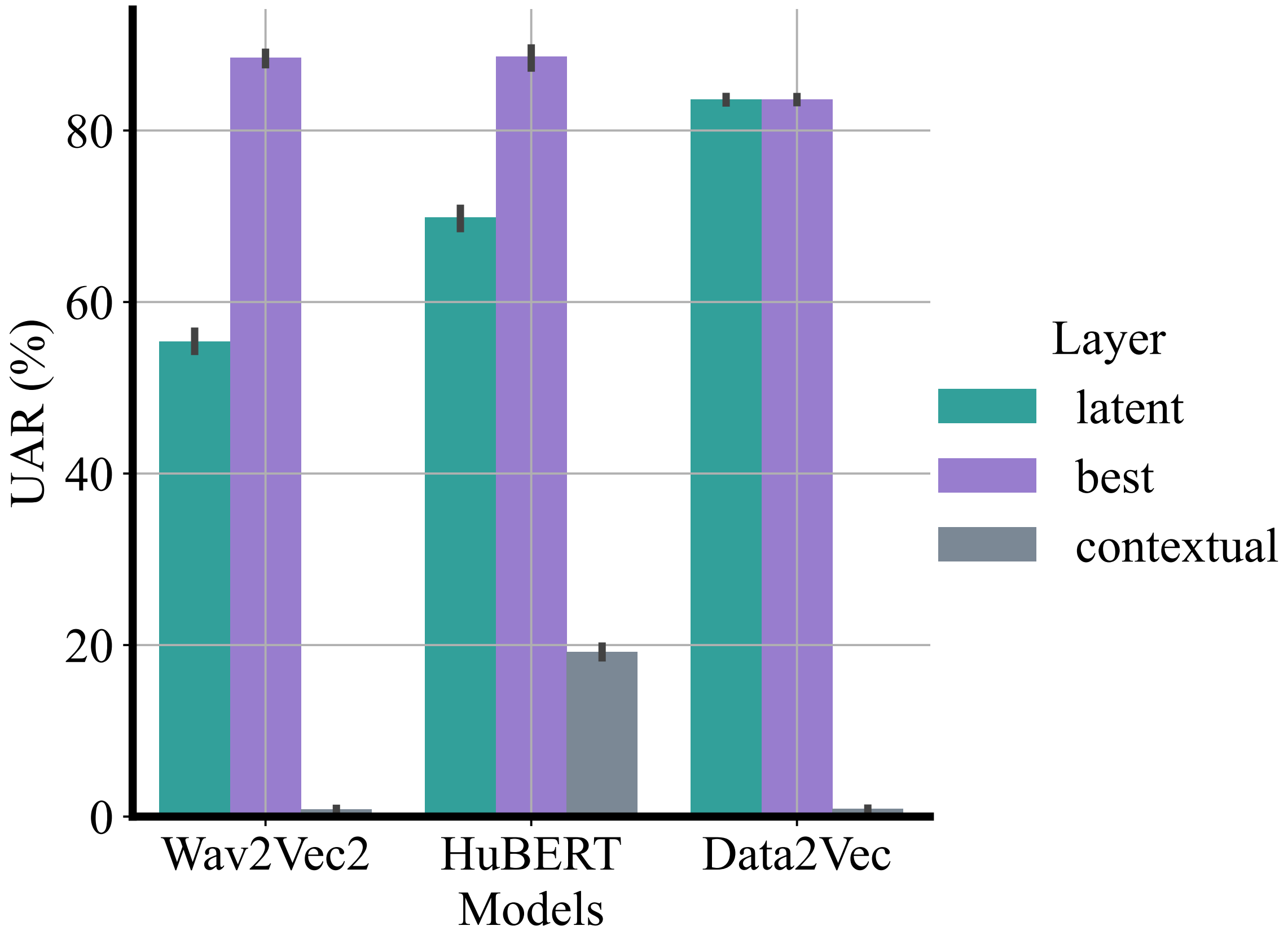

Figure 5.3: Comparing latent, best and contextual/final representations of Wav2Vec2, HuBERT, and Data2Vec on TIMIT corpus.

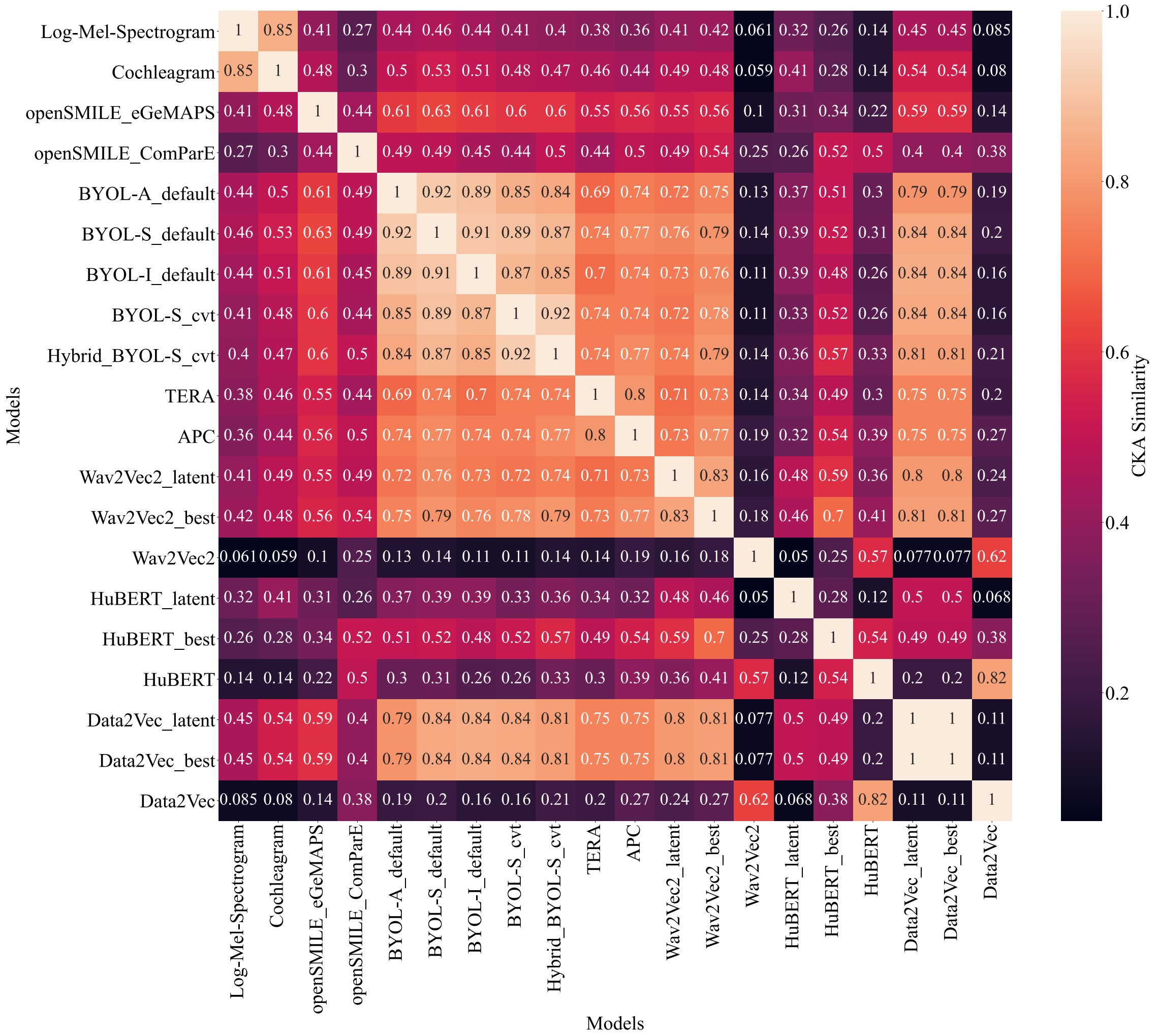

Figure 5.4: CKA analysis across all models' representations on TIMIT corpus.

# Dimensionality Reduction Methods

We used different methods of dimensionality reduction such as PCA, t-SNE, UMAP, and PaCMAP to get a better idea of how the speakers' samples are clustered together in 2D. However, one constraint is that these methods are sensitive to their hyperparameters (except PCA) which could affect our interpretation of the results. Thus, a grid search across the hyperparameters for each method is implemented.

Another issue would be quantifying the ability of these methods to preserve the distances amongst samples in the original space in a lower dimensional space. To address this, we are using two metrics KNN and CPD that represent the goodness of the algorithm to preserve local and global structures of the original embedding space, respectively. Both metrics are adopted from [161] in which they define both metrics as follows:

**KNN**: The fraction of k-nearest neighbours in the original high-dimensional data that are preserved as k-nearest neighbours in the embedding. KNN quantifies preservation of the local, or microscopic structure. The value of K used here is the min number of samples a speaker would have in the original space.

**CPD**: Spearman correlation between pairwise distances in the high-dimensional space and in the reduced embedding. CPD quantifies preservation of the global, or macroscropic structure. Computed across all pairs among 1000 randomly chosen points without replacement.

Consequently, we present the results from dimensionality reduction methods in two ways, one optimizing local structure metric (KNN) and the other optimizing global structure metric (CPD). However, all plots presented in this document are resulted from only optimizing CPD.

# Distance and Correlation Metrics

$$EuclideanDistance\left(x,y\right)=\sqrt{\sum_{i=1}^{n}\left(x_i-y_i\right)^2} \tag{5.1}$$

$$CosineDistance\left(x,y\right)=\frac{x\cdot y}{\|x\|_2\,\|y\|_2} \tag{5.2}$$

$$SpearmanCorrelation=1-\frac{6\sum d_i^2}{n(n^2-1)} \tag{5.3}$$

# Signal Detection analysis (d')

We used detection theory to identify if subjects responded properly to the trials or gave random/single response for all trials. Our discrimination experiment follows the AX design in which a pair of stimuli is prompted and the subject is required to respond if stimuli X is the same as stimuli A or not. Given that design, we first listed all possible outcomes in the format of a signal detection 4-cell table, as shown in Figure 5.5.

In this scheme, we define the HITs as correctly responding to a *Different* trial (signal). Thus, the Correct Rejections means responding correctly to *Same* trials (no signal). That way we can compute the Hit Rate ($H = P(''different''|''Different'')$) and the False Alarm rate ($H = P(''different''|''Same'')$). After computing both variables, we compute subject's sensitivity

| | **Response:** ***different*** | **Response:** ***same*** |
|---|---|---|
| **Stimuli:** ***Different*** | **HIT** | **Miss** |
| **Stimuli:** ***Same*** | **False Alarm** | **Correct Rejection** |

Figure 5.5: Signal Detection Table for AX Discrimination Experiment Design.

using the statistic d-prime (d’) which is the distance between the Signal and the Signal+Noise, calculated as shown in Equation 5.4. where $z$ is the z-transform computed for each rate.

$$d' = z(H) - z(F) \tag{5.4}$$

# fMRIPrep Boilerplate

Results included in this manuscript come from preprocessing performed using *fMRIPrep* 22.1.0+0.gce344b39.dirty ([162]; [163]; RRID:SCR_016216), which is based on *Nipype* 1.8.5 ([164]; [165]; RRID:SCR_002502).

**Preprocessing of B0 inhomogeneity mappings** A total of 1 fieldmaps were found available within the input BIDS structure for this particular subject. A *B0* nonuniformity map (or *fieldmap*) was estimated from the phase-drift map(s) measure with two consecutive GRE (gradient-recalled echo) acquisitions. The corresponding phase-map(s) were phase-unwrapped with `prelude` (FSL 6.0.5.1:57b01774).

**Anatomical data preprocessing** A total of 1 T1-weighted (T1w) images were found within the input BIDS dataset.The T1-weighted (T1w) image was corrected for intensity non-uniformity (INU) with `N4BiasFieldCorrection` [166], distributed with ANTs 2.3.3 [167, RRID:SCR_004757], and used as T1w-reference throughout the workflow. The T1w-reference was then skull-stripped with a *Nipype* implementation of the `antsBrainExtraction.sh` workflow (from ANTs), using OASIS30ANTs as target template. Brain tissue segmentation of cerebrospinal fluid (CSF), white-matter (WM) and gray-matter (GM) was performed on the brain-extracted T1w using `fast` [FSL 6.0.5.1:57b01774, RRID:SCR_002823, 168]. Brain surfaces were reconstructed using `recon-all` [FreeSurfer 7.2.0, RRID:SCR_001847, 169], and the brain mask estimated previously was refined with a custom variation of the method to reconcile ANTs-derived and FreeSurfer-derived segmentations of the cortical gray-matter of Mindboggle [RRID:SCR_002438, 170]. Volume-based spatial normalization to two standard spaces (MNI152NLin6Asym, MNI152NLin2009cAsym) was performed through nonlinear registration with `antsRegistration` (ANTs 2.3.3), using brain-extracted versions of both T1w reference and the T1w template. The following templates were selected for spatial normalization: *FSL's MNI ICBM 152 non-linear 6th Generation Asymmetric Average Brain Stereotaxic Registration Model* [[171], RRID:SCR_002823; TemplateFlow ID: MNI152NLin6Asym], *ICBM 152 Nonlinear Asymmetrical template version 2009c* [[172], RRID:SCR_008796; TemplateFlow ID: MNI152NLin2009cAsym].

**Functional data preprocessing** For each of the 16 BOLD runs found per subject (across all tasks and sessions), the following preprocessing was performed. First, a reference volume and its skull-stripped version were generated using a custom methodology of *fMRIPrep*. Head-motion parameters with respect to the BOLD reference (transformation matrices, and six corresponding rotation and translation parameters) are estimated before any spatiotemporal filtering using `mcflirt` [FSL 6.0.5.1:57b01774, 173]. BOLD runs were slice-time corrected to 0.976s (0.5 of slice acquisition range 0s-1.95s) using `3dTshift` from AFNI [174, RRID:SCR_005927]. The BOLD time-series (including slice-timing correction when applied) were resampled onto their original, native space by applying the transforms to correct for head-motion. These resampled BOLD time-series will be referred to as *preprocessed BOLD in original space*, or just *preprocessed BOLD*. The BOLD reference was then co-registered to the T1w reference using `bbregister` (FreeSurfer) which implements boundary-based registration [175]. Co-registration was configured with six degrees of freedom. Several confounding time-series were calculated based on the *preprocessed BOLD*: framewise displacement (FD), DVARS and three region-wise global signals. FD was computed using two formulations following Power (absolute sum of relative motions, [176]) and Jenkinson (relative root mean square displacement between affines, [173]). FD and DVARS are calculated for each functional run, both using their implementations in *Nipype* [[]]compcor. Principal components are estimated after high-pass filtering the *preprocessed BOLD* time-series (using a discrete cosine filter with 128s cut-off) for the two *CompCor* variants: temporal (tCompCor) and anatomical (aCompCor). tCompCor components are then calculated from the top 2% variable voxels within the brain mask. For aCompCor, three probabilistic masks (CSF, WM and combined CSF+WM) are generated in anatomical space. The implementation differs from that of Behzadi et al. in that instead of eroding the masks by 2 pixels on BOLD space, a mask of pixels that likely contain a volume fraction of GM is subtracted from the aCompCor masks. This mask is obtained by dilating a GM mask extracted from the FreeSurfer's *aseg* segmentation, and it ensures components are not extracted from voxels containing a minimal fraction of GM. Finally, these masks are resampled into BOLD space and binarized by thresholding at 0.99 (as in the original implementation). Components are also calculated separately within the WM and CSF masks. For each CompCor decomposition, the *k* components with the largest singular values are retained, such that the retained components' time series are sufficient to explain 50 percent of variance across the nuisance mask (CSF, WM, combined, or temporal). The remaining components are dropped from consideration. The head-motion estimates calculated in the correction step were also placed within the corresponding confounds file. The confound time series derived from head motion estimates and global signals were expanded with the inclusion of temporal derivatives and quadratic terms for each [177]. Frames that exceeded a threshold of 0.5 mm FD or 1.5 standardized DVARS were annotated as motion outliers. Additional nuisance timeseries are calculated by means of principal components analysis of the signal found within a thin band (*crown*) of voxels around the edge of the brain, as proposed by [178]. The BOLD time-series were resampled into standard space, generat-ing a *preprocessed BOLD run in MNI152NLin6Asym space*. First, a reference volume and its skull-stripped version were generated using a custom methodology of *fMRIPrep*. The BOLD time-series were resampled onto the following surfaces (FreeSurfer reconstruction nomenclature): *fsaverage*. Automatic removal of motion artifacts using independent component analysis [ICA-AROMA, 179] was performed on the *preprocessed BOLD on MNI space* time-series after removal of non-steady state volumes and spatial smoothing with an isotropic, Gaussian kernel of 6mm FWHM (full-width half-maximum). Corresponding "non-aggresively" denoised runs were produced after such smoothing. Additionally, the "aggressive" noise-regressors were collected and placed in the corresponding confounds file. *Grayordinates* files [180] containing 91k samples were also generated using the highest-resolution `fsaverage` as intermediate standardized surface space. All resamplings can be performed with *a single interpolation step* by composing all the pertinent transformations (i.e. head-motion transform matrices, susceptibility distortion correction when available, and co-registrations to anatomical and output spaces). Gridded (volumetric) resamplings were performed using `antsApplyTransforms` (ANTs), configured with Lanczos interpolation to minimize the smoothing effects of other kernels [181]. Non-gridded (surface) resamplings were performed using `mri_vol2surf` (FreeSurfer).

**Functional data preprocessing** For each of the 16 BOLD runs found per subject (across all tasks and sessions), the following preprocessing was performed. First, a reference volume and its skull-stripped version were generated using a custom methodology of *fMRIPrep*. Head-motion parameters with respect to the BOLD reference (transformation matrices, and six corresponding rotation and translation parameters) are estimated before any spatiotemporal filtering using `mcflirt` [FSL 6.0.5.1:57b01774, 173]. BOLD runs were slice-time corrected to 0.975s (0.5 of slice acquisition range 0s-1.95s) using `3dTshift` from AFNI [174, RRID:SCR_005927]. The BOLD time-series (including slice-timing correction when applied) were resampled onto their original, native space by applying the transforms to correct for head-motion. These resampled BOLD time-series will be referred to as *preprocessed BOLD in original space*, or just *preprocessed BOLD*. The BOLD reference was then co-registered to the T1w reference using `bbregister` (FreeSurfer) which implements boundary-based registration [175]. Co-registration was configured with six degrees of freedom. Several confounding time-series were calculated based on the *preprocessed BOLD*: framewise displacement (FD), DVARS and three region-wise global signals. FD was computed using two formulations following Power (absolute sum of relative motions, [176]) and Jenkinson (relative root mean square displacement between affines, [173]). FD and DVARS are calculated for each functional run, both using their implementations in *Nipype* [following the definitions by 176]. The three global signals are extracted within the CSF, the WM, and the whole-brain masks. Additionally, a set of physiological regressors were extracted to allow for component-based noise correction [*CompCor*, 182]. Principal components are estimated after high-pass filtering the *preprocessed BOLD* time-series (using a discrete cosine filter with 128s cut-off) for the two *CompCor* variants: temporal (tCompCor) and anatomical (aCompCor). tCompCor components are then calculated from the top 2% variable voxels within the brain mask. For aCompCor, three probabilistic masks (CSF, WM and combined CSF+WM) are generated in anatomical space. The implementation differs from that of Behzadi et al. in that instead of eroding the masks by 2 pixels on BOLD space, a mask of pixels that likely contain a volume fraction of GM is subtracted from the aCompCor masks. This mask is obtained by dilating a GM mask extracted from the FreeSurfer's *aseg* segmentation, and it ensures components are not extracted from voxels containing a minimal fraction of GM. Finally, these masks are resampled into BOLD space and binarized by thresholding at 0.99 (as in the original implementation). Components are also calculated separately within the WM and CSF masks. For each CompCor decomposition, the *k* components with the largest singular values are retained, such that the retained components' time series are sufficient to explain 50 percent of variance across the nuisance mask (CSF, WM, combined, or temporal). The remaining components are dropped from consideration. The head-motion estimates calculated in the correction step were also placed within the corresponding confounds file. The confound time series derived from head motion estimates and global signals were expanded with the inclusion of temporal derivatives and quadratic terms for each [177]. Frames that exceeded a threshold of 0.5 mm FD or 1.5 standardized DVARS were annotated as motion outliers. Additional nuisance timeseries are calculated by means of principal components analysis of the signal found within a thin band (*crown*) of voxels around the edge of the brain, as proposed by [178]. The BOLD time-series were resampled into standard space, generating a *preprocessed BOLD run in MNI152NLin6Asym space*. First, a reference volume and its skull-stripped version were generated using a custom methodology of *fMRIPrep*. The BOLD time-series were resampled onto the following surfaces (FreeSurfer reconstruction nomenclature): *fsaverage*. Automatic removal of motion artifacts using independent component analysis [ICA-AROMA, 179] was performed on the *preprocessed BOLD on MNI space* time-series after removal of non-steady state volumes and spatial smoothing with an isotropic, Gaussian kernel of 6mm FWHM (full-width half-maximum). Corresponding "non-aggresively" denoised runs were produced after such smoothing. Additionally, the "aggressive" noise-regressors were collected and placed in the corresponding confounds file. *Grayordinates* files [180] containing 91k samples were also generated using the highest-resolution `fsaverage` as intermediate standardized surface space. All resamplings can be performed with *a single interpolation step* by composing all the pertinent transformations (i.e. head-motion transform matrices, susceptibility distortion correction when available, and co-registrations to anatomical and output spaces). Gridded (volumetric) resamplings were performed using `antsApplyTransforms` (ANTs), configured with Lanczos interpolation to minimize the smoothing effects of other kernels [181]. Non-gridded (surface) resamplings were performed using `mri_vol2surf` (FreeSurfer).

Many internal operations of *fMRIPrep* use *Nilearn* 0.9.1 [183, RRID:SCR_001362], mostly within the functional processing workflow. For more details of the pipeline, see the section corresponding to workflows in *fMRIPrep*'s documentation.

### Copyright Waiver

# Supplementary Figures and Tables

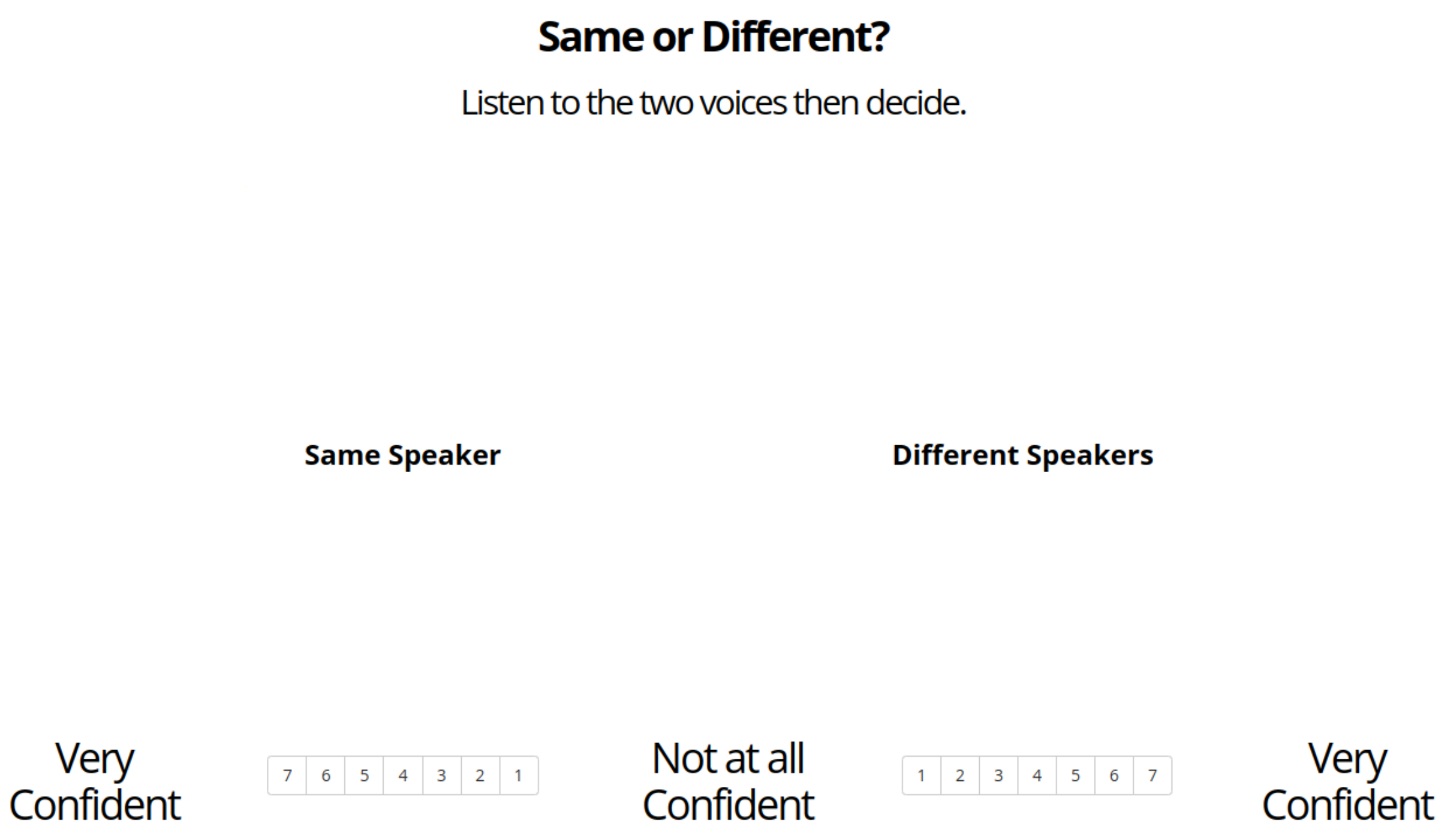

Figure 5.6: Screenshot from the online behavioral experiment trial.

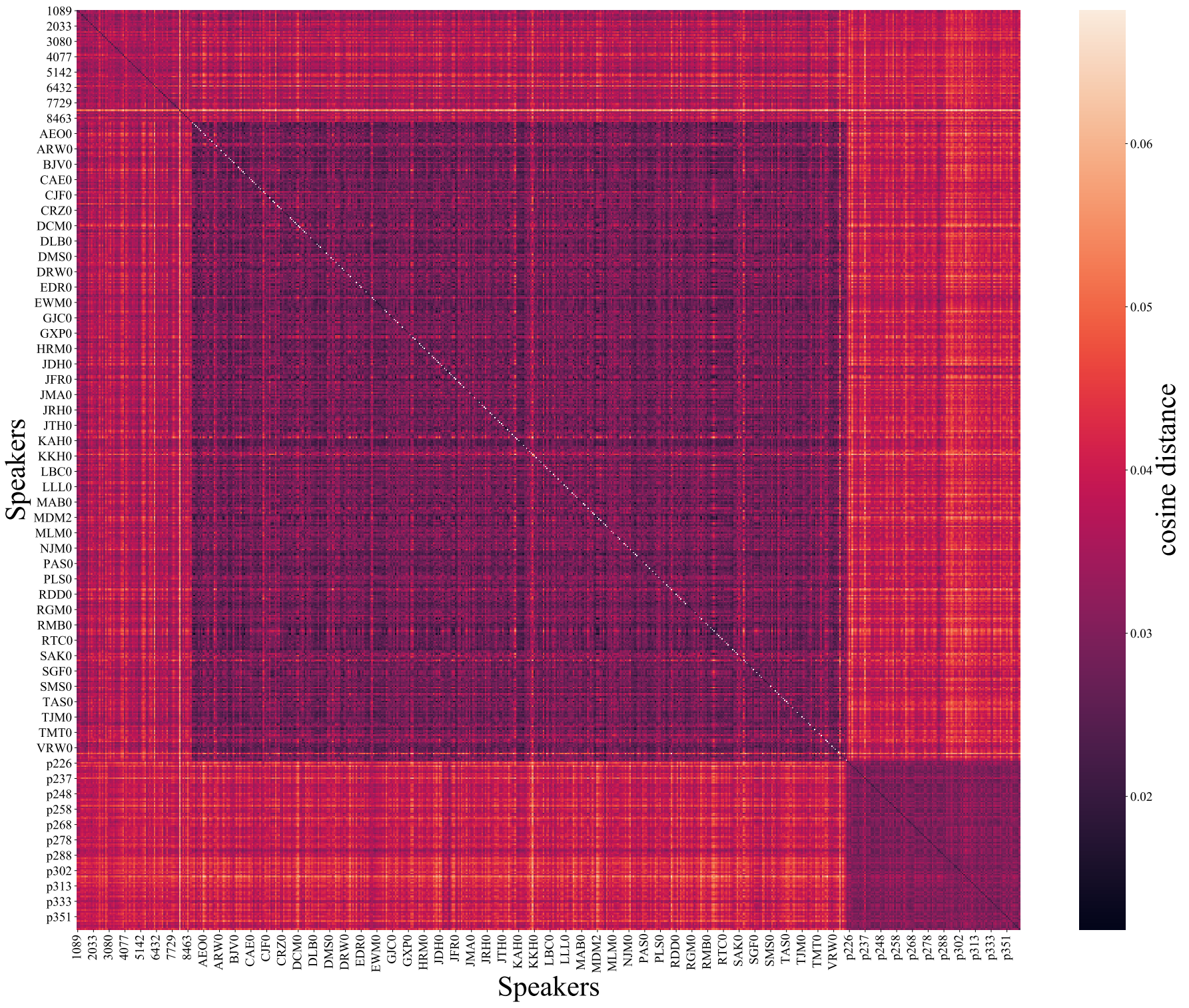

Figure 5.7: Heatmap showing pairwise cosine distances between 599 speakers.

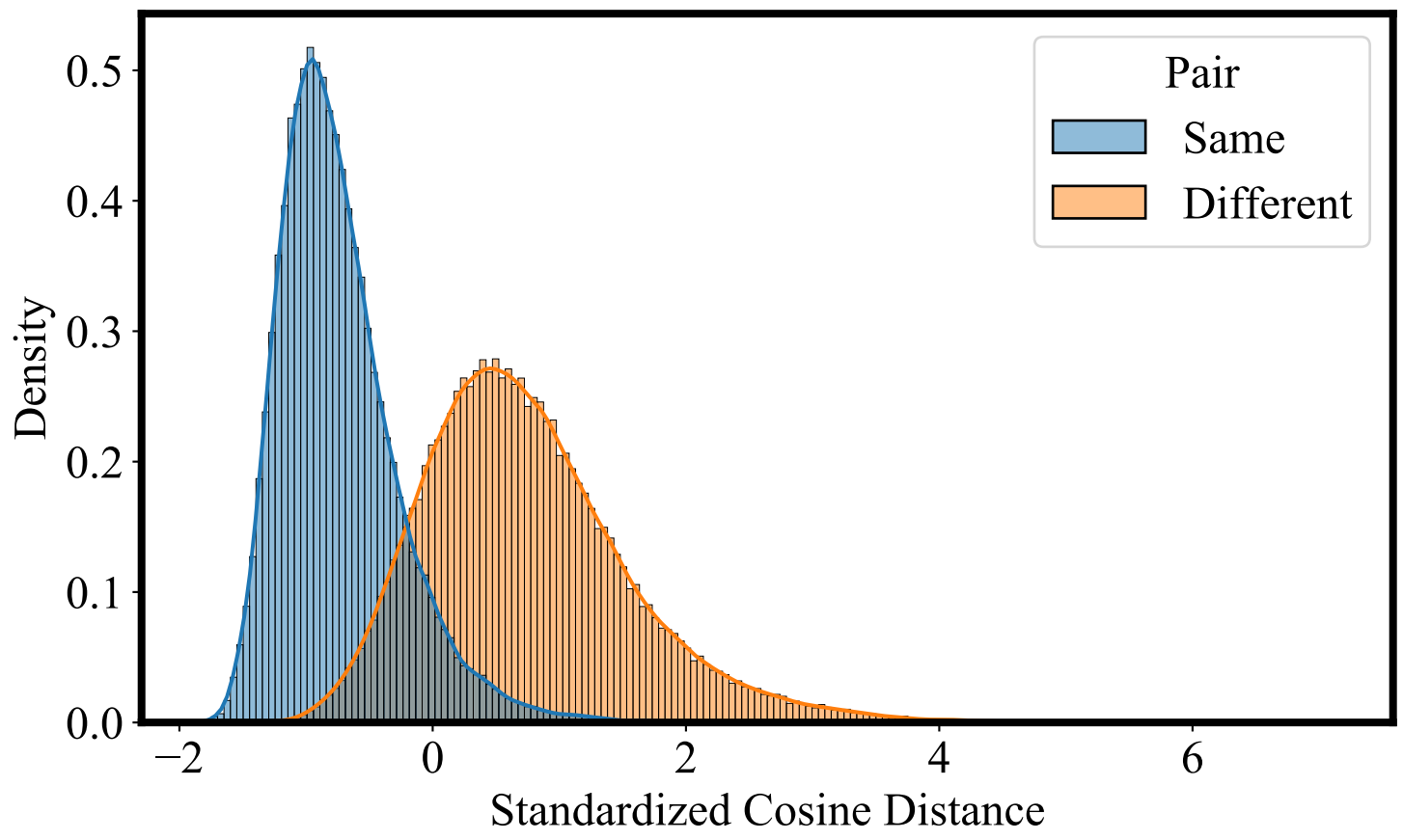

Figure 5.8: Distribution of standardized cosine distances for *same* and *different* pairs.

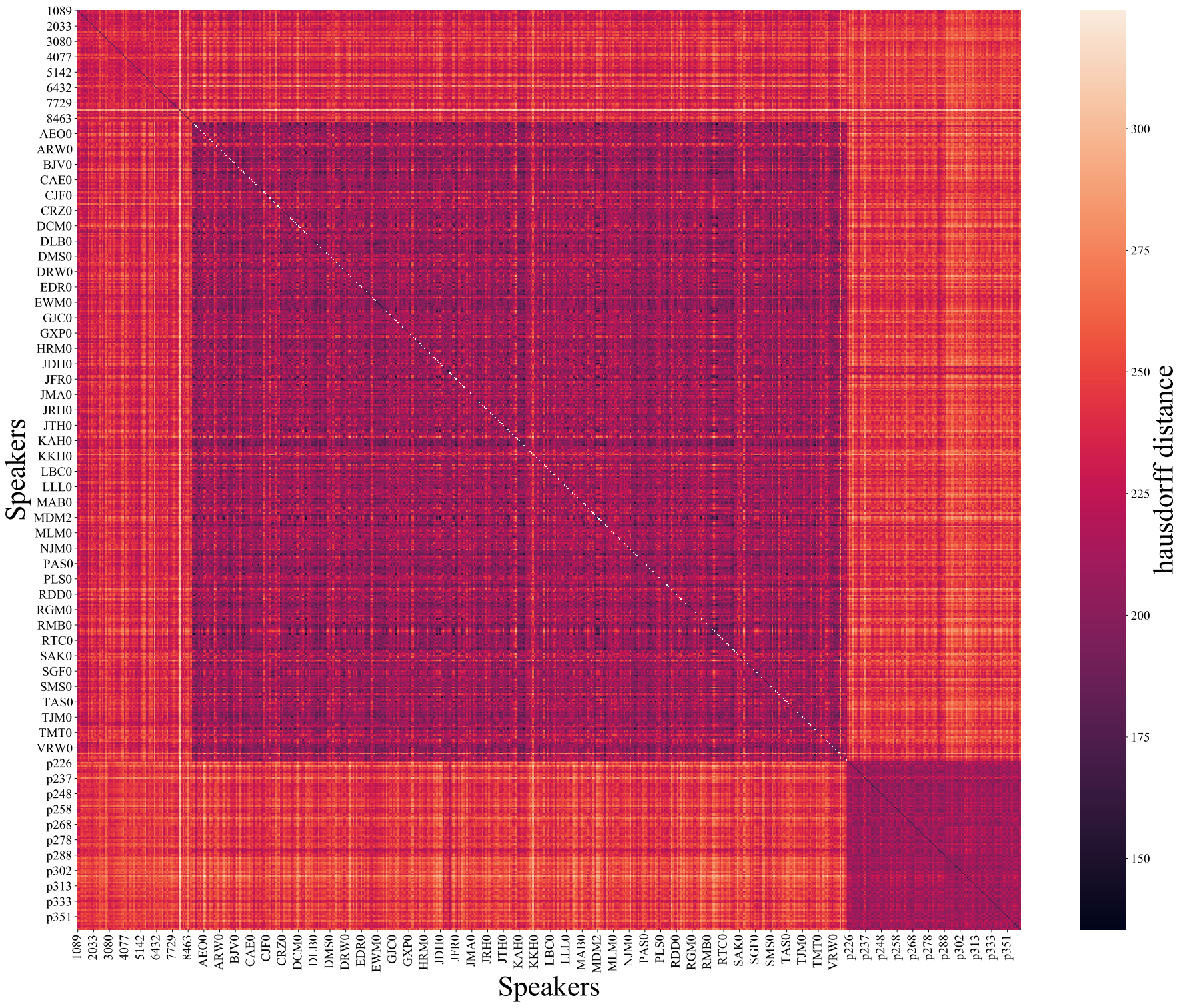

Figure 5.9: Heatmap showing pairwise hausdorff distances between 599 speakers.

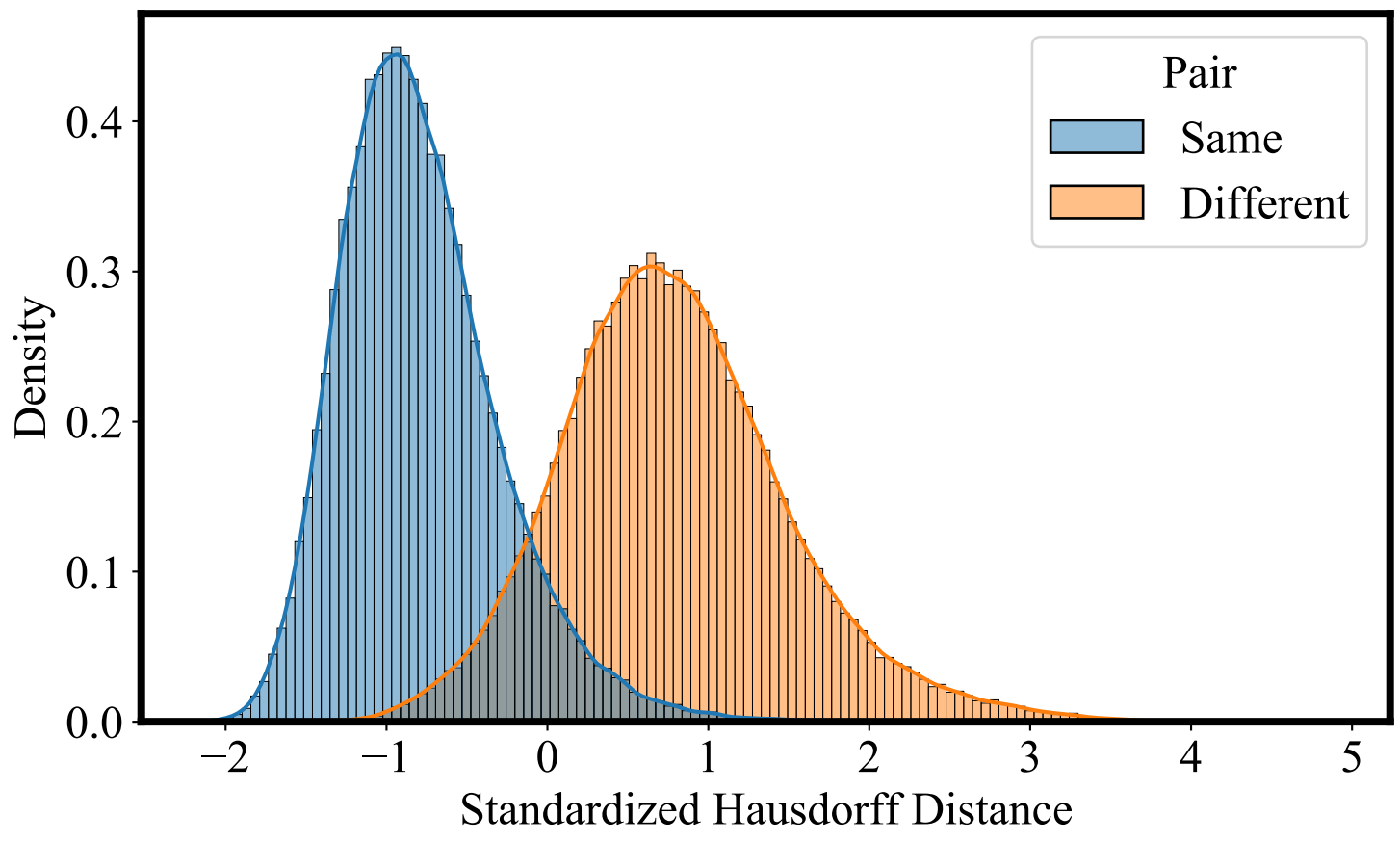

Figure 5.10: Distribution of standardized hausdorff distances for *same* and *different* pairs.

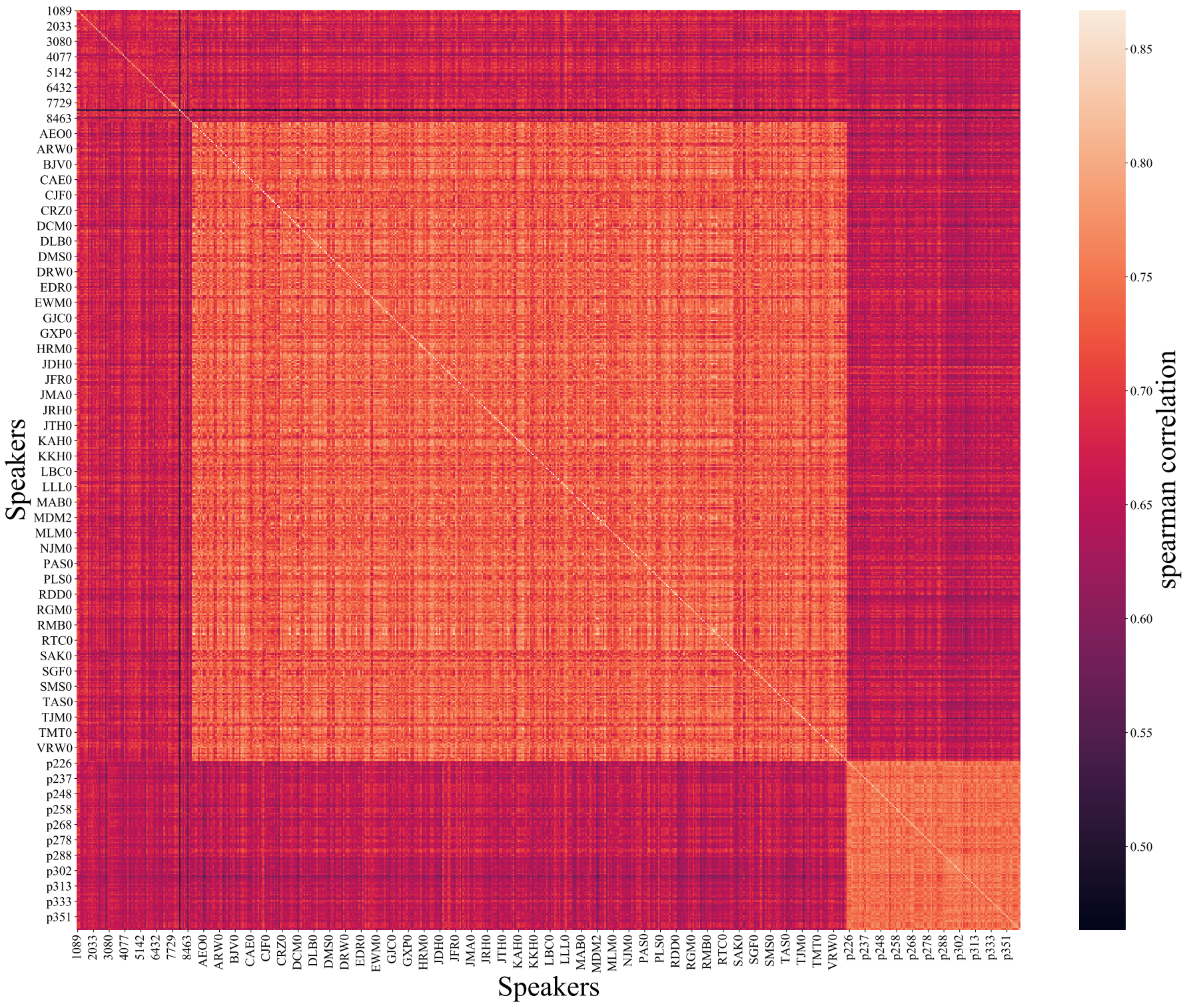

Figure 5.11: Heatmap showing pairwise spearman correlation between 599 speakers.

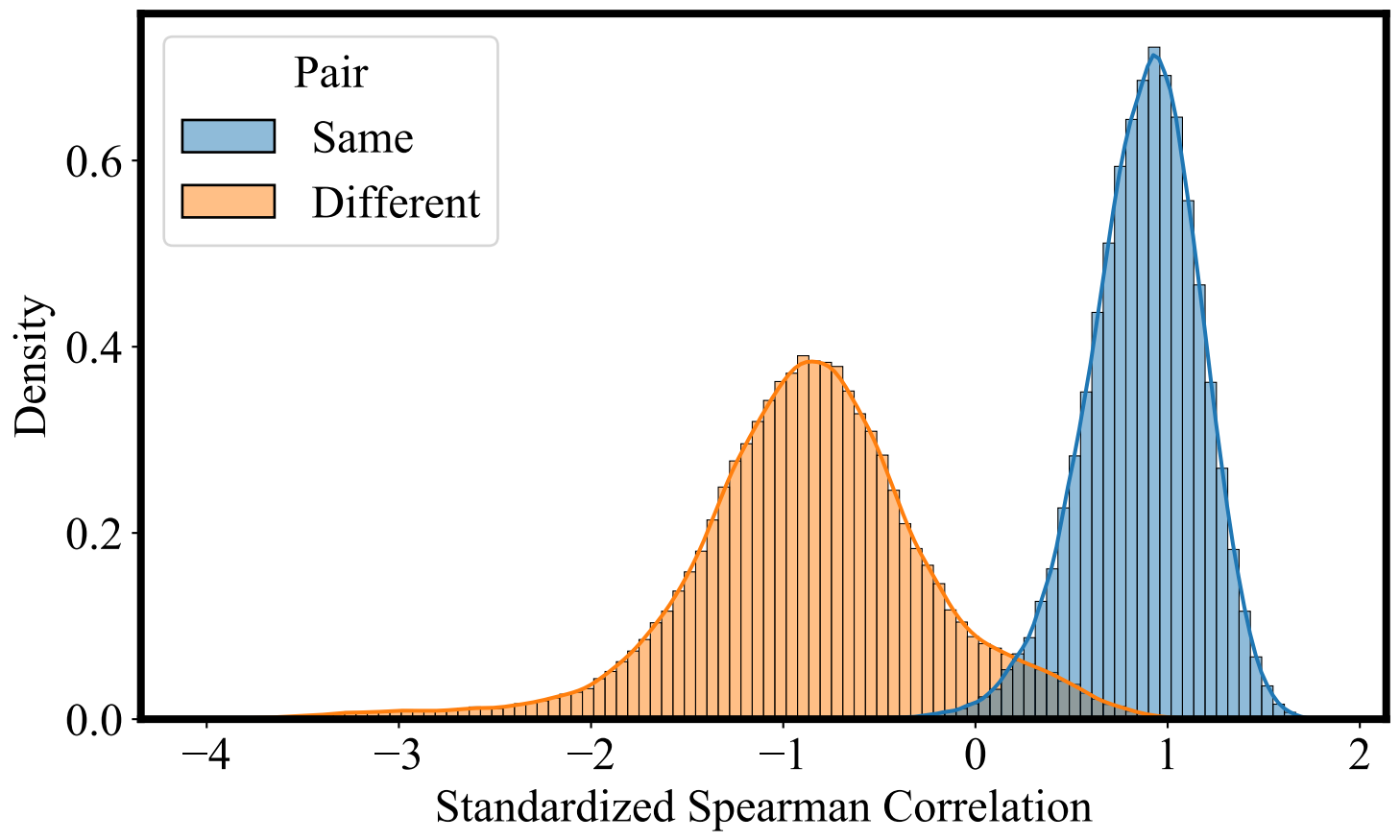

Figure 5.12: Distribution of standardized spearman correlation for *same* and *different* pairs.

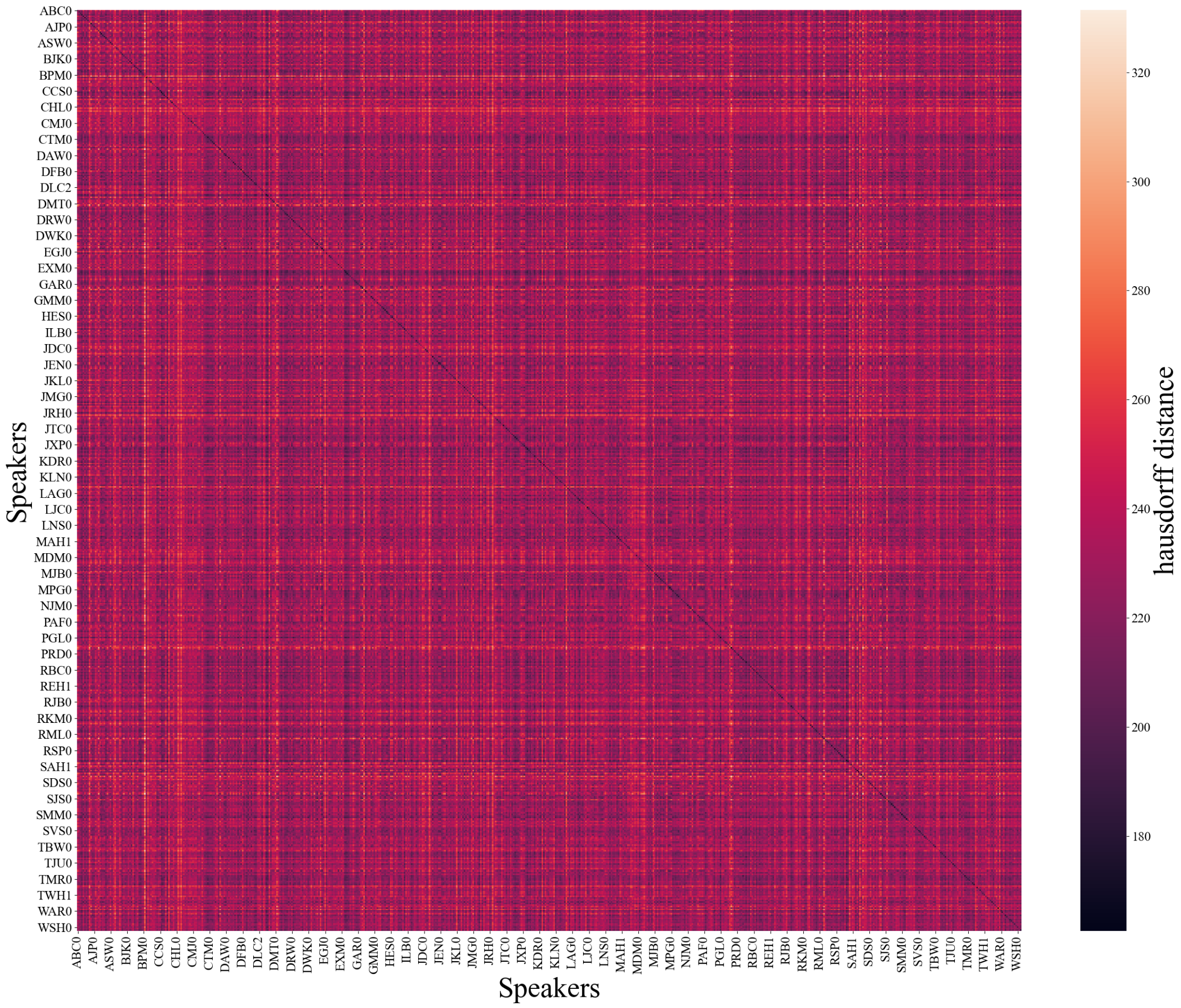

Figure 5.13: Hausdorff Distance Matrix across speakers.

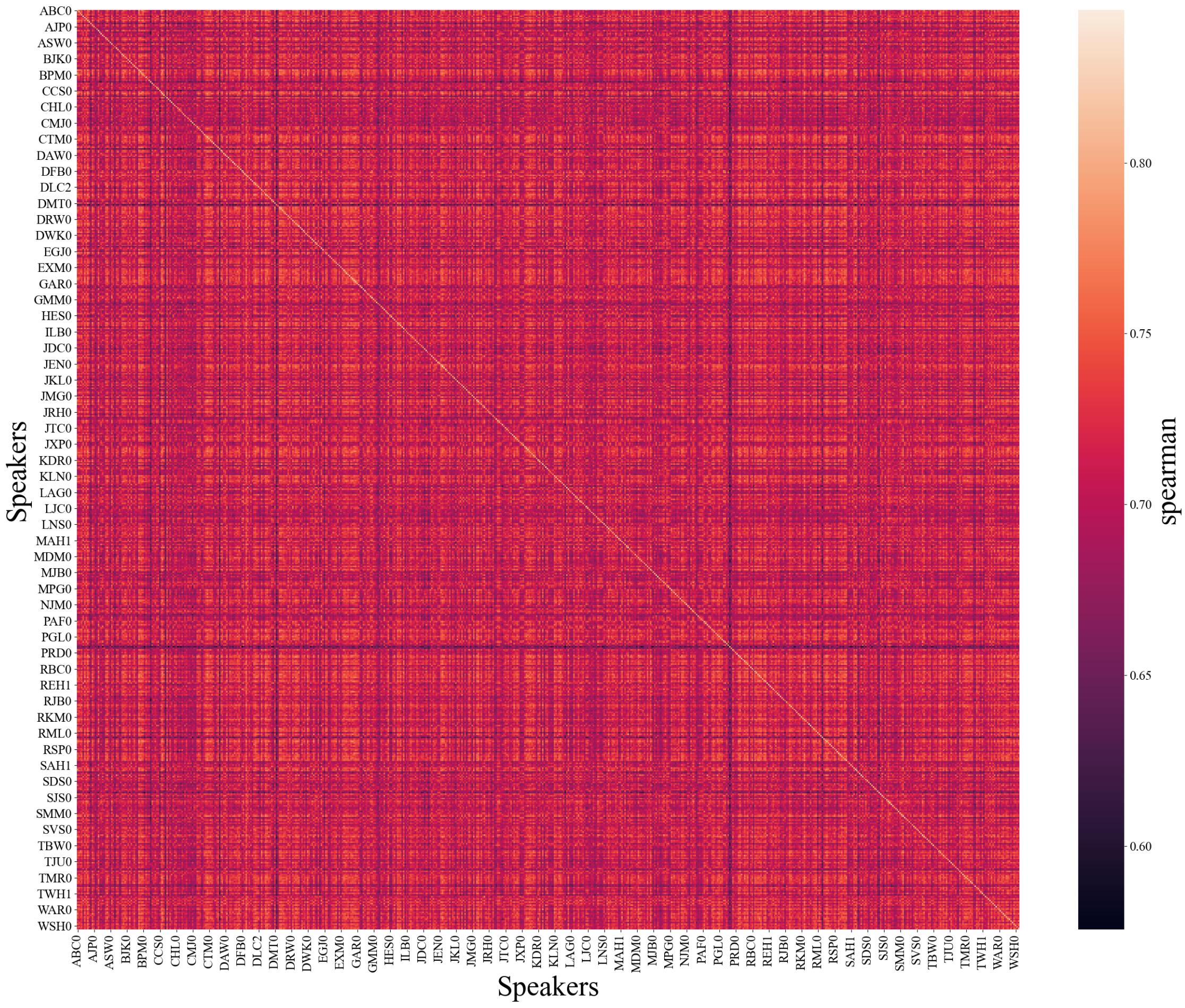

Figure 5.14: Spearman Correlation Matrix across speakers.

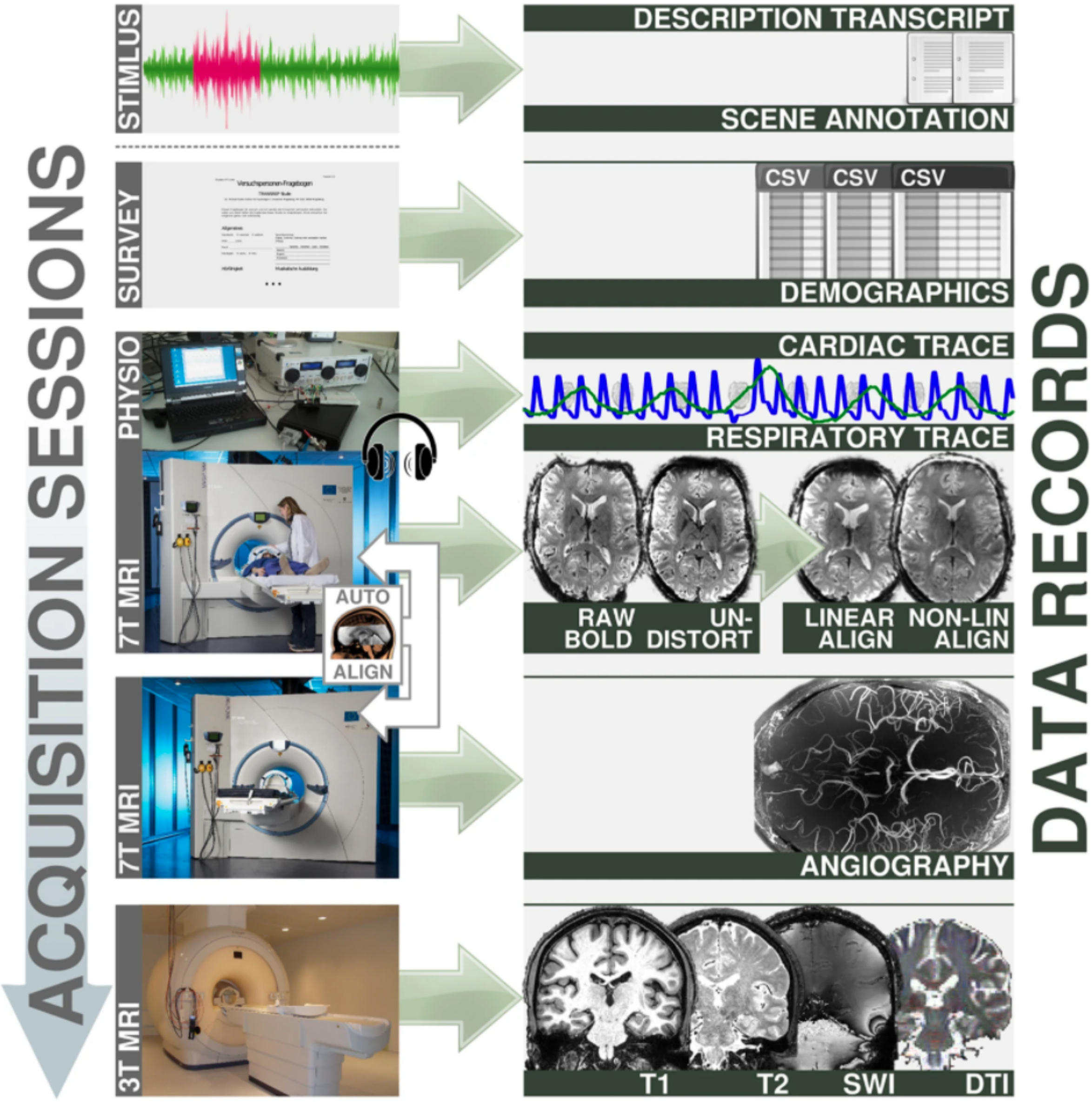

Figure 5.15: StudyForrest scans acquisition. Figure is taken from [143].

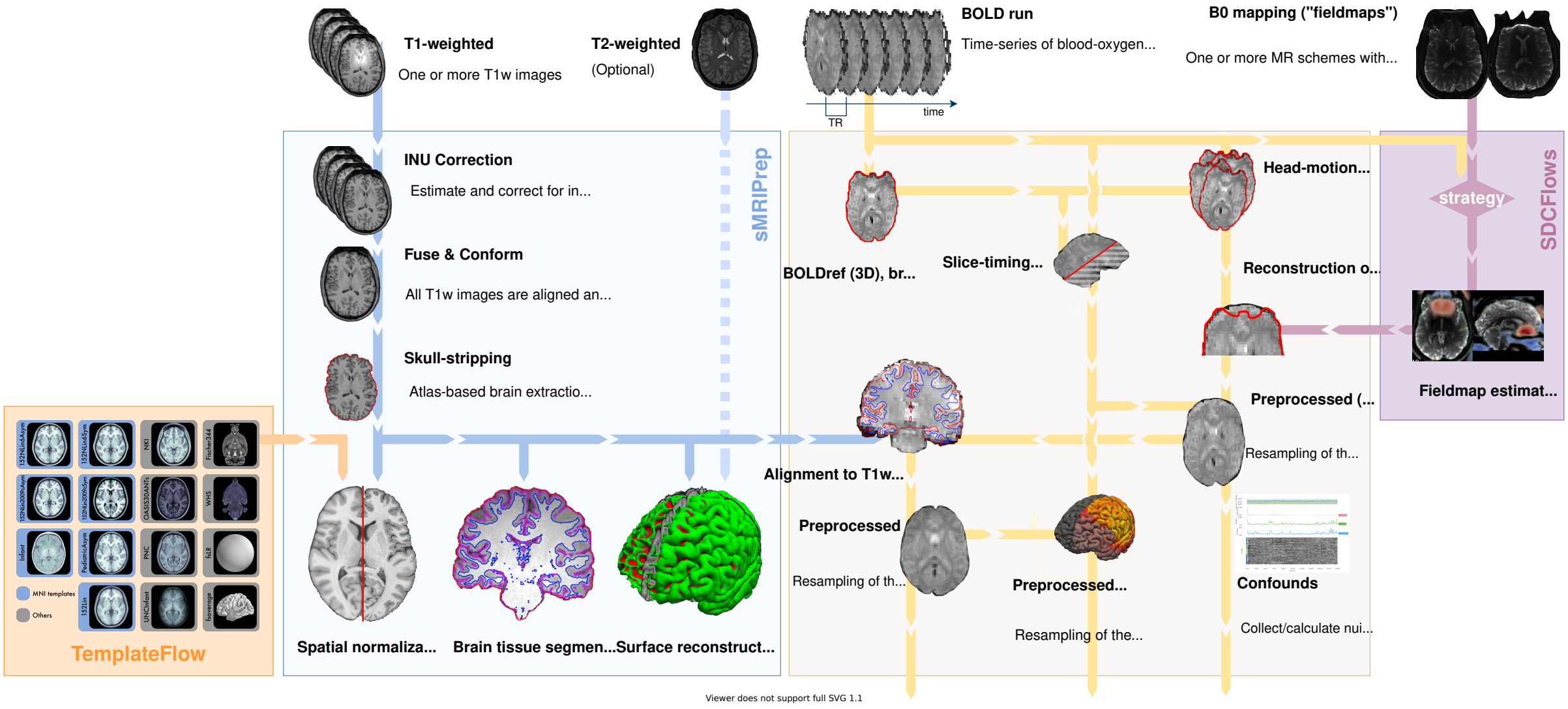

Figure 5.16: fMRIPrep Software Pipeline for preprocessing neuroimaging data. Figure is taken from [144].

| stimuli_1 | Sex_1 | stimuli_2 | Sex_2 | Pair | Distance_seuc | Distance_cos | Distance_haus | Distance_spear | Distance_1-spear | Score | Dataset | Order |
|---|---|---|---|---|---|---|---|---|---|---|---|---|
| p250_328_mic1_norm.flac | F | p250_336_mic1_norm_noise.flac | F | Same | -0.155928877559874 | -0.010805171250699 | -0.004142465670555 | 0.013593370974214 | 0.986406629025786 | 0.203882528636165 | VCTK | C-N |
| p259_365_mic1_norm.flac | M | p259_448_mic1_norm_noise.flac | M | Same | -0.092802876467936 | -0.034297204546524 | -0.009083271228798 | 0.078839685449652 | 0.921160314550347 | 0.196244240576772 | VCTK | C-N |
| p282_085_mic1_norm.flac | F | p282_255_mic1_norm_noise.flac | F | Same | -0.003462189481625 | -0.033150547547096 | -0.019005266271797 | 0.228393752908951 | 0.771606247091049 | 0.178997060947633 | VCTK | C-N |
| p276_109_mic1_norm_noise.flac | F | p276_354_mic1_norm.flac | F | Same | -0.048144258582387 | -0.10215691411808 | -0.059222193834555 | 0.099973804713679 | 0.900026195286321 | 0.172625707187825 | VCTK | N-C |
| p283_158_mic1_norm_noise.flac | F | p283_426_mic1_norm.flac | F | Same | -0.120538411129061 | -0.041693126764261 | -0.007461296040942 | 0.160424857561744 | 0.839575142438256 | 0.167470577125998 | VCTK | N-C |
| p314_116_mic1_norm.flac | F | p314_264_mic1_norm_noise.flac | F | Same | -0.12943433923571 | -0.005918830268755 | -0.0671137723299 | 0.129198700792812 | 0.870801299207188 | 0.167083589343206 | VCTK | C-N |
| p282_255_mic1_norm_noise.flac | F | p282_369_mic1_norm.flac | F | Same | -0.002067937988239 | -0.066953833491652 | -0.073299064325975 | 0.189912426600649 | 0.810087573399352 | 0.166941684398371 | VCTK | N-C |
| p244_119_mic1_norm.flac | F | p244_312_mic1_norm_noise.flac | F | Same | -0.012579094138601 | -0.123825547369799 | -0.056323666562197 | 0.154562416776836 | 0.845437583223164 | 0.163177318788142 | VCTK | C-N |
| p294_123_mic1_norm_noise.flac | F | p294_184_mic1_norm.flac | F | Same | -0.190198488544435 | -0.006997371950295 | -0.119570696688486 | 0.034321377607997 | 0.965678622392003 | 0.162228016302196 | VCTK | N-C |
| p239_080_mic1_norm_noise.flac | F | p239_418_mic1_norm.flac | F | Same | -0.23366657792744 | -0.033837296405819 | -0.033283737405908 | 0.078489404875053 | 0.921510595124947 | 0.155180745846445 | VCTK | N-C |
| p335_202_mic1_norm.flac | F | p335_394_mic1_norm_noise.flac | F | Same | -0.01788380949032 | -0.001239314267075 | -0.085816097765192 | 0.274828201280424 | 0.725171798719576 | 0.155058144299247 | VCTK | C-N |
| p253_062_mic1_norm.flac | F | p253_113_mic1_norm_noise.flac | F | Same | -0.083664584873587 | -0.085833618275435 | -0.09651368274146 | 0.117543065960467 | 0.882456934039533 | 0.154111262037263 | VCTK | C-N |
| p310_131_mic1_norm_noise.flac | F | p310_139_mic1_norm.flac | F | Same | -0.119580883219085 | -0.010627975911722 | -0.013523822066501 | 0.249284159629417 | 0.750715840370583 | 0.151745789793319 | VCTK | N-C |
| p264_228_mic1_norm_noise.flac | F | p264_285_mic1_norm.flac | F | Same | -0.024537736914866 | -0.003351186674755 | -0.055436770031961 | 0.324747665345231 | 0.675252334654769 | 0.147981660258297 | VCTK | N-C |
| p345_098_mic1_norm.flac | M | p345_364_mic1_norm_noise.flac | M | Same | -0.163865517548903 | -0.047612112976822 | -0.016183111546313 | 0.182565527616837 | 0.817434472383163 | 0.147443432577781 | VCTK | C-N |
| p316_180_mic1_norm_noise.flac | M | p316_181_mic1_norm.flac | M | Same | -0.114392889574519 | -0.166393824797316 | -0.03933928809111 | 0.116358602014906 | 0.883641397985094 | 0.140878848880537 | VCTK | N-C |
| p270_339_mic1_norm_noise.flac | M | p270_398_mic1_norm.flac | M | Same | -0.104159077550112 | -0.014412990724 | -0.014557855177408 | 0.304393770955437 | 0.695606229044564 | 0.140619076398261 | VCTK | N-C |
| p374_030_mic1_norm_noise.flac | M | p374_250_mic1_norm.flac | M | Same | -0.116865566011308 | -0.080150832744769 | -0.01029848538642 | 0.235997629233433 | 0.764002370766567 | 0.13917187165601‌8 | VCTK | N-C |
| p288_026_mic1_norm.flac | F | p288_196_mic1_norm_noise.flac | F | Same | -0.132752335882262 | -0.111920041935289 | -0.036940690005522 | 0.162283652831705 | 0.837716347168295 | 0.139025819836306 | VCTK | C-N |
| p308_209_mic1_norm_noise.flac | F | p308_298_mic1_norm.flac | F | Same | -0.173756798475711 | -0.020768110790649 | -0.017167189136852 | 0.232765290452289 | 0.767234709547711 | 0.138885652786125 | VCTK | N-C |
| p335_032_mic1_norm.flac | F | p335_257_mic1_norm_noise.flac | F | Same | -0.13389541814611 | -0.061901951223737 | -0.093090956734412 | 0.176487327787369 | 0.823512672212631 | 0.133656086527093 | VCTK | C-N |
| p336_039_mic1_norm.flac | F | p336_055_mic1_norm_noise.flac | F | Same | -0.092340871720781 | -0.058445141473852 | -0.129192910378366 | 0.185689199980957 | 0.814310800019043 | 0.133582969111511 | VCTK | C-N |
| p315_057_mic1_norm.flac | M | p315_193_mic1_norm_noise.flac | M | Same | -0.065293956980147 | -0.192502855955902 | -0.090645021595694 | 0.123222872222416 | 0.876777127777584 | 0.13208382331146 | VCTK | C-N |
| p285_126_mic1_norm.flac | M | p285_195_mic1_norm_noise.flac | M | Same | -0.093610096262978 | -0.059682644633332 | -0.067657691521272 | 0.254728796776457 | 0.745271203223543 | 0.13108019270149 | VCTK | C-N |
| p244_070_mic1_norm_noise.flac | F | p244_423_mic1_norm.flac | F | Same | -0.068780878105526 | -0.13510329384836 | -0.104936608398102 | 0.17550573280104 | 0.82449426719896 | 0.128918371711743 | VCTK | N-C |
| p363_193_mic1_norm_noise.flac | M | p363_313_mic1_norm.flac | M | Same | -0.149178484232113 | -0.008527678438286 | -0.012364011241689 | 0.327249016380409 | 0.672750983619591 | 0.125670202426876 | VCTK | N-C |
| p246_146_mic1_norm_noise.flac | M | p246_155_mic1_norm.flac | M | Same | -0.002556105924979 | -0.107264161115571 | -0.035625629472174 | 0.360675255911114 | 0.639324744088886 | 0.123469711894041 | VCTK | N-C |
| p333_111_mic1_norm.flac | F | p333_262_mic1_norm_noise.flac | F | Same | -0.125260280253036 | -0.181715709364861 | -0.047612884265658 | 0.156772466872034 | 0.843227533127966 | 0.122159664811103 | VCTK | C-N |
| p239_213_mic1_norm.flac | F | p239_375_mic1_norm_noise.flac | F | Same | -0.092515627143705 | -0.034607291516726 | -0.090274857933884 | 0.305175344792087 | 0.694824655207913 | 0.1193567196534 | VCTK | C-N |
| p313_271_mic1_norm.flac | F | p313_320_mic1_norm_noise.flac | F | Same | -0.046720039121275 | -0.06232018575347 | -0.013168685466031 | 0.409421436798569 | 0.590578563201431 | 0.117092413215164 | VCTK | C-N |
| p329_344_mic1_norm_noise.flac | F | p329_413_mic1_norm.flac | F | Same | -0.186392391828378 | -0.019090293128014 | -0.036235548060832 | 0.299177333284147 | 0.700822666715853 | 0.114776108424657 | VCTK | N-C |
| p329_057_mic1_norm_noise.flac | F | p329_087_mic1_norm.flac | F | Same | -0.102765632840731 | -0.152157798414806 | -0.150789209243701 | 0.137675880398385 | 0.862324119601615 | 0.114152869775594 | VCTK | N-C |
| p239_245_mic1_norm_noise.flac | F | p239_459_mic1_norm.flac | F | Same | -0.053240367713744 | -0.03778601441634 | -0.063567845054523 | 0.401463832551081 | 0.598536167448919 | 0.110985485066078 | VCTK | N-C |
| p351_098_mic1_norm.flac | F | p351_131_mic1_norm_noise.flac | F | Same | -0.04190387609881 | -0.075087530449758 | -0.122577200072093 | 0.317055806694136 | 0.682944193305864 | 0.110843896671301 | VCTK | C-N |
| p264_266_mic1_norm_noise.flac | F | p264_318_mic1_norm.flac | F | Same | -0.150649646771593 | -0.049596048429443 | -0.020481462186449 | 0.338524618522597 | 0.661475381477403 | 0.110187056022479 | VCTK | N-C |
| p273_100_mic1_norm.flac | M | p273_361_mic1_norm_noise.flac | M | Same | -0.065442609049081 | -0.206505265552994 | -0.088594399859487 | 0.202133201743874 | 0.797866798256126 | 0.109331130948641 | VCTK | C-N |
| p339_010_mic1_norm.flac | F | p339_121_mic1_norm_noise.flac | F | Same | -0.142571666683656 | -0.104735061671914 | -0.031754668674454 | 0.284691275528935 | 0.715308724471065 | 0.10906183186026 | VCTK | C-N |
| p294_299_mic1_norm_noise.flac | F | p294_409_mic1_norm.flac | F | Same | -0.123487424866757 | -0.126606088643917 | -0.06434031542945 | 0.251074382643119 | 0.74892561735688‌1 | 0.108622947104189 | VCTK | N-C |
| p318_277_mic1_norm.flac | F | p318_322_mic1_norm_noise.flac | F | Same | -0.145596515662598 | -0.037927261942526 | -0.088885707074456 | 0.294501230003715 | 0.705498769996285 | 0.108272321329176 | VCTK | C-N |
| p247_260_mic1_norm_noise.flac | M | p247_391_mic1_norm.flac | M | Same | -0.211588985692912 | -0.069789205820292 | -0.004957629251445 | 0.280636894038592 | 0.719363105961409 | 0.10825682129919 | VCTK | N-C |
| p292_266_mic1_norm.flac | M | p292_349_mic1_norm_noise.flac | M | Same | -0.051549655492373 | -0.046070556064692 | -0.112541003002078 | 0.359080257736271 | 0.640919742263729 | 0.107689631926146 | VCTK | C-N |
| p294_126_mic1_norm_noise.flac | F | p294_227_mic1_norm.flac | F | Same | -0.105502825212546 | -0.058449258793072 | -0.082362400959793 | 0.330988340203738 | 0.669011659796262 | 0.105674293707713 | VCTK | N-C |
| p294_095_mic1_norm.flac | F | p294_107_mic1_norm_noise.flac | F | Same | -0.02575209423463 | -0.139802305180964 | -0.14743631890662 | 0.264384969535559 | 0.735615030464441 | 0.105656078035557 | VCTK | C-N |
| p239_073_mic1_norm.flac | F | p239_364_mic1_norm_noise.flac | F | Same | -0.206319113117081 | -0.145936778813885 | -0.126769245571426 | 0.104500922589254 | 0.895499077410746 | 0.104118484977088 | VCTK | C-N |
| p283_216_mic1_norm.flac | F | p283_233_mic1_norm_noise.flac | F | Same | -0.068079899166371 | -0.051147033987731 | -0.005887806780632 | 0.461831924210427 | 0.538168075789573 | 0.10326333396371 | VCTK | C-N |
| p234_044_mic1_norm_noise.flac | F | p234_129_mic1_norm.flac | F | Same | -0.073961474008387 | -0.211720521553961 | -0.163698943100941 | 0.141783545904247 | 0.858216454095753 | 0.102208878858116 | VCTK | N-C |
| p363_087_mic1_norm_noise.flac | M | p363_270_mic1_norm.flac | M | Same | -0.09268544415834 | -0.221326104935054 | -0.101229312887692 | 0.175947143281228 | 0.824052856718772 | 0.102202998684422 | VCTK | N-C |
| p294_236_mic1_norm_noise.flac | F | p294_395_mic1_norm.flac | F | Same | -0.159491493399037 | -0.163625748637004 | -0.106349482598067 | 0.163717524939925 | 0.836282475060075 | 0.101703937606492 | VCTK | N-C |
| p330_142_mic1_norm_noise.flac | F | p330_386_mic1_norm.flac | F | Same | -0.142274881916602 | -0.107623292011746 | -0.116422997248696 | 0.231213833786614 | 0.768786166213386 | 0.100616248759086 | VCTK | N-C |
| p233_109_mic1_norm.flac | F | p233_160_mic1_norm_noise.flac | F | Same | -0.013857917659947 | -0.032301437004343 | -0.157689478349546 | 0.397186122838168 | 0.602813877161832 | 0.099741261036999 | VCTK | C-N |
| p249_127_mic1_norm.flac | F | p297_270_mic1_norm_noise.flac | F | Different | -0.208020899773291 | -0.183184343658378 | -0.231599562970033 | 0.496031044264421 | 0.503968955735579 | -0.029708962666531 | VCTK | C-N |
| p256_208_mic1_norm.flac | M | p271_402_mic1_norm_noise.flac | M | Different | -0.167831185583688 | -0.239409058713388 | -0.241328648192312 | 0.445981540853886 | 0.554018459146114 | -0.023637608335819 | VCTK | C-N |
| p247_286_mic1_norm.flac | M | p334_185_mic1_norm_noise.flac | M | Different | -0.244908836786225 | -0.220471288256563 | -0.232634946058306 | 0.393578622620042 | 0.606421377379958 | -0.022898423430284 | VCTK | C-N |
| p244_136_mic1_norm_noise.flac | F | p318_336_mic1_norm.flac | F | Different | -0.205084563525083 | -0.220431992872181 | -0.243152020333158 | 0.407249907074645 | 0.592750092925354 | -0.018979620951267 | VCTK | N-C |
| p246_124_mic1_norm_noise.flac | M | p316_119_mic1_norm.flac | M | Different | -0.207311599727139 | -0.157905978529332 | -0.238841337858656 | 0.467969928040436 | 0.532030071959564 | -0.018007211038891 | VCTK | N-C |
| p227_119_mic1_norm.flac | M | p259_077_mic1_norm_noise.flac | M | Different | -0.190036756139256 | -0.249842098494998 | -0.131920724301683 | 0.493739007231137 | 0.506260992768863 | -0.016384646541769 | VCTK | C-N |
| p232_071_mic1_norm_noise.flac | M | p316_369_mic1_norm.flac | M | Different | -0.24830431317588 | -0.238651958608622 | -0.184422320658974 | 0.367741345884863 | 0.632258654115137 | -0.009779984582085 | VCTK | N-C |
| p244_185_mic1_norm.flac | F | p262_204_mic1_norm_noise.flac | F | Different | -0.137858643076819 | -0.239637195303016 | -0.240457421985377 | 0.415821064395356 | 0.584178935604645 | -0.008443581190142 | VCTK | C-N |
| p238_299_mic1_norm_noise.flac | F | p300_366_mic1_norm.flac | F | Different | -0.231220940031984 | -0.20286943792479 | -0.211355199611754 | 0.381309734483901 | 0.618690265516099 | -0.006688828013107 | VCTK | N-C |
| p240_186_mic1_norm_noise.flac | F | p333_076_mic1_norm.flac | F | Different | -0.224404700862595 | -0.170543892496648 | -0.204141825862628 | 0.405285967678421 | 0.594714032321579 | -0.001094096725073 | VCTK | N-C |
| p247_435_mic1_norm.flac | M | p376_133_mic1_norm_noise.flac | M | Different | -0.191678799234321 | -0.135959698595577 | -0.169012451429878 | 0.499068532918488 | 0.500931467081512 | 0.001070129455434 | VCTK | C-N |
| p234_257_mic1_norm.flac | F | p264_164_mic1_norm_noise.flac | F | Different | -0.214886201628635 | -0.119444453979998 | -0.171044333823613 | 0.483930476661129 | 0.516069523338871 | 0.002673633476656 | VCTK | C-N |
| p252_131_mic1_norm_noise.flac | M | p274_047_mic1_norm.flac | M | Different | -0.231344793954168 | -0.202709443690456 | -0.132230263977589 | 0.408645783408087 | 0.591354216591913 | 0.006267428742425 | VCTK | N-C |
| p267_397_mic1_norm_noise.flac | F | p276_367_mic1_norm.flac | F | Different | -0.245431386250733 | -0.21086067927285 | -0.152308263583852 | 0.364558693886124 | 0.635441306113876 | 0.00671024425161 | VCTK | N-C |
| p233_237_mic1_norm.flac | F | p244_264_mic1_norm_noise.flac | F | Different | -0.14692926913293 | -0.164850890306904 | -0.230937146299808 | 0.429568265457165 | 0.570431734542835 | 0.006928607200798 | VCTK | C-N |
| p269_337_mic1_norm_noise.flac | F | p351_046_mic1_norm.flac | F | Different | -0.165042146806329 | -0.23001114094955 | -0.230343183537644 | 0.33710880752231 | 0.66289119247769 | 0.009373680296042 | VCTK | N-C |
| p279_341_mic1_norm_noise.flac | M | p334_233_mic1_norm.flac | M | Different | -0.243789099411408 | -0.181513422415441 | -0.210223841104561 | 0.324298835571744 | 0.675701164428256 | 0.010043700374212 | VCTK | N-C |
| p302_032_mic1_norm_noise.flac | M | p363_211_mic1_norm.flac | M | Different | -0.196293124110672 | -0.180266452111414 | -0.188732196444384 | 0.377609080916561 | 0.622390919083438 | 0.014274786604242 | VCTK | N-C |
| p333_035_mic1_norm.flac | F | p341_067_mic1_norm_noise.flac | F | Different | -0.183870883863867 | -0.17675655408395 | -0.176294425469006 | 0.39832427240737 | 0.601675727592629 | 0.016188466043952 | VCTK | C-N |
| p263_409_mic1_norm_noise.flac | M | p284_138_mic1_norm.flac | M | Different | -0.139512281390261 | -0.242097435427741 | -0.226664589477574 | 0.32555546900668 | 0.67444453099332 | 0.016542556174436 | VCTK | N-C |
| p261_091_mic1_norm.flac | F | p323_193_mic1_norm_noise.flac | F | Different | -0.224357033770237 | -0.237082827858524 | -0.211674740383226 | 0.255127639998112 | 0.744872360001888 | 0.017939439497475 | VCTK | C-N |
| p231_397_mic1_norm.flac | F | p333_190_mic1_norm_noise.flac | F | Different | -0.138293388751035 | -0.200014240677078 | -0.12497820691958 | 0.460258284607206 | 0.539741715392794 | 0.019113969761276 | VCTK | C-N |
| p263_407_mic1_norm.flac | M | p270_443_mic1_norm_noise.flac | M | Different | -0.197417156833667 | -0.238198950524266 | -0.246589217708872 | 0.237472719637766 | 0.762527280362234 | 0.020080488823857 | VCTK | C-N |
| p226_032_mic1_norm.flac | M | p326_166_mic1_norm_noise.flac | M | Different | -0.112495659759123 | -0.094432923329226 | -0.214848964457419 | 0.49672658427573 | 0.50327341572427 | 0.020373967044626 | VCTK | C-N |
| p237_043_mic1_norm_noise.flac | M | p243_314_mic1_norm.flac | M | Different | -0.139012070965101 | -0.166662857220772 | -0.20296633199281 | 0.402949339942978 | 0.597050660057022 | 0.022102349969585 | VCTK | N-C |
| p272_255_mic1_norm_noise.flac | M | p376_102_mic1_norm.flac | M | Different | -0.212000653300102 | -0.19191552959005 | -0.206778087336469 | 0.300322152723084 | 0.699677847276916 | 0.022245894262574 | VCTK | N-C |
| p286_428_mic1_norm_noise.flac | M | p376_188_mic1_norm.flac | M | Different | -0.172142681494576 | -0.183421479727566 | -0.18506533148631 | 0.365639587494908 | 0.634360412505092 | 0.02343272994916 | VCTK | N-C |
| p278_299_mic1_norm.flac | M | p304_384_mic1_norm_noise.flac | M | Different | -0.138118944099624 | -0.173056924362678 | -0.210142677365857 | 0.381823089626391 | 0.618176910373609 | 0.02421459113636‌3 | VCTK | C-N |
| p259_479_mic1_norm.flac | M | p376_213_mic1_norm_noise.flac | M | Different | -0.18524678746307 | -0.238387722384569 | -0.207956254838519 | 0.270183484490314 | 0.729816515509686 | 0.024556437705882 | VCTK | C-N |
| p234_245_mic1_norm.flac | F | p341_184_mic1_norm_noise.flac | F | Different | -0.191194779478372 | -0.118187770175251 | -0.115132536053387 | 0.476564092664609 | 0.523435907335391 | 0.024730205407095 | VCTK | C-N |
| p277_216_mic1_norm_noise.flac | F | p313_050_mic1_norm.flac | F | Different | -0.205043454584697 | -0.222356646201897 | -0.221784843780366 | 0.246360283587908 | 0.753639716412092 | 0.026113692961283 | VCTK | N-C |
| p252_228_mic1_norm_noise.flac | M | p270_335_mic1_norm.flac | M | Different | -0.165034412752182 | -0.203256894409508 | -0.166883493637916 | 0.354686687140102 | 0.645313312859898 | 0.027534628015073 | VCTK | N-C |
| p323_083_mic1_norm_noise.flac | F | p361_268_mic1_norm.flac | F | Different | -0.24631925232075 | -0.210594904634933 | -0.178850667293736 | 0.253871231390245 | 0.746128768609754 | 0.027590986090084 | VCTK | N-C |
| p247_359_mic1_norm.flac | M | p254_149_mic1_norm_noise.flac | M | Different | -0.195576473192583 | -0.223214948377629 | -0.217160674516599 | 0.251299584424607 | 0.748700415575393 | 0.028187079872146 | VCTK | C-N |
| p276_230_mic1_norm_noise.flac | F | p312_239_mic1_norm.flac | F | Different | -0.130290492418483 | -0.201287082338075 | -0.195921335348835 | 0.359405957217919 | 0.640594042782081 | 0.028273783169172 | VCTK | N-C |
| p304_402_mic1_norm.flac | M | p345_327_mic1_norm_noise.flac | M | Different | -0.210986022562072 | -0.160545833102436 | -0.13565925082424 | 0.378031306153466 | 0.621968693846534 | 0.028694396839447 | VCTK | C-N |
| p240_235_mic1_norm.flac | F | p261_362_mic1_norm_noise.flac | F | Different | -0.18228602234762 | -0.24107269877442 | -0.221856563914844 | 0.238520713552016 | 0.761479286447984 | 0.029066000352775 | VCTK | C-N |
| p227_266_mic1_norm_noise.flac | M | p255_101_mic1_norm.flac | M | Different | -0.0746030792046 | -0.222876042716001 | -0.227302466632859 | 0.350261266042391 | 0.649738733957609 | 0.031239286351037 | VCTK | N-C |
| p231_354_mic1_norm.flac | F | p239_342_mic1_norm_noise.flac | F | Different | -0.088776471202077 | -0.188122181345228 | -0.193975012624143 | 0.403062053247257 | 0.596937946752743 | 0.031516070395324 | VCTK | C-N |
| p269_070_mic1_norm_noise.flac | F | p283_219_mic1_norm.flac | F | Different | -0.138473913261742 | -0.143011670979598 | -0.149541891543664 | 0.435318142822506 | 0.564681857177494 | 0.033413595348122 | VCTK | N-C |
| p278_129_mic1_norm.flac | M | p292_120_mic1_norm_noise.flac | M | Different | -0.121309763430003 | -0.188592843840118 | -0.223119911221515 | 0.33230005627075 | 0.66769994372925 | 0.033669356309403 | VCTK | C-N |
| p268_178_mic1_norm_noise.flac | F | p312_177_mic1_norm.flac | F | Different | -0.135146640972752 | -0.122424951271921 | -0.221251191615052 | 0.384236008680821 | 0.615763991319179 | 0.034235301864863 | VCTK | N-C |
| p308_230_mic1_norm.flac | F | p339_046_mic1_norm_noise.flac | F | Different | -0.113598275755433 | -0.13770290026453 | -0.144853226210684 | 0.466684291913505 | 0.533315708086495 | 0.034290326463962 | VCTK | C-N |
| p228_290_mic1_norm.flac | F | p351_282_mic1_norm_noise.flac | F | Different | -0.185659943179594 | -0.087412317624628 | -0.158881927108374 | 0.429626570610574 | 0.570373429389426 | 0.034604810369208 | VCTK | C-N |
| p292_124_mic1_norm.flac | M | p363_309_mic1_norm_noise.flac | M | Different | -0.095414241292613 | -0.245430472772909 | -0.199929213142817 | 0.32074484419621 | 0.67925515580379 | 0.034620307148863 | VCTK | C-N |
| p239_408_mic1_norm_noise.flac | F | p265_162_mic1_norm.flac | F | Different | -0.142908886298163 | -0.148814609585289 | -0.166150805233384 | 0.39355464106594 | 0.60644535893406 | 0.037142764454306 | VCTK | N-C |
| p300_306_mic1_norm_noise.flac | F | p340_112_mic1_norm.flac | F | Different | -0.217894361381151 | -0.044625506462687 | -0.085674285255125 | 0.498755273868031 | 0.501244726131969 | 0.038262643258252 | VCTK | N-C |
| p313_327_mic1_norm_noise.flac | F | p351_201_mic1_norm.flac | F | Different | -0.11235223349578 | -0.150413416871061 | -0.124937430388247 | 0.454467413772822 | 0.545532586227178 | 0.039457376368023 | VCTK | N-C |
| p228_163_mic1_norm_noise.flac | F | p269_110_mic1_norm.flac | F | Different | -0.111640159490055 | -0.218158095096293 | -0.1427721226385 | 0.366446416905723 | 0.633553583094277 | 0.040245801467357 | VCTK | N-C |
| p244_180_mic1_norm.flac | F | p323_292_mic1_norm_noise.flac | F | Different | -0.248817025565603 | -0.184732150257035 | -0.221841528726421 | 0.17787600965458 | 0.82212399034542 | 0.04168332144909 | VCTK | C-N |

Table 5.2: Stimuli data used for the online behavioral experiment.

Table 5.3: Hyperparameter Tuning with **Hybrid BYOL-S/CvT** model on an ASpD task

| **Tunable Hyperparameters** | | | **Validation Accuracy** |
|---|---|---|---|
| **Learning Rate** | **Batch Size** | **Decoder (MLP) Layers** | % |
| 0.0001 | 64 | [4096] | 86.3 |
| | | [4096,256] | 86.4 |
| | | [4096,256,128] | 86.0 |
| | | [4096,256,128,64] | 85.4 |
| | 128 | [4096] | 84.1 |
| | | [4096,256] | 84.2 |
| | | [4096,256,128] | 83.7 |
| | | [4096,256,128,64] | 86.9 |
| | 256 | [4096] | 85.8 |
| | | [4096,256] | 86.3 |
| | | [4096,256,128] | 84.3 |
| | | [4096,256,128,64] | 85.8 |
| 0.001 | 64 | [4096] | 82.2 |
| | | [4096,256] | 86.5 |
| | | [4096,256,128] | 86.7 |
| | | [4096,256,128,64] | 83.9 |
| | 128 | [4096] | 81.3 |
| | | [4096,256] | 84.3 |
| | | [4096,256,128] | 84.4 |
| | | [4096,256,128,64] | 84.8 |
| | 256 | [4096] | 84.0 |
| | | [4096,256] | 85.1 |
| | | [4096,256,128] | 83.6 |
| | | [4096,256,128,64] | 83.7 |